\documentclass[10pt,aps,pre,twocolumn,superscriptaddress,citesort]{revtex4-2}

\newcommand{\bzq}[1]{\textcolor{black}{#1}}

\usepackage{graphicx}
\usepackage{amsmath}
\usepackage{amsfonts}

\usepackage{bm}
\usepackage{xcolor}

\definecolor{tab_blue}{HTML}{1F77B4}
\usepackage{titlesec}

\usepackage[colorlinks=true,linkcolor=blue,citecolor=blue,urlcolor=blue]{hyperref}

\hypersetup{%
  breaklinks=true,
  colorlinks=true,
  linkcolor=tab_blue,
  filecolor=tab_blue,
  urlcolor=tab_blue,
  citecolor=tab_blue,
  pdfborder=0 0 0,
}
\begin{document}

\title{Stochastic resonance of rotating particles in turbulence}

\author{Ziqi Wang}
\affiliation{Fluids and Flows group and J.M. Burgers Center for Fluid Mechanics, Department of Applied Physics and Science Education, Eindhoven University of Technology, 5600 MB Eindhoven, Netherlands}
\author{Xander M. de Wit}
\affiliation{Fluids and Flows group and J.M. Burgers Center for Fluid Mechanics, Department of Applied Physics and Science Education, Eindhoven University of Technology, 5600 MB Eindhoven, Netherlands}
\author{Roberto Benzi}
\affiliation{Sino-Europe Complex Science Center, School of Mathematics \\North University
of China, Shanxi, Taiyuan 030051, China}
\affiliation{Department of Physics and Istituto Nazionale di Fisica Nucleare, University of Rome Tor Vergata, Rome I-00133, Italy}
\author{Chunlai Wu}
\affiliation{Fluids and Flows group and J.M. Burgers Center for Fluid Mechanics, Department of Applied Physics and Science Education, Eindhoven University of Technology, 5600 MB Eindhoven, Netherlands}
\author{Rudie P. J. Kunnen}
\affiliation{Fluids and Flows group and J.M. Burgers Center for Fluid Mechanics, Department of Applied Physics and Science Education, Eindhoven University of Technology, 5600 MB Eindhoven, Netherlands}
\author{Herman J. H. Clercx}
\affiliation{Fluids and Flows group and J.M. Burgers Center for Fluid Mechanics, Department of Applied Physics and Science Education, Eindhoven University of Technology, 5600 MB Eindhoven, Netherlands}
\author{Federico Toschi}
\email{f.toschi@tue.nl}
\affiliation{Fluids and Flows group and J.M. Burgers Center for Fluid Mechanics, Department of Applied Physics and Science Education, Eindhoven University of Technology, 5600 MB Eindhoven, Netherlands}
\affiliation{Consiglio Nazionale delle Ricerche - Istituto per le Applicazioni del Calcolo, Rome I-00185, Italy}

\date{\today}

\begin{abstract}
The chaotic dynamics of small-scale vorticity plays a key role in understanding and controlling turbulence, with direct implications for energy transfer, mixing, and coherent structure evolution. However, measuring or controlling its dynamics remains a major conceptual and experimental challenge due to its transient and chaotic nature. 
Here we use a combination of experiments, theory and simulations to show that small magnetic particles of different densities, exploring flow regions of distinct vorticity statistics, can act as effective probes for measuring and forcing turbulence at its smallest scale. The interplay between the magnetic torque, from an externally controllable magnetic field, and hydrodynamic stresses, from small-scale turbulent vorticity, uncovers an extremely rich phenomenology. 
Notably, we present the first \bzq{experimental and numerical} observation of stochastic resonance for particles in turbulence, where turbulent fluctuations, remarkably acting as an effective noise, enhance the particle rotational response to external forcing.
We identify a pronounced resonant peak in the particle rotational phase lag when the applied rotating magnetic field matches the characteristic intensity of small-scale turbulent vortices. Furthermore, we reveal a novel symmetry-breaking mechanism: an oscillating magnetic field with zero-mean angular velocity can counterintuitively induce a net particle rotation in turbulence with zero-mean vorticity, as turbulent fluctuations aid the particle in ``surfing'' the magnetic field.
Leveraging this stochastic resonance, our findings pave the way to the development of techniques for flexibly manipulating particle dynamics in complex flows. Moreover, the discovered mechanism introduces the possibility to develop a novel magnetic resonance-based approach for measuring turbulent vorticity, where particles acting as probes emit a detectable magnetic field that can be used to characterize turbulence even under optically inaccessible conditions.
\end{abstract}
\maketitle	

Turbulent flows, characterized by chaotic multi-scale fluctuations, are central to a wide range of natural and industrial processes \cite{frisch1995turbulence, Pope_2000}. A key challenge in turbulence research is the direct measurement and control of the vorticity, which is primarily concentrated in small-scale vortex filaments \cite{ jimenez1998characteristics, sreenivasan1997phenomenology}. Due to their transient nature and rapid chaotic evolution, capturing the rotational dynamics of these filaments remains a major conceptual and experimental challenge. 
Several methods have been developed to infer small-scale turbulence characteristics, including particle tracking methods (e.g., tracer particles, deformable particles, and pattern-coated particles), as well as thermal and optical anemometry \cite{wallace1995measurement, marcus2014measurements, mathai2020bubbly}. 
However, these approaches often suffer from inherent limitations such as insufficient spatial and temporal resolution, signal occlusion in optically dense environments, and difficulties in directly resolving vorticity at the smallest turbulence scales. As a result, achieving real-time, high-fidelity measurements of rotational motion in turbulent flows remains an ongoing challenge, particularly in complex and optically inaccessible flows.

\begin{figure*}[!t]
    \centering
    \includegraphics[width=0.99\textwidth]{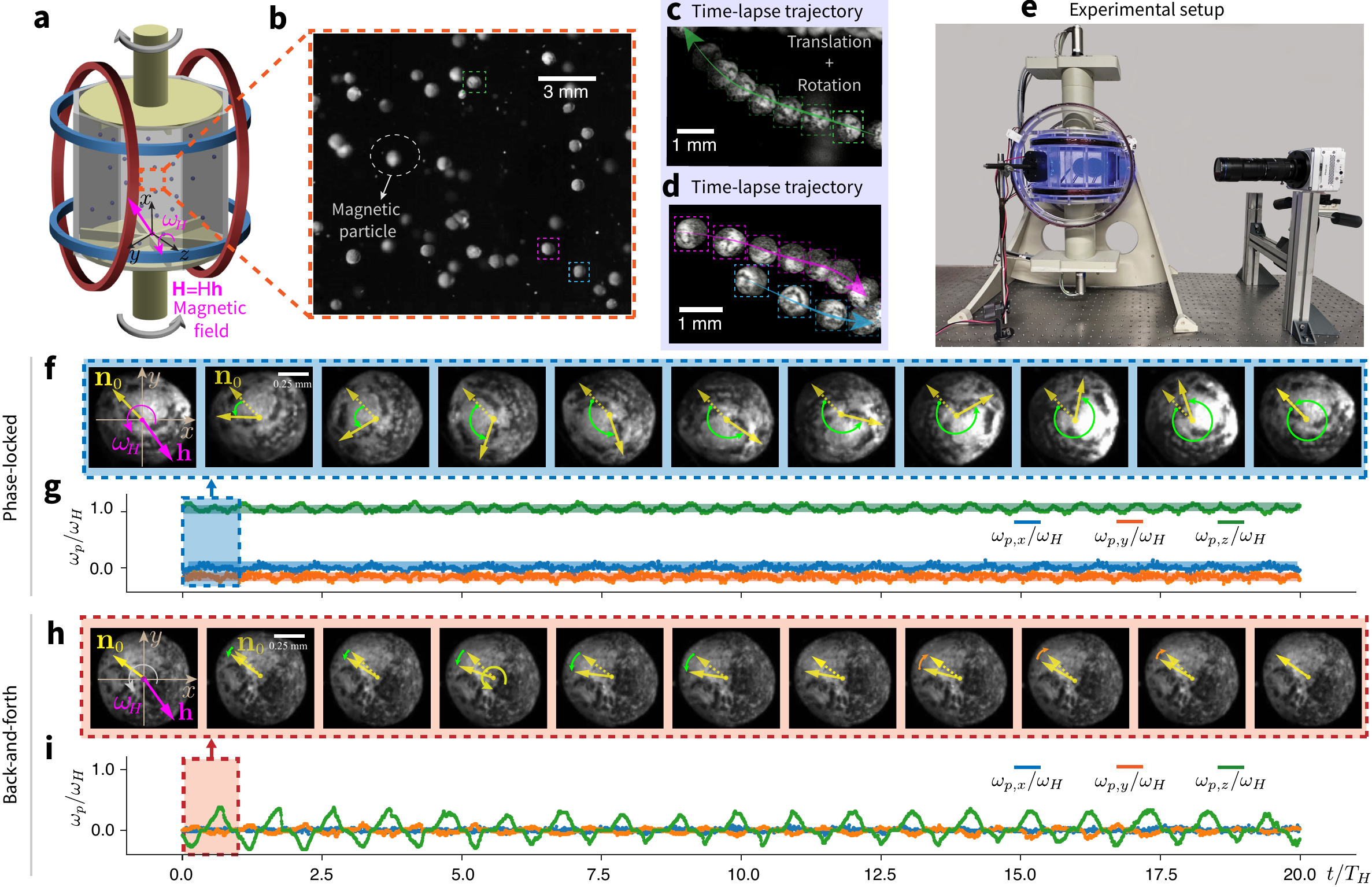}
    \caption{\textbf{Rotational dynamics of magnetic particles in a rotating magnetic field: ``phase-locked'' versus ``back-and-forth'' regimes.} 
    \textbf{a}, Magnetic particles in turbulence under a rotating magnetic field: 
    The experimental setup consists of a turbulence generator, a magnetic field generator, and magnetic particles.
    A Von K\'{a}rm\'{a}n-type turbulent flow is generated in an octagonal water-filled container with an internal diameter of $2R = 150$ mm and a height of 220 mm, driven by two counter-rotating bladed disks. 
    A uniform rotating magnetic field $\bm{H} = H \bm{h}$, with angular frequency $\omega_H$, is generated using a system of two pairs of perpendicular Helmholtz coils. 
    \textbf{b}, Magnetic particles consist of a Styrofoam core coated with magnetic paint. Surface patterns enable tracking rotational motion, as illustrated in the stroboscopic time-lapse trajectories (\textbf{c}-\textbf{d}).
    \textbf{e}, Photograph of the setup. 
    \textbf{f}-\textbf{i}, In a quiescent fluid, particles exhibit two rotational regimes: ``phase-locked'' (low $\omega_H$) and ``back-and-forth'' (high $\omega_H$). \textbf{f} (Supplementary Movie 1) and \textbf{h} (Supplementary Movie 2) show rotational trajectories within one rotation period, viewed in a plane perpendicular to the rotation plane of the magnetic field. 
    The particle initial preferred magnetization direction $\bm{n}_0$ (dashed arrows) and instantaneous orientation $\bm{n}$ (solid arrows) are marked over time.
    During rotation within a quiescent fluid, $\bm{n}$ and $\bm{h}$ remain within the same plane \cite{tierno2009overdamped}. The corresponding normalized particle angular velocity, $\omega_{p,i}/\omega_H$ ($i=x,y,z$), is shown in \textbf{g} and \textbf{i}. Detailed information about the experiments and simulations can be found in Methods. 
    }
    \label{fig:fig1}
\end{figure*}

\begin{figure*}[!t]
    \centering
    \includegraphics[width=0.99\textwidth]{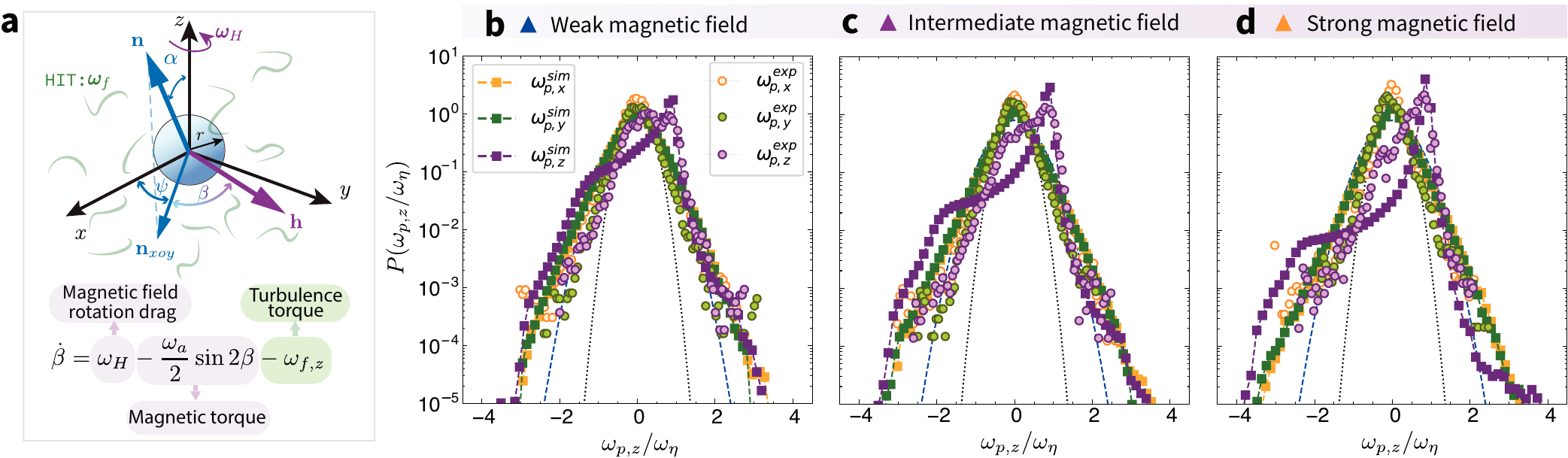}
    \caption{\textbf{Experimental validation of the theoretical model for particle rotational dynamics with turbulence.}
    \textbf{a}, Schematic of a magnetic particle in turbulence under a rotating magnetic field. 
    The particle with preferred magnetization direction $\bm{n}$ is subjected to a magnetic field $\bm{h}$ in the $xoy$-plane, rotating around the $z$-axis with frequency $\omega_H$, and experiences turbulent vorticity $\bm{\omega}_f$ in homogeneous isotropic turbulence. The phase lag $\beta$ (defined as the angle between the projection of $\bm{n}$, i.e., $\bm{n}_{xoy}$, and $\bm{h}$) and the rotation angle $\psi$ (angle of $\bm{n}_{xoy}$ relative to the $x$-axis) describe the particle dynamics.
    The dynamical evolution of $\beta$ is governed by the interplay of the magnetic field rotation drag, magnetic torque, and turbulence torque.
    \textbf{b}-\textbf{d}, Comparison of the probability density distribution function (PDF) of particle angular velocity in experimental (circles) and numerical (squares) studies.
    The experimental results (circles) are shown for varying magnetic field strengths with fixed turbulence intensity.
    \textbf{b} weak (1.2mT, Supplementary Movie 3), \textbf{c} intermediate (1.4mT, Supplementary Movie 4), and \textbf{d} strong (1.6mT, Supplementary Movie 5), with constant rotational frequency of the magnetic field (parameter settings can be found in the Methods and Supplementary Information). 
    }
    \label{fig:fig2}
\end{figure*}

\begin{figure*}[htbp]
    \centering
    \includegraphics[width=0.99\textwidth]{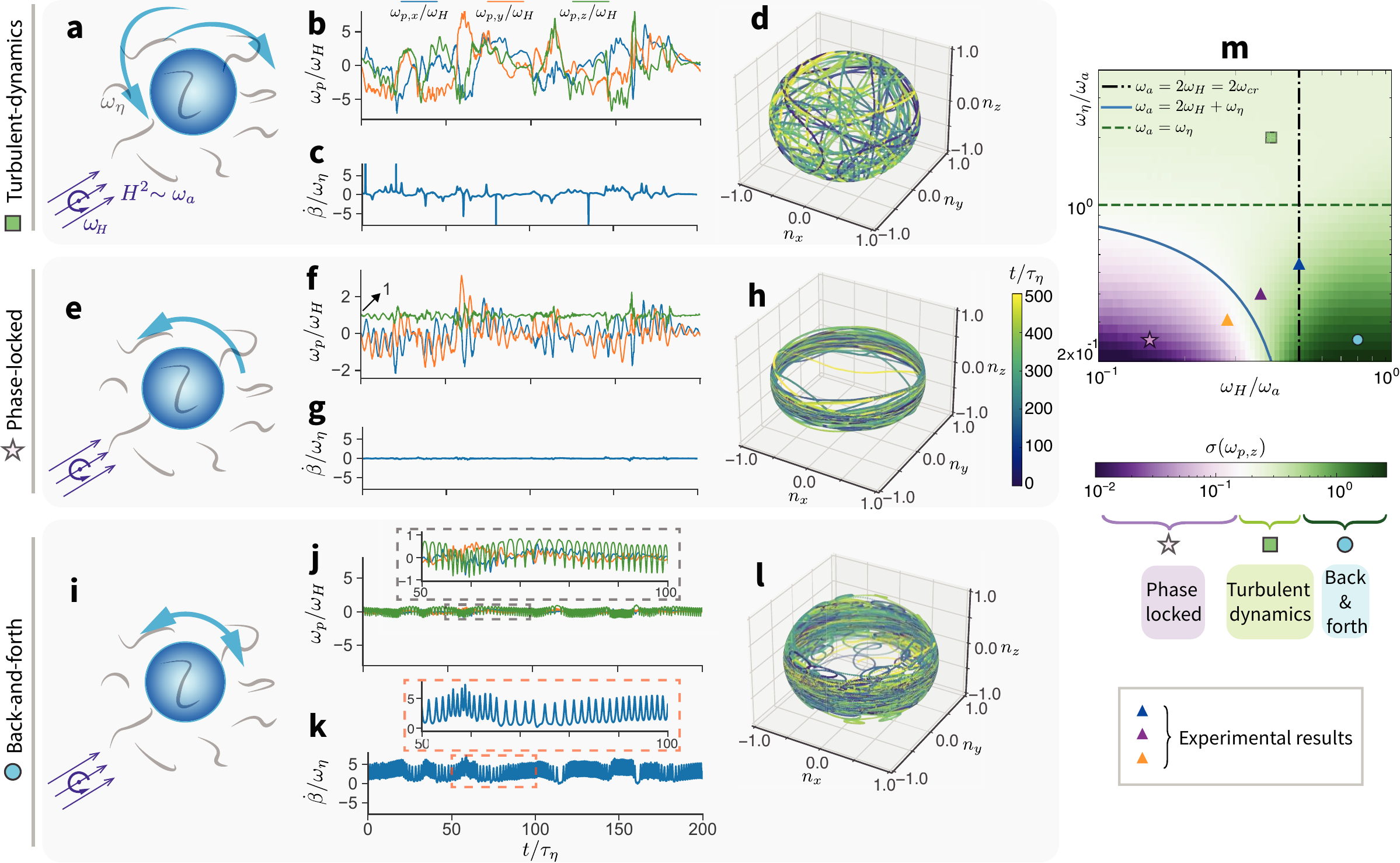}
    \caption{\textbf{Three distinct regimes of magnetic particle rotation dynamics in turbulence under a rotating magnetic field.}
    \textbf{a}, Turbulence-dominated regime ($\omega_a \ll \omega_\eta$).
    \textbf{b}, The normalized particle angular velocity, $\omega_{p}/\omega_H$, exhibits turbulent fluctuations with a zero average and the phase lag derivative, $\dot{\beta}$ is randomized (\textbf{c}).
    \textbf{d}, The evolution of the tip of preferred magnetization direction of the particle, $\bm{n}$, is visualized in space, with color indicating time progression (Supplementary Movie 6). The random distribution of orientations confirms the turbulent-dominated nature.  
    \textbf{e}, ``Phase-locked'' regime ($\omega_a \gg \omega_\eta$, $\omega_H<\omega_{cr}$): The angular velocity of the particle along the magnetic field rotation axis ($z$-axis) is locked to $\omega_H$ (\textbf{f}, green line), with zero net drift for other components (\textbf{f}, blue and orange lines).
    The phase lag derivative $\dot{\beta}$ is effectively zero (\textbf{g}) and the particle orientation follows a 2D trajectory (\textbf{h}, Supplementary Movie 7).
    \textbf{i}, ``Back-and-forth'' regime ($\omega_a \gg \omega_\eta$, $\omega_H>\omega_{cr}$): 
    The angular velocity of the particle along the $z$-axis exhibits periodic acceleration and deceleration oscillations (\textbf{j}, inset), and $\dot{\beta}$ varies similarly (\textbf{k}, inset).
    The orientation vector oscillates periodically (back-and-forth looping in \textbf{l}, Supplementary Movie 8).
    \textbf{m}, Phase diagram of particle rotation regime, colored by the normalized variance $\sigma(\omega_{p,z})$ of the particle angular velocity along the $z$-direction, $\sigma(\omega_{p,z})=\langle \omega_{p,z}^2 - \langle \omega_{p,z}\rangle^2\rangle/\omega_\eta^2$.
    Black dot-dashed line: the critical frequency, $\omega_{cr} = \omega_a/2$. 
    Green dashed line: the condition where the turbulence fluctuation intensity is balanced by the magnetic strength, i.e., $\omega_a = \omega_\eta$. 
    The phase-locked regime is bounded with the blue thick line, predicted by the scaling analysis, i.e., $\omega_a = 2\omega_H + \omega_\eta$. 
    Symbols (square, star, and circle): representative cases from the three distinct regimes shown in earlier panels.
    Triangle markers: experimental settings from Fig.~\ref{fig:fig2}\textbf{b}-\textbf{d}.
    % \bzq{The diamond symbols denote the experimental results at $\omega_H/\omega_a = 0.12 \pm 0.01$, providing evidence of stochastic resonance, which will be discussed in Fig.~\ref{fig:fig4}\textbf{a} and (\textbf{c}).}
    }
    \label{fig:fig3}
\end{figure*}

\begin{figure*}[htbp]
    \centering
    \includegraphics[width=0.99\textwidth]{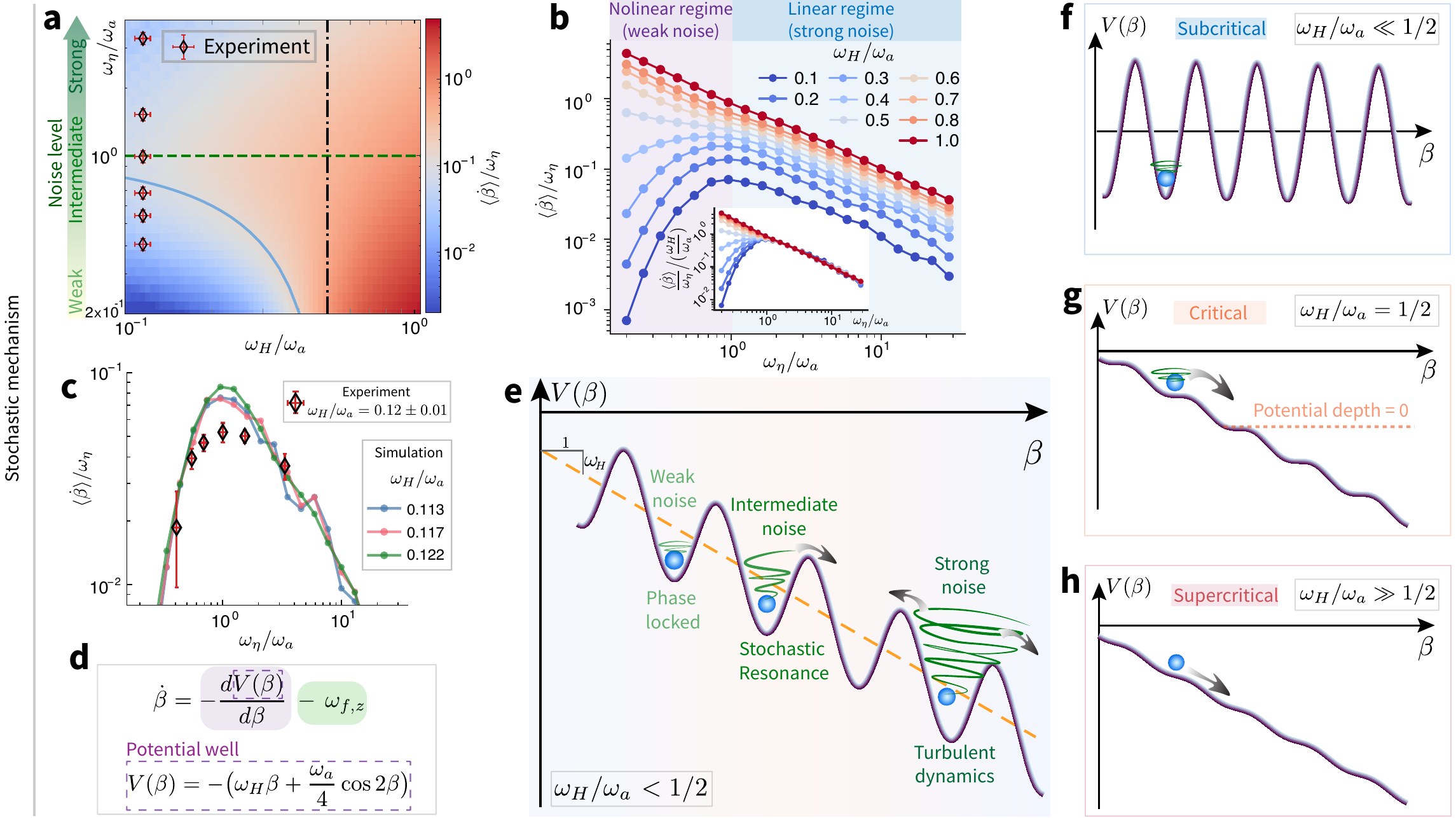}
    \caption{\textbf{Stochastic resonance.} 
    \textbf{a}, Phase diagram of stochastic resonance. The color represents the normalized time-averaged phase lag derivative, $\langle \dot{\beta}\rangle/\omega_\eta$. 
    The vertical axis, $\omega_\eta/\omega_a$, quantifies the relative intensity of turbulent fluctuations (noise) to the magnetic field strength.
    The horizontal axis, $\omega_H/\omega_a$, distinguishes subcritical, critical ($\omega_H/\omega_a=1/2$, marked by the black dot-dashed line) and supercritical regimes. 
    For a fixed $\omega_H/\omega_a$ in the subcritical (supercritical) range, $\langle \dot{\beta}\rangle/\omega_\eta$ exhibits a non-monotonic (monotonic) dependence on noise. 
    A pronounced resonant peak appears at $\omega_\eta/\omega_a \approx 1$ when $\omega_H/\omega_a<1/2$ (\textbf{b}), with a perfect collapse in the linear regime when normalized by $\omega_H/\omega_a$ (\textbf{b}, inset).
    \bzq{The diamond symbols in panel \textbf{a} denote the parameter settings for experiments shown in panel \textbf{c}. 
    % providing evidence of stochastic resonance, as shown in \textbf{c}.
    }
    % =============================================================
    \bzq{\textbf{c}, Experimental evidence of stochastic resonance (diamonds) at $\omega_H/\omega_a = 0.12 \pm 0.01$. For comparison, simulation results at a similar value of $\omega_H/\omega_a$ (circles) are also shown. The error bars correspond to the standard deviation calculated from 20 independent measurements at the same parameter value.}
    % =============================================================
    \textbf{d}, Effective potential well framework: the magnetic field contributes a cosine potential modulated by a linear term, and turbulence acts as a stochastic noise, see Eq.~\eqref{eq:main_beta_eqn}. 
    When $\omega_H/\omega_a<1/2$ (\textbf{e}), for weak noise, the phase lag remains near the local minimum of the potential well. As noise increases, the particle phase lag exhibits unidirectional escape, signaling stochastic resonance. At even higher noise levels, escape becomes bidirectional, leading to a subsequent decline in $\langle\dot{\beta}\rangle$.
    For $\omega_H/\omega_a \ll 1/2$, the potential well is sufficiently deep to strongly localize the particle phase lag (\textbf{f}). 
    At the critical threshold $\omega_H/\omega_a = 1/2$, the potential well vanishes, making the phase lag sensitive to small perturbations (\textbf{g}). In the supercritical range ($\omega_H/\omega_a \gg 1/2$), the potential well inherently favors persistent phase slipping, regardless of noise intensity (\textbf{h}).
    }
    \label{fig:fig4}
\end{figure*}

\begin{figure*}[htbp]
    \centering
    \includegraphics[width=0.99\textwidth]{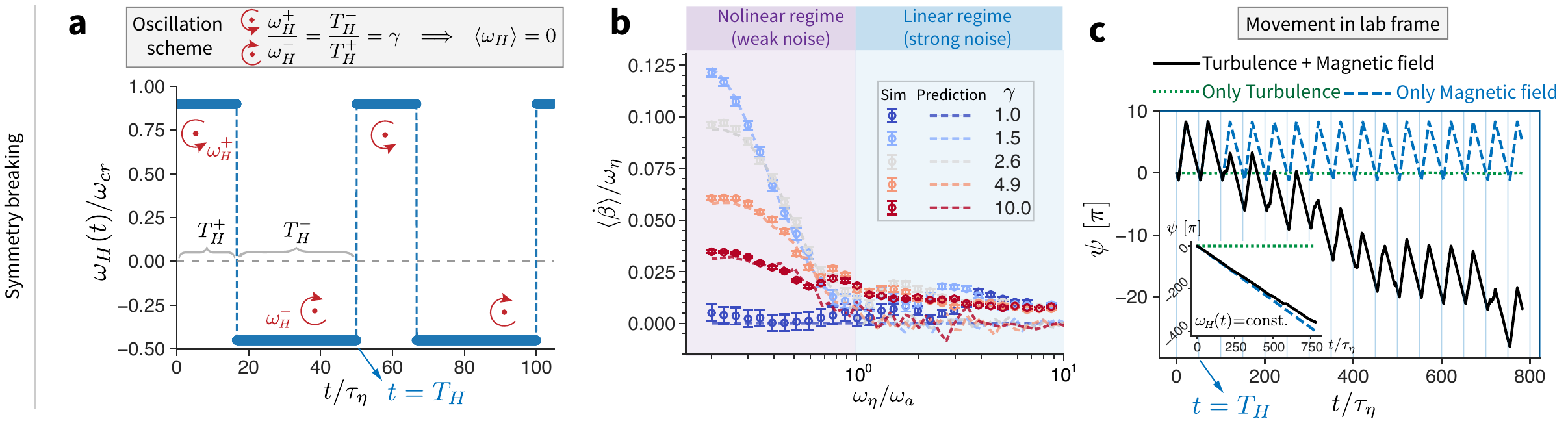}
    \caption{\textbf{Symmetry breaking.}
    \textbf{a}, The applied magnetic field oscillates with zero-average angular velocity, $\omega_H(t)$.
    Without loss of generality, we set $\omega_H^{+}/\omega_H^{-} = T_H^{-}/T_H^{+} = \gamma \ge 1$ and $\omega_H^{+}=0.9 \omega_{cr}$.
    \textbf{b}, The normalized phase lag derivative, $\langle \dot{\beta}\rangle/\omega_\eta$, shows distinct regimes as noise intensity $\omega_\eta/\omega_a$ increases for fixed $\gamma$. In the nonlinear regime (low noise), $\langle \dot{\beta}\rangle/\omega_\eta$ decreases monotonically, while in the linear regime (strong noise, where system responses collapse, see Fig.~\ref{fig:fig4}b, inset), it remains nearly unchanged and close to zero. When $\gamma=1$, the particle rotation motions cancel each other out in $T_H^{-}$ and $T_H^{+}$, resulting in a zero response. The observed deviation from a strictly zero value in the simulations is attributable to the finite periods of the applied oscillation cycles of $\omega_H(t)$. Theoretically, with an infinite simulation time, the value would converge to zero.
    The error bars represent the standard deviation of the results obtained by sampling over one full period of $\omega_H(t)$ at the final stage of the simulation.
    The predicted trends by the superposition model of Eq.~\eqref{eq:superpositionmodel} (\textbf{b}, dashed lines) exhibit excellent agreement with the numerical data (\textbf{b}, symbols).
    \textbf{c}, Particle motion in laboratory frame. With turbulence alone (green dotted line), the particle shows no net angular movement in terms of $\psi$; with only the magnetic field (blue dashed line), the motion is periodic with zero net angular displacement. When both effects act together, a net angular movement emerges (black solid line), revealing symmetry-breaking mechanism. This is different from the case of constant $\omega_H$ (inset), where the particle follows $\psi = \omega_H t$ (black solid line) when it is purely driven by a magnetic field. Parameter settings are $T_H = 50 \tau_\eta \approx T_L$ with $T_L$ the integral time scale of turbulence, $\omega_\eta/\omega_a = 0.25$, and $\gamma = 2$. Inset: $\omega_\eta/\omega_a = 0.25$ and $\omega_H/\omega_a = 0.4$.
    }
    \label{fig:fig5}
\end{figure*}

A promising alternative is the use of magnetic particles, which respond to both stochastic vorticity fluctuations of turbulence (hydrodynamic torques \cite{guazzelli2011physical}) and deterministic torques imposed by external magnetic fields.  
The dynamics of buoyant magnetic particles, in particular, are closely linked to the local vorticity. Particles with different densities exhibit distinct rotational behaviors as they explore different regions of the flow, leading to preferential concentration: light (heavy) particles tend to accumulate in high- (low-) vorticity regions \cite{mathai2020bubbly, calzavarini2008dimensionality, wang2024localization}.
These particles offer a unique dual role: they can serve as candidates for probing rotational fluctuations, while also enabling active flow manipulation through external forcing. This ability to modulate particle rotation and measure its response in real time provides a new pathway for investigating turbulence at small scales and suggests an elegant strategy for measuring turbulent vorticity. 
However, in order to unlock these novel directions, an accurate understanding of the complex interplay between hydrodynamic and magnetic torques on particle dynamics is required.

Previous studies have extensively explored magnetic particles in quiescent fluids under rotating magnetic fields, revealing rich phenomena such as self-assembly into chain-like structures \cite{martin2009theory,shanko2019microfluidic}, synchronization-selected structures \cite{yan2012linking}, and more complex arrays \cite{piet2013viscosity}, enhanced mixing  \cite{martin2009theory, wittbracht2012enhanced, shanko2019microfluidic}, \bzq{turbulence design \cite{falcon2017dissipated,cazaubiel2021three,gorce2024freely}} and phase locking, where particles follow the rotating magnetic field with a constant phase lag with respect to the magnetic field when the rotating magnetic angular velocity $\omega_H$ is below a threshold $\omega_{cr}$ \cite{cimurs2019stability}. 
Investigations have also extended to self-propelling active magnetic particles (e.g., magnetotactic bacteria) \cite{cebers2006dynamics,erglis2007dynamics}, magnetic Janus particles \cite{yan2012linking,Sinn2011Magnetically}, 
particles with different aspect ratios \cite{cimurs2013dynamics,morozov2017dynamics}, \bzq{colloidal spinners and active swimmers \cite{francois2014three,kokot2017active,bourgoin2020kolmogorovian}}. 

The competition and interplay between effects of turbulence and magnetic fields may lead to complex particle dynamics, yet to be understood, presenting a fundamental challenge in predicting and controlling particle behavior in turbulent environments.

Here, we explore this question by investigating the rotational dynamics of magnetic particles in turbulence under a rotating magnetic field. 

While deterministic synchronized and oscillation (back-and-forth) rotational regimes have been observed in zero-noise systems (e.g., compasses or rolling particles in quiescent fluids \cite{croquette1981cascade, poye2018bipolar, kaiser2017flocking}), our work extends this understanding into the complex environment of fully developed turbulence. We systematically explore the complex interplay between deterministic external magnetic forcing and turbulent hydrodynamic torques, revealing a significantly richer phase space of particle rotational dynamics. 
We show that the effect of turbulence (vorticity) acts as an effective noise (Supplementary Information) and reveal the first signature of the emergence of stochastic resonance (SR)\textemdash{}a counterintuitive phenomenon where internal fluctuations non-linearly cooperate with external forcing to amplify the system response \cite{benzi1981mechanism,nicolis1982stochastic,gammaitoni1998stochastic,simonotto1997visual,benzi2010stochastic,budrikis2021forty,lu2015effects,wu2024nonlinearity,peters2021extremely}\textemdash{}of particles in turbulence, which can be used as the mechanism of magnetic particles acting as effective vorticity probes for a novel type of turbulence ``microscope''.
While the concept of SR, initially observed in bistable systems subjected to noise and weak periodic signals, has since been extended to a wide range of physical and biological systems, as well as signal processing and computer science \cite{douglass1993noise,braun1994oscillation,wiesenfeld1995stochastic,mcdonnell2009stochastic,mcdonnell2011benefits,pisarchik2023coherence}, its role in turbulence-driven rotational dynamics remains mostly unexplored. 

To identify potential SR signatures, we set out to leverage turbulent vorticity fluctuations as a source of noise and examine the response of the magnetic particle dynamics to the rotating magnetic field, by means of experiments, simulations, and theoretical modeling.
Our experimental platform (Fig.~\ref{fig:fig1}a and Fig.~\ref{fig:fig1}e) involves a dilute collection of approximately spherical light magnetic particles (produced by Styrofoam core coated with magnetic paint, Fig.~\ref{fig:fig1}b, with a mean density of 20\% with respect to water and with a mean diameter of 0.7 mm, see Methods for details) which are immersed in a Von K\'{a}rm\'{a}n-type turbulent flow (producing homogeneous and isotropic turbulence (HIT) in the center region of the cell  \cite{mordant2004experimental,volk2008acceleration,voth2002measurement,volk2011dynamics}) in an octagonal water-filled chamber. The flow is generated by two counter-rotating bladed disks and the chamber is placed in a system of Helmholtz coils, which is programmed to generate a uniform magnetic field in a restricted measurement volume, $\bm{H} = H\bm{h}$, rotating in the $xoy$-plane with angular velocity vector $\bm{\omega}_H = \omega_H \hat{\bm{z}}$ where $\bm{\hat{z}}$ is the unit vector along the $z$-axis (Methods and Supplementary Information).
As the induced magnetic dipole, $\bm{m}$, of a particle tends to align with the magnetic field direction, $\bm{h}$, the particle experiences a magnetic torque balanced by the rotational drag and spins around an axis in the particle body frame parallel to $\hat{\bm{z}}$ along its trajectory (as shown by typical stroboscopic time-lapse trajectories in Fig.~\ref{fig:fig1}c-d). The rotational motion is measured by tracking particle surface patterns (Methods).

\subsection*{Theoretical representation}

The particle rotation is characterized by a preferred magnetization direction, $\bm{n}$ (Fig.~\ref{fig:fig2}a), whose trajectory can be tracked in turbulence under the influence of a rotating magnetic field.
By balancing the magnetic torque and drag torque induced by turbulence \cite{cimurs2019stability,tierno2009overdamped} (Methods and Supplementary Information), the equations governing the particle rotational motion can be obtained, in the overdamped limit, as
\begin{subequations}\label{eq:full}
\begin{align}
    \frac{d\bm{n}}{d t} &= \omega_a (\bm{n} \cdot \bm{h}) \left( \bm{h} - (\bm{n} \cdot \bm{h})\bm{n} \right) + \bm{\omega}_f \times \bm{n},\\
    \frac{d \left(\xi_{r}\bm{\omega}_{p,\parallel}\right)}{d t} &=  -\xi_{r} \left(\bm{\omega}_{p, \parallel} - \left(\bm{\omega}_f \cdot \bm{n}\right)\bm{n}\right),
\end{align}
\end{subequations}
where $\omega_a = \mu \mathcal{V}_p H^2 (\chi_\parallel-\chi_\bot)/\xi_{r}$ is the magnetic field strength normalized by the rotational drag coefficient $\xi_{r}$, with $\mu$, $\mathcal{V}_p$, and $\chi_{{\parallel},{\bot}}$ the particle permeability, volume, and anisotropic magnetic susceptibilities, respectively \cite{osborn1945demagnetizing,cimurs2019stability, tierno2009overdamped} (Methods).
The term $\bm{\omega}_{p,\parallel}$ represents the particle spinning rate. The perpendicular rotational component, $\bm{\omega}_{p,\bot}$, can be obtained by $\bm{\omega}_{p,\bot} = \bm{n} \times \dot{\bm{n}}$ which represents the particle tumbling rate. 
The turbulent vorticity experienced by the particle is $\bm{\omega}_{f}$, which is defined as $\bm{\omega}_f = \frac{1}{2}[\bm{\nabla} \times \bm{u}_f]_O$ with the subscript \textit{O} implying evaluation at the center of mass of the particle.
The total particle angular velocity is thus $\bm{\omega}_p = \bm{\omega}_{p,\parallel} +  \bm{\omega}_{p,\bot}$.

In a noiseless system, i.e., $\bm{\omega}_f=\bm{0}$, the critical frequency is $\omega_{cr}=\omega_a/2$ (details in Supplementary Information). For $\omega_H\leq\omega_{cr}$, the particle is synchronized with the rotating magnetic field (``phase-locked''), i.e. $\omega_{p,z}=\omega_H$. Otherwise, when $\omega_H > \omega_{cr}$, the particle undergoes back-and-forth motion (``back-and-forth'') when the maximum magnetic torque cannot balance the response drag torque: the phase lag angle $\beta$ between $\bm{n}_{xoy}$ (the projection of $\bm{n}$ on $xoy$-plane) and $\bm{h}$ changes periodically over time \cite{cebers2006dynamics}. These two regimes can be observed experimentally (Fig.~\ref{fig:fig1}f-i).
In the ``phase-locked'' regime (Supplementary Movie 1), the particle rotates synchronously with the magnetic field, as evidenced by the normalized angular velocity, $\omega_{p,z}/\omega_{H} \approx 1$, in Fig.~\ref{fig:fig1}g. In the ``back-and-forth'' regime (Supplementary Movie 2), the particle oscillates periodically (Fig.~\ref{fig:fig1}i).

As illustrated in the schematic diagram of Fig.~\ref{fig:fig2}a, the dynamics of the particle preferred magnetization direction, $\bm{n}$, is fundamentally characterized by the phase lag angle, $\beta$, with respect to $\bm{h}$ on the $xoy$ plane. The governing equations, Eqs.~\eqref{eq:full}, can be simplified and explicitly written as (derivations in Methods)
\begin{align}\label{eq:main_beta_eqn}
    \dot{\beta} & = \omega_H - \frac{\omega_{a}}{2} \sin{2 \beta} - \omega_{f,z}.
\end{align}
Geometrically, we have $\beta = \omega_H t-\psi$. The phase lag angle $\beta$ represents the particle rotational displacement in the co-rotating frame of the magnetic field, while $\psi$ is the rotation angle of the projection $\bm{n}$ on the $xoy$-plane with respect to the $x$-axis, which describes the particle rotational displacement in the laboratory frame. The turbulent vorticity signal, $\bm{\omega}_f$, represents the vorticity experienced by the particle center along its trajectory. This simplified representation, Eq.~\eqref{eq:main_beta_eqn}, separates different contributions and clearly shows that the rotational motion of the particle results from the interplay between the magnetic field rotational drag, the magnetic torque, and the turbulence drag torque (Fig.~\ref{fig:fig2}a).

\subsection*{Switching on turbulence}
Now we wonder how this picture will change when turbulence is taken into consideration.
The turbulence induces fluctuations on the small particles of a typical magnitude $\omega_\eta = 1/\tau_\eta$, the Kolmogorov characteristic frequency.
Experimentally, we turn on the turbulence and vary the intensity of the magnetic field. By statistically analyzing the probability density function (PDF) of the particle angular velocity $\bm{\omega}_p$, a dominant peak is observed in the $z$-direction at approximately $\omega_{p,z} \approx \omega_{H}$ (Fig.~\ref{fig:fig2}b-d, purple circles), while the angular velocities in the $x$ and $y$ directions are similar to each other but significantly different from that in the $z$-direction (Fig.~\ref{fig:fig2}b-d, yellow and green circles). 

To understand the influence of turbulence on the particle dynamics further, we carry out a systematic numerical investigation. To model the motion of light particles in HIT, we first perform direct numerical simulations using a pseudo-spectral method \cite{peyret2002spectral, bec2006acceleration, wang2024localization} to obtain $\bm{\omega}_f$ (see Methods for details). The resulting vorticity data is then incorporated into the theoretical model, Eqs.~\eqref{eq:full}, which is integrated using a third-order Runge-Kutta method. 

The simulated particle rotation is validated against the experimental results (Fig.~\ref{fig:fig2}b-d, Supplementary Movies 3-5, respectively). Notably, the theoretical model (squares) accurately captures the essential statistical physics (peak positions) of the particle stochastic rotational dynamics in the experiments (circles).
The slight discrepancies in the PDF shapes arise from polydispersity in particle shape and magnetic properties in experiments which influences how individual particles respond to an external magnetic field, whereas the simulations assume idealized spherical particles with the same magnetic characteristics for each particle (Methods and Supplementary Information). Further comparisons between simulation and experimental results all demonstrate good qualitative agreement (Supplementary Information, Supplementary Movies 9 and 10).

Additionally, to gain deeper insight into the stochastic nature of particle rotation in turbulence, we analyze the waiting time, $\tau$\textemdash{}the duration between successive transitions of $\beta$ between its metastable states\textemdash{}to characterize the nature of turbulence-induced noise in the absence of external forcing (i.e., $\omega_H = 0$). The Poisson-like distribution of $\tau$ confirms that, within the investigated parameter space, stochastic transitions induced by turbulent (vorticity) fluctuations can be effectively modeled using white noise with an equivalent intensity (Supplementary Information). 

\subsection*{Three distinct regimes of particle dynamics}
With the theoretical model capturing the core dynamics of this complex system, and with simulations allowing exploration of a broader parameter space and finer details of particle rotation, we systematically investigate how turbulence-induced vorticity fluctuations and the deterministic magnetic forcing shape particle rotational dynamics.

Beyond the previously identified ``phase-locked'' and ``back-and-forth'' regimes, we uncover a third regime: the turbulence-dominated regime, observed at small $\omega_a$ w.r.t. $\omega_\eta$, when the effect from the magnetic field is negligible. Here, the particle rotation is primarily driven by turbulence, and its normalized angular velocity, $\omega_{p}/\omega_H$, fluctuates randomly around zero with no net angular drift (Fig.~\ref{fig:fig3}b), as evidenced by the random orientation distribution of the particle orientation, $\bm{n}$ (Fig.~\ref{fig:fig3}d, Supplementary Movie 6). Consequently, the phase lag $\beta$ and its derivative are also randomized (Fig. \ref{fig:fig3}c).
In the ``phase-locked'' regime ($\omega_a \gg \omega_\eta$, $\omega_H<\omega_{cr}$), the particle phase lag, $\beta$, remains constant, meaning $\dot{\beta}=0$ (Fig.~\ref{fig:fig3}g). The particle angular velocity along the magnetic field rotation axis ($z$-axis) is locked to $\omega_H$ (Fig.~\ref{fig:fig3}f, green line).
The particle orientation is strongly confined, exhibiting 2D behavior (Fig.~\ref{fig:fig3}h, Supplementary Movie 7).
In the ``back-and-forth'' regime ($\omega_a \gg \omega_\eta$, $\omega_H>\omega_{cr}$), the particle undergoes periodic acceleration and deceleration (Fig.~\ref{fig:fig3}j-k, inset), forming distinct looping patterns in its orientation trajectory (Fig.~\ref{fig:fig3}l, Supplementary Movie 8).
These regimes are summarized in a phase diagram, characterized by the normalized variance of the particle angular velocity in the $z$-direction, $\langle \omega_{p,z}^2 - \langle \omega_{p,z}\rangle^2\rangle/\omega_\eta^2$, where $\langle \cdot\rangle$ represents a time average. The noiseless critical frequency $\omega_{cr} = \omega_a/2$ is marked by a black dot-dashed line. Unlike in a quiescent fluid, the phase-locked regime (Fig.~\ref{fig:fig3}m, purple region) is bounded by the blue line, predicted by the scaling argument based on the simplified governing equation of Eq.~\eqref{eq:main_beta_eqn}: $\omega_a = 2\omega_H + \omega_\eta$ (detailed reasoning can be found in Methods).
This shift in the subcritical boundary, relative to the noiseless case ($\omega_a = 2\omega_H$), accounts for the influence of turbulence fluctuations, which have a standard deviation on the order of $\omega_\eta$. 

\subsection*{Stochastic resonance}
To determine whether SR occurs, we focus on the $\beta$ and its behavior across different parameter settings. 
We summarize the time-averaged value of $\langle \dot{\beta}\rangle/\omega_\eta$ in one phase diagram (Fig.~\ref{fig:fig4}a), with $\omega_\eta/\omega_a$ (vertical axis) quantifying the relative turbulent fluctuation intensity (noise) to the applied magnetic field strength, and $\omega_H/\omega_a$ (horizontal axis) distinguishing noiseless subcritical, critical (i.e., $\omega_H/\omega_a=1/2$, black dot-dashed line), and supercritical states. 

For a fixed $\omega_H/\omega_a$ in the subcritical regime ($\omega_H/\omega_a<1/2$), the particle rotation in the co-rotating frame $\langle \dot{\beta}\rangle/\omega_\eta$ exhibits a non-monotonic dependence on noise intensity (Fig.~\ref{fig:fig4}b), with a pronounced resonant peak at $\omega_\eta/\omega_a \approx 1$, indicating that resonance arises at the transition from the ``phase-locked'' regime to the ``turbulent-dynamics'' regime. This is a clear suggestion that the observed resonance behavior results from the cooperation between turbulent fluctuations and magnetic forcing.
\bzq{Stochastic resonance is also clearly observed experimentally, as shown in Fig.\ref{fig:fig4}c: diamonds denote the experimental results at $\omega_H/\omega_a = 0.12 \pm 0.01$ (corresponding to the points marked in the phase diagram in Fig.~\ref{fig:fig4}a), while circles represent the simulation results at a comparable value of $\omega_H/\omega_a$ for direct comparison. The experimental and simulation results agree reasonably well; at the resonance peak, the experimental curve appears flatter which can be attributed to effects of polydispersity of the particles used in experiments.}
The resonance mechanism can be interpreted via an effective potential well framework (Fig.~\ref{fig:fig4}d), where the magnetic field induces a cosine-type potential (i.e., $-\frac{\omega_a}{4}\cos{2\beta}$) modulated by a linear term (i.e., $-\omega_H\beta$), while turbulence acts as a stochastic noise. 

For $\omega_H/\omega_a<1/2$ (Fig.~\ref{fig:fig4}e), the phase lag angle is initially trapped in the potential minimum at weak noise levels (small $\omega_\eta/\omega_a$), exhibiting only mild fluctuations locally. 
As the turbulence-induced fluctuations become comparable to the magnetic field strength ($\omega_\eta/\omega_a \approx 1$), preferential unidirectional escape occurs. This clearly identifies the observed resonant behavior as an example of stochastic resonance: a cooperative interplay between stochastic forcing and the external deterministic forcing. This phenomenon can be thought of as the particle ``surfing'' the rotating magnetic field, aided by the turbulent vorticity.
At even higher noise levels, randomization overrides the cooperative effect, causing bidirectional escape and a decline in $\langle\dot{\beta}\rangle$. 

%For $\omega_H/\omega_a \ll 1/2$, the deep potential well strongly confines $\beta$ (Fig.~\ref{fig:fig4}e). 
At the critical threshold $\omega_H/\omega_a = 1/2$, the potential well vanishes, rendering $\beta$ critically sensitive to any perturbations (Fig.~\ref{fig:fig4}g). In the supercritical range ($\omega_H/\omega_a \gg 1/2$), the potential well inherently favors persistent phase slipping, regardless of noise intensity (Fig.~\ref{fig:fig4}h), such that the SR mechanism vanishes in this regime.

In strongly subcritical regime ($\omega_H/\omega_a \ll 1/2$), the resonance peak is purely set by the turbulent noise, making it sensitive to the preferential sampling effect \cite{mathai2020bubbly, calzavarini2008dimensionality, wang2024localization} (see Supplementary Information for neutral and heavy particles results).
Here we introduce the relaxation time $\tau^* = 1/\left(\omega_a \sqrt{1-4\big(\frac{\omega_H}{\omega_a}\big)^2} \right)$. Physically, $\tau^*$ represents the characteristic time it takes for a perturbed particle rotational motion to return to its phase-locked state, i.e., $\dot{\beta}=0$, after a small disturbance (detailed derivations are in Supplementary Information).
Close to the critical ratio $\omega_H/\omega_a=1/2$, the significantly increasing particle relaxation time $\tau^*$ shifts the resonance position, making it the dominant factor. 
Therefore, a longer $\tau^*$ allows particles to escape the potential well even at lower noise levels.

\subsection*{Symmetry breaking mechanism}
A striking consequence of SR is the symmetry-breaking effect, manifesting as a ``zero plus zero is greater than zero'' phenomenon.
This leverages the fact that in the weak noise regime ($\omega_\eta < \omega_a$), the response of the system to the noise is non-linear (Fig.~\ref{fig:fig4}b, purple-shaded part).

To reveal this effect, instead of applying a constant-frequency rotating magnetic field ($\omega_H$), we now impose a time-dependent field rotation with a zero-average, $\omega_H(t)$ (Fig.~\ref{fig:fig5}a). Each oscillation cycle consists of a forward and a reverse rotation phase, characterized by angular velocities $\omega_H^{+}$ and $\omega_H^{-}$ with corresponding durations $T_H^{+}$ and $T_H^{-}$. These parameters are scaled as $\omega_H^{+}/\omega_H^{-} = T_H^{-}/T_H^{+} = \gamma$, ensuring that the net angular velocity over a full cycle remains zero. 
%This setup provides a framework to explore the nonlinear response to noise (Fig.~\ref{fig:fig4}b, the purple-shaded part).
Two distinct response regimes emerge as $\omega_\eta/\omega_a$ increases for a fixed $\gamma$ (Fig.~\ref{fig:fig5}b), highlighting the ability of this oscillation framework to probe the nonlinearity of the system in response to noise. Specifically, the response $\langle \dot{\beta}\rangle/\omega_\eta$ decreases monotonically in the nonlinear regime (weak noise limit), while it remains nearly constant and close to zero in the linear regime (strong noise limit, characterized by response collapse, see also Fig.~\ref{fig:fig4}b, inset).
This behavior aligns well with a superposition model which predicts the response as 
\begin{equation}
\label{eq:superpositionmodel}
    \langle\dot{\beta}\rangle = \frac{\langle\dot{\beta}(\omega_H^{+})\rangle T_H^{+} - \langle\dot{\beta}(\omega_H^{-})\rangle T_H^{-}}{T_H},
\end{equation}
with $T_H = T_H^{+}+T_H^{-}$, where $\langle\dot{\beta}(\omega_H^{+})\rangle$ and $\langle\dot{\beta}(\omega_H^{-})\rangle$ are interpolated from the numerical results in Fig.~\ref{fig:fig4}b. The predicted trends (Fig.~\ref{fig:fig5}b, dashed lines) show excellent agreement with the frequency-modulated field results (Fig.~\ref{fig:fig5}b, symbols).

We emphasize that when combining turbulence and the frequency-modulated magnetic field in this way, this leads to particle motion directly in the lab frame. To evidence this, we consider the lab-frame angular particle motion $\psi(t)$ (Fig.~\ref{fig:fig5}c). With turbulence alone, the particle exhibits no net angular movement (green dotted line). Under a purely frequency-oscillatory magnetic field, the particle undergoes periodic motion with zero net angular displacement (blue dashed line). However, when both effects act concurrently, a net, non-zero mean angular drift remarkably emerges (black solid line), as the turbulent vorticity aids the particle in ``surfing'' the magnetic field, which reveals an unexpected symmetry-breaking mechanism in turbulent flows. This stands in contrast to the case of a constant $\omega_H$ (Fig.~\ref{fig:fig5}c, inset), where a purely magnetic-field-driven particle, instead of showing zero net angular movement as in the oscillating $\omega_H$ scheme, follows $\psi = \omega_H t$ (black solid line).

\subsection*{Conclusions and outlook}

We have revealed the first \bzq{experimental and numerical} evidence of stochastic resonance (SR) for magnetic particles suspended in turbulence, demonstrating how the interplay between turbulent fluctuations and external forcing induces coherent rotational motion of magnetic particles. 
Three distinct regimes: ``phase-locked'', ``back-and-forth'', and ``turbulent-dynamics'' are identified, with SR occurring at the transition between ``phase-locked'' and ``turbulent-dynamics'' regimes. Remarkably, even when both turbulence and the magnetic field have zero mean angular velocity, their collective effect is capable of generating a nonzero particle rotational response, as the turbulent vorticity helps the particle to ``surf'' the rotating magnetic field.

Different particle properties influence their response, and when combined with the stochastic resonance mechanism, they enable a novel magnetic resonance-based approach to measure turbulent vorticity. This method leverages activated magnetic particles as probes, effectively functioning as a turbulent vorticity ``microscope'' that operates in various conditions. 
In the present work, we validate this concept using optical measurements of particle rotation. However, in principle, once the angular response is well-characterized, the same protocol could be applied even in optically inaccessible environments by measuring the magnetic fields emitted by the spinning magnetic particle directly. In such scenarios, the activated magnetic particles would function as remotely readable vorticity probes.
The protocol is straightforward: by dispersing magnetic particles into turbulence and varying the magnetic frequency $\omega_H$ and intensity $\omega_a$ while maintaining a constant ratio $\omega_H / \omega_a$, one can measure the resonance curve of the averaged particle angular response. The location of the resonance peak directly reveals the turbulent vorticity magnitude.

While the present work focuses on understanding particle angular dynamics, it lays essential groundwork for subsequent investigations into how these actively tunable particles can shape turbulent structures, 

opening possibilities for flexible manipulation of turbulence, where propeller-like particles could be tuned to navigate specific flow structures, inject energy into turbulence, and influence flow properties \bzq{such as intermittency} \cite{freitas2025statistical}.
More broadly, our study provides fundamental mechanisms that externally tunable rotational dynamics could be leveraged to design active materials \cite{scholz2018rotating,yang2021motion,jiang2023magnetic}.
If organized coherently, their collective behavior may also exhibit hydrodynamic properties analogous to systems with odd viscosity, which may be promising experimental realizations of chiral fluids \cite{chen2025self,fruchart2023odd}. 

This research paves the way for innovative experimental techniques, offering a simple yet underutilized approach to studying and controlling turbulence dynamics.

\section*{Data availability}
The data generated during the course of this study is publicly available on Zenodo at https://zenodo.org/records/17076195 \cite{wang_2025_17076195}

\section*{Code availability}
The code used for processing the data and generating the figures is publicly available on Zenodo at https://zenodo.org/records/17076195 \cite{wang_2025_17076195} under the CC-BY-4.0 license.

\section*{Acknowledgments}
We thank F. van Uittert, G. Oerlemans and J. van der Veen for technical support. This work is supported by the Netherlands Organization for Scientific Research (NWO) through the use of supercomputer facilities (Snellius) under Grant No. 2023.026. This publication is part of the project “Shaping turbulence with smart particles” with Project No. OCENW.GROOT.2019.031 of the research program Open Competitie ENW XL which is (partly) financed by the Dutch Research Council (NWO).

\section*{Author contributions}
F.T. designed the research. Z.W. planned and carried out the simulations. F.T., C.W., H.J.H.C., and R.P.J.K. conceived and planned the experiments. C.W. carried out the experiments. Z.W., X.M.d.W., R.B., and F.T. developed the theoretical formalism and contributed to the interpretation of the results. Z.W. took the lead in writing the manuscript. All authors contributed to and approved the final version of the manuscript. 

\clearpage
\section*{\Large{Methods}}

\subsection*{Experimental setup}

The experimental setup comprises three fundamental components: a turbulence generator, a magnetic field generator, and magnetic particles. Detailed specifications of the experimental apparatus, along with a comprehensive description of the particle rotation tracking methodology, are provided in our dedicated instrumentation and methodology paper \cite{wu2025tracking}. 
Here, we present only the essential information pertinent to the current study.

{\textit{Turbulence generator:}}
The experiments are conducted in a confined Von K\'{a}rm\'{a}n-type turbulent flow, which serves as a model system for studying homogeneous and isotropic turbulence, particularly in the context of Eulerian and Lagrangian statistics \cite{zocchi1994measurement,voth2002measurement,mordant2001measurement,la2001fluid,volk2011dynamics}.

The working fluid is water, contained within an octagonal-shaped vessel with an internal diameter of $2R = 150$ mm and a height of 220 mm. This specific geometry is chosen to enhance optical accessibility for visualization. The flow is driven by two counter-rotating disks, each with a diameter of 150 mm, separated by a distance of 150 mm, and rotating at the same angular velocity magnitude in opposite directions. Each disk is equipped with six straight blades, each 10 mm in height, designed to enhance inertial stirring. The disks are powered by two calibrated DC motors operating at a constant voltage, ensuring that the angular velocity $\Omega$ remains stable over time with a precision of approximately 1\%. The motor rotation frequency can be adjusted within the range of $0.25$ to $8$ Hz.
Under these conditions, a fully developed turbulent flow with up to $Re_\lambda = 447$ can be generated in the central region of the container \bzq{with small-scale statistics close to local isotropy} \cite{mordant2004experimental,volk2008acceleration,zocchi1994measurement,voth2002measurement,mordant2001measurement,la2001fluid,volk2011dynamics}. This region, measuring approximately $20$ mm $\times$ $20$ mm $\times$ $20$ mm, serves as the primary domain for measurements. 
% \bzq{Since our small particles ($r \sim 5\eta$) primarily interact with dissipative-scale velocity gradients in this central region, the observed stochastic resonance and symmetry breaking are governed by particle-scale turbulence and magnetic forcing rather than large-scale flow anisotropy.}
To enable particle rotational motion tracking, a Photron NOVA S6 high-speed camera, operating at 3000 frames per second with a resolution of $1024 \times 1024$ pixels, is employed to record particle motion within the flow field.

% The corresponding Taylor Reynolds number, $\text{Re}_\lambda$, ranges from $xx$ to $xx$, depending on the imposed rotation rate.

{\textit{Magnetic field generator:}}
A uniform rotating planar magnetic field within a restricted measurement volume, $\bm{H} = H \bm{h}$, with an angular frequency $\omega_H$, is generated using a system of two pairs of mutually perpendicular Helmholtz coils. The vertical pair (marked in red in Fig.~\ref{fig:fig1}a) has an interior diameter of 300 mm, while the horizontal pair (marked in blue in Fig.~\ref{fig:fig1}a) has an interior diameter of 245 mm. For each pair, the separation between the two coils is equal to the corresponding coil diameter.  

Each Helmholtz coil pair generates a magnetic field along a single axis. By driving the two perpendicular coil pairs with AC currents that have a phase difference of $\pi/2$, a resultant magnetic field is produced that rotates at a constant angular velocity $\omega_H$. This system is capable of generating rotating magnetic fields with frequencies of up to 50 Hz.

% \bzq{
% The rotating planar magnetic field was calibrated using a SENIS F3A magnetic-field-to-voltage transducer with an integrated 3D Hall probe. Across the frequency range of 1–50 Hz, the maximum magnetic flux density remained stable at $1.60 \pm 0.02$ mT. The driving frequency is set to be 20 Hz for Fig.~\ref{fig:fig2} and 10 Hz for Fig.~\ref{fig:fig4}, which illustrates the stochastic resonance. Further details of the field characterization are reported in our experimental study \citep{wu2025tracking}.
% }

{\textit{Magnetic particles:}}
The magnetic particles are produced by coating spherical Styrofoam cores with a layer of magnetic paint, which is paramagnetic. The Styrofoam cores are initially arranged in a fixed, evenly spaced configuration on a substrate. The magnetic paint is then applied uniformly from the top of the substrate to the Styrofoam spheres through a spraying process, ensuring that all particles are coated in a consistent manner; however, at the single-particle level the coating is not perfectly homogeneous, and small variations in thickness or distribution lead to slight differences in magnetic properties between particles. Finally, the coated particles are detached from the substrate.

The particle volume fraction used in the experiments is $0.17\%$, ensuring that the system remains in the dilute regime. This low concentration minimizes inter-particle interactions, allowing us to focus on the dynamics of particle response to the complex interplay between turbulent fluctuations and the external magnetic field, without significant effects from collective dynamics.

These particles exhibit anisotropy in three key aspects:
(1) Shape: The particles are approximately spherical, with a mean diameter of $0.762 \pm 0.066$ mm.
(2) Density: The magnetic particles have a mean density of $\rho_\text{particle}= 208 \pm 14$ kg/m$^3$.
(3) Magnetic Properties: 
Due to the non-uniform distribution of the magnetic paint on the particle surface, the magnetic properties vary between particles. Specifically, the magnetic moment magnitude $|\bm{m}|$, the orientation of the magnetic moment, and the spatial distribution of magnetization across the surface differ from particle to particle. This variability influences how individual particles respond to an external magnetic field.
All these three aspects of particle polydispersity can introduce minor differences between the simulation results and the experimental observations. However, these differences remain small, and overall, the simulations show good agreement with the experiments. More importantly, our theoretical modeling and the simulations successfully capture key physical features observed in the experiments, such as the location of peak responses.

{\textit{Particle rotation tracking:}}
The particle rotational motion is tracked using an efficient particle rotation tracking algorithm.
The algorithm initiates by pre-processing the experimental images through the application of Gaussian filtering to mitigate noise, contrast enhancement to optimize image quality, and central cropping to ensure precise target particle localization. Subsequently, a set of discrete candidate rotation angles is defined, and corresponding theoretical rotated images are generated utilizing 3D rotation matrices. For each temporal frame, the algorithm computes the cross-correlation coefficient between the actual experimental image and each theoretical rotated image to determine the optimal matching rotation angle. By tracking these optimal angles across the time series, the rotational motion trajectory of the particle is reconstructed and visualized. More details of the tracking methodology are reported in our dedicated instrumentation and methodology paper \cite{wu2025tracking}.

\subsection*{Theoretical model}

A paramagnetic particle moves in a homogeneous and isotropic turbulent flow while subjected to an external magnetic field $\bm{H}$. The magnetic field rotates in the $xoy$-plane (laboratory frame) with an angular velocity $\omega_H$ and intensity ${H}$, expressed as $\bm{H} = H\bm{h} = H(\cos(\omega_{H} t), \sin{(\omega_{H} t)}, 0)^T$. 
The particle is modeled as a sphere with intrinsic magnetic anisotropy (same as that of a prolate spheroid of aspect ratio 1.2), characterized by a preferred magnetization direction $\bm{n}$, which dynamically evolves over time. This instantaneous unit vector is given by $\bm{n} = (\sin \alpha \cos \psi, \sin \alpha \sin \psi, \cos \alpha)^T$, where $\alpha$ and $\psi$ (defined in Fig.~\ref{fig:fig2}a) describe the orientation of $\bm{n}$.

When the particle is placed in an external magnetic field, $\bm{H}$, its induced magnetic moment takes the form \cite{cimurs2019stability}
\begin{equation}
    \bm{m} = \mathcal{V}_p H [\chi_{\bot} \bm{h} + (\chi_{\parallel} - \chi_{\bot})(\bm{n}\cdot \bm{h})\bm{n}],
\end{equation}
where $\mathcal{V}_p=4\pi r^3/3$ is the particle volume, with $r$ denoting its radius. The anisotropic magnetic susceptibilities $\chi_{{\parallel},{\bot}}$, arising due to the anisotropy of the particle shape (e.g., spheroidal) and the demagnetizing field factors $N_{\parallel,\bot}$, and $\chi_{\parallel,\bot}=\chi_0/(1+\chi_0 N_{\parallel,\bot})$ \cite{osborn1945demagnetizing}, causes the particle to respond differently to an external magnetic field depending on the direction. The subscripts parallel ($\parallel$) and perpendicular ($\bot$) refer to directions relative to the preferred magnetization direction $\bm{n}$. 
Due to the tendency of the magnetic moment $\bm{m}$ to align with the magnetic field direction $\bm{h}$, the rotating magnetic field induces the particle rotational motion.
The magnetic torque acting on the particle is given by $\bm{T}_{M} = \mu_{w} \bm{m} \times \bm{H}$, where $\mu_{w}$ is the magnetic permeability. The magnetic torque tries to align the particle’s preferred magnetic axis ($\bm{n}$) with the magnetic field direction ($\bm{h}$) and is balanced by the rotational drag torque (approximated in the Stokes regime) $\bm{T}_D = -\xi_r(\bm{\omega}_p - \bm{\omega}_f)$ \cite{naso2010interaction}, where $\xi_r=8\pi\mu r^3$ is the rotational drag coefficient and $\mu$ is the dynamic viscosity of water.

The particle rotational velocity $\bm{\omega}_p$ naturally decomposes into two components: a perpendicular component $\bm{\omega}_{p,\bot}$ (tumbling) and a parallel component $\bm{\omega}_{p,\parallel}$ (spinning) along $\bm{n}$.

The particle rotational motion is governed by the balance of torques, and we consider the overdamped condition \cite{tierno2009overdamped}
\begin{equation}
\label{eq:drag_balance}
    \bm{T}_M + \bm{T}_D = \bm{0},
\end{equation}
with $\bm{T}_{M}$ the magnetic torque and $\bm{T}_{D}$ the rotational drag torque.

By projecting this balance along and perpendicular to $\bm{n}$, we obtain the equations governing tumbling and spinning.
The tumbling dynamics, $\dot{\bm{n}}$, obey
\begin{equation}\label{eq:bot_eqn}
   \frac{d\bm{n}}{d t} = \omega_a (\bm{n} \cdot \bm{h}) \left( \bm{h} - (\bm{n} \cdot \bm{h})\bm{n} \right) + \bm{\omega}_f \times \bm{n},
\end{equation}
where $\omega_a = \mu \mathcal{V}_p H^2 (\chi_{\parallel}-\chi_{\bot})/\xi_{r}$ is the normalized magnetic field intensity which is also called anisotropic frequency \cite{tierno2009overdamped, cimurs2019stability}, with $\omega_a>0$ ($\omega_a<0$) for a prolate (oblate) particle. 

The particle spinning rate $\bm{\omega}_{p,\parallel}$ satisfies
\begin{equation}\label{eq:parallel_eqn}
   \frac{d \left(\xi_{r}\bm{\omega}_{p,\parallel}\right)}{d t} =  -\xi_{r} \left(\bm{\omega}_{p, \parallel} - \left(\bm{\omega}_f \cdot \bm{n}\right)\bm{n}\right),
\end{equation}
where for the overdamped particle, we have $\frac{d \left(\xi_{r}\bm{\omega}_{p,\parallel}\right)}{d t} =\bm{0}$. $\bm{\omega}_{f}$ is half of the vorticity of the background flow field, which is obtained through performing Direct Numerical Simulations (DNS) using a pseudo-spectral method and details can be found later. 
The validity of the overdamped model is further discussed in the Supplementary Information. 

The total particle angular velocity is thus $\bm{\omega}_p = \bm{\omega}_{p,\parallel} +  \bm{\omega}_{p,\bot} 
= (\bm{n} \cdot \bm{\omega}_p) \bm{n} + \bm{n} \times \dot{\bm{n}}$, with $\dot{\bm{n}} = d\bm{n}/dt$ (see Eq.~\eqref{eq:bot_eqn}). Equations of motion \eqref{eq:bot_eqn} and \eqref{eq:parallel_eqn} are numerically integrated using a third-order Runge-Kutta method. 

The governing equation \eqref{eq:drag_balance} can be reformulated in terms of the phase lag angle $\beta$ and the orientation angle $\alpha$, where $\alpha$ denotes the angle between the unit vector $\bm{n}$ and the $z$-axis, $\bm{\hat{z}}$.  

To derive the governing equation for $\beta$, we project Eq.~\eqref{eq:drag_balance} onto $\bm{\hat{z}}$-direction, yielding  
\begin{equation}
\label{eq:drag_balance_z_proj}
    \bm{T}_M \cdot \bm{\hat{z}} + \bm{T}_D\cdot \bm{\hat{z}} = 0.
\end{equation}
Here, the magnetic torque component along $\bm{\hat{z}}$ is given by $ \bm{T}_M \cdot \bm{\hat{z}} = \omega_a \xi_r \sin^2\alpha \sin\beta \cos\beta$, while the drag torque contribution is expressed as $\bm{T}_D\cdot \bm{\hat{z}} = -\xi_r(\omega_{p,z}-\omega_{f,z})$.  

Rewriting Eq.~\eqref{eq:drag_balance_z_proj} explicitly in terms of $\beta$ and $\alpha$, we obtain  
\begin{equation}
\label{eq:beta}
    \dot{\beta}  = \omega_H - \frac{\omega_{a}}{2} \kappa \sin{2 \beta} - \omega_{f,z},
\end{equation}
with the coefficient $\kappa=\sin^2\alpha$.

To establish the evolution equation for $\alpha$, we introduce a unit vector $\bm{N}=(-\sin\psi, \cos\psi,0)^T$, which denotes the direction perpendicular to $\bm{n}_{xoy}$, the projection of $\bm{n}$ onto the $xoy$-plane (see Fig.~\ref{fig:fig2}a for the definition). Projecting Eq.~\eqref{eq:drag_balance} along $\bm{N}$ leads to  
\begin{equation}
\label{eq:drag_balance_N_proj}
     \bm{T}_M \cdot \bm{N} + \bm{T}_D\cdot \bm{N} = 0.
\end{equation}
Here, the magnetic torque component along $\bm{N}$ is given by $\bm{T}_M \cdot \bm{N}=\frac{1}{2}\omega_a\xi_r\cos^2\beta\sin2\alpha$, while the drag torque contribution is $\bm{T}_D\cdot \bm{N}=-\xi_r(\bm{\omega}_p \cdot \bm{N}-\bm{\omega}_f\cdot\bm{N})$. Noting that $\bm{\omega}_p \cdot \bm{N} = \dot{\alpha}$, Eq.~\eqref{eq:drag_balance_N_proj} can be rewritten explicitly as  
\begin{equation}
\label{eq:alpha}
    \dot{\alpha}  = \bm{\omega}_{f} \cdot \bm{N} + \frac{\omega_{a}}{2} \sin{2 \alpha} \cos^2{\beta}.
\end{equation}

To capture the essential physics of the system, we assume $\kappa \approx 1$, which is reasonable because $\kappa$ follows a probability density function that is sharply peaked at $1$ and decays rapidly to zero for values smaller than unity within the investigated parameter space (further details are provided in the Supplementary Information).  
The equation for $\beta$ can be reduced to
\begin{align}\label{eq:simplified_angle_eqn}
    \dot{\beta} & = \omega_H - \frac{\omega_{a}}{2} \sin{2 \beta} - \omega_{f,z},
\end{align}
which is controlled closely by the magnetic field.

Eq.~\eqref{eq:simplified_angle_eqn} is used to predict the phase transition between the phase-locked regime and the turbulent-dynamics and back-and-forth regimes (blue solid line in Fig.~\ref{fig:fig3}m), and is ultimately describing the dynamics that gives rise to the mechanism of stochastic resonance (Figs.~\ref{fig:fig4}c-g).
Similar dynamical equations as Eq.~\eqref{eq:simplified_angle_eqn} have been widely applied in multiple physical systems, including condensed matter physics, chemical and biological systems, optics and laser physics, electrical engineering, and financial econometrics \cite{coffey2012langevin,bouchaud1998langevin}, with a range of related physical problems, such as Josephson junctions \cite{jia2001effects, berdichevsky1997josephson,de2022generation,pountougnigni2023detection,wand2024estimating}, vortex diffusion and flux-line dynamics in superconductors \cite{inui1989pinning,jia2001effects,soroka2007guiding}, reaction-rate characteristics relevant to chemistry, engineering, and biology \cite{hanggi1990reaction}, Brownian motors \cite{reimann2002brownian}, and decision-making processes under uncertainty which resemble financial market fluctuations \cite{bouchaud1998langevin,wand2024estimating}. The universality of this equation suggests that our findings have broad implications beyond the specific context of magnetic particles in turbulence, offering deeper theoretical insights into nonlinear dynamical systems and facilitating the development of new applications in diverse scientific and technological domains.

Now we show the details of predicting the boundary (Fig.~\ref{fig:fig3}m, blue solid line) that separates the phase-locked regime (Fig.~\ref{fig:fig3}m, purple region) from the back-and-forth and turbulent-dynamics regimes. The phase-locked regime exists if the equation $\dot{\beta}=0$ admits fixed points. Based on the simplified governing equation, Eq.~\eqref{eq:main_beta_eqn} or Eq.~\eqref{eq:simplified_angle_eqn}, this condition translates to ensuring that the following equation has real solutions for any imposed $\omega_H$:
\begin{equation}
\label{eq:predict_boundary}
    \frac{\omega_a}{2} \sin{2\beta}=\omega_H-\omega_{f,z}.
\end{equation}

The left-hand side of Eq.~\eqref{eq:predict_boundary} is bounded within $[-\omega_a/2, \omega_a/2]$, while the right-hand side varies typically over $\left[\omega_H-\sqrt{\sigma(\omega_{f,z})}, \omega_H+\sqrt{\sigma(\omega_{f,z})}\right]$, where $\sigma(\omega_{f,z})$ is the variance of $\omega_{f,z}$. 
For Eq.~\eqref{eq:predict_boundary} to have real solutions for any imposed $\omega_H$, the following condition must hold:
\begin{equation}
\label{eq:bound_full}
    \omega_H+\sqrt{\sigma(\omega_{f,z})} \leq \frac{\omega_a}{2},
\end{equation}
which gives 
\begin{equation}
\label{eq:bound_full_rearrange1}
\omega_a \geq 2 \omega_H + 2\sqrt{\sigma(\omega_{f,z})}.
\end{equation}
Notice that the standard deviation of $\omega_{f,z}$ is approximated to be of the order of the Kolmogorov frequency scale, i.e., $\sqrt{\sigma(\omega_{f,z})} \sim \omega_\eta$, where $\omega_\eta \sim 1/\tau_\eta$. In our simulations, we observe that $2 \sqrt{\sigma(\omega_{f,z})} \sim \omega_\eta$, which justifies this approximation. Based on this, we obtain the final criterion for the phase-locked regime from Eq.~\eqref{eq:bound_full_rearrange1} as
\begin{equation}
    \omega_a \geq 2\omega_H + \omega_\eta.
\end{equation}

\subsection*{Direct numerical simulation of the Navier-Stokes equation}
To accurately integrate the governing equations \eqref{eq:bot_eqn} and \eqref{eq:parallel_eqn}, it is essential to correctly incorporate the Lagrangian vorticity signal of the particles $\bm{\omega}_f$.  

We begin by performing DNS of statistically stationary homogeneous isotropic turbulence (HIT) to obtain the Lagrangian vorticity dynamics ($\bm{\omega}_f$) of small, passively advected particles. The computational domain consists of a cubic box of size $L = 2\pi$ with periodic boundary conditions. The simulations employ a pseudo-spectral method  \cite{peyret2002spectral} for spatial discretization and an Adams-Bashforth scheme for temporal integration, with the standard $2/3$ dealiasing rule applied to ensure numerical accuracy  \cite{bec2006acceleration}. The numerical framework has been validated against multiple integration schemes, interpolation methods, and large-scale forcing techniques \cite{Calzavarinibook}.  

The suspended particles are modeled as point-like, dilute tracers that interact with the turbulent flow through hydrodynamic forces alone. Their dynamics are characterized by the Stokes number ($\text{St}$) and the density contrast ($\beta_p$), which describe their response time relative to the fluid and their mass relative to the surrounding medium, respectively. To capture the essential physics while simplifying the problem, we adopt a one-way coupled model, which has been extensively validated in both experimental and numerical studies  \cite{maxey1983equation, auton1988force, biferale2010measurement, toschi2009lagrangian, calzavarini2008dimensionality, dewit2024, wang2024localization}. 

The governing equation for particle motion is 
\begin{equation}
    \ddot{\bm{x}}_p = \beta_p \frac{\mathrm{D} [\bm{u}_f]_O}{\mathrm{D}t} - \frac{1}{\text{St}} \left(\dot{\bm{x}}_p - [\bm{u}_f]_O\right),
\label{eq1}
\end{equation}  
where $[\bm{u}_f]_O$ denotes the fluid velocity with the subscript $O$ indicating evaluation at the center of the particle, $\bm{x}_p$ represents the particle position, and $\bm{v}_p = \dot{\bm{x}}_p$ is the particle velocity. The Lagrangian vorticity is computed as $\bm{\omega}_f = \frac{1}{2} [\bm{\nabla} \times \bm{u}_f]_O$. The density contrast is defined as $\beta_p = \frac{3}{1+2\rho_p / \rho_f}$, where $\rho_p / \rho_f$ is the particle-to-fluid density ratio. The Stokes number is given by $\text{St} = d_p^2/(12 \beta_p \nu \tau_{\eta})$, where $d_p$ is the particle diameter and $\nu$ is the fluid kinematic viscosity.  

In this study, we perform three-dimensional DNS at a resolution of $N^3=128^3$ with a Taylor-scale Reynolds number $\text{Re}_{\lambda}\approx60$ \cite{peyret2002spectral, bec2006acceleration, Calzavarinibook, wang2024localization} and collect 512 independent samples for three particle types: light ($\beta_p \to 3.0$, representing bubbles where $\rho_p \to 0$), neutral ($\beta_p =1.0$), and heavy ($\beta_p =0.01$) particles, all with a Stokes number of $\text{St} = 1.0$. This specific Stokes number is chosen because light particles in HIT tend to accumulate preferentially in vortex filaments, leading to strong dynamic localization \cite{maxey1994simulation, calzavarini2008dimensionality, mathai2020bubbly, wang2024localization}. 

A key consideration in extending our results to higher Reynolds number turbulence is the effect of intermittency, particularly the heavy-tailed distribution of vorticity fluctuations in fully developed turbulence. In principle, for very large Reynolds numbers, the intermittency of the small-scale vorticity field should be accounted for, as extreme vorticity events become more pronounced. However, previous studies have shown that while intermittency affects the fine-scale structure of turbulence, its influence on the rotational dynamics of small particles (compared to the Kolmogorov length scale), particularly in the parameter regimes relevant to our study, remains relatively weak, because the intense vorticity events, although frequent, are spatially localized and short-lived; thus, their cumulative effect on particle rotation is averaged out \cite{toschi2009lagrangian,wang2019rotational,zimmermann2011tracking}. Our results and the underlying mechanisms in the main paper provide valuable insights into the fundamental principles governing particle rotation in turbulent flows. Future research could explore the effects of finite-sized particles and beyond the dilute regime.

\clearpage
\onecolumngrid 

\renewcommand{\figurename}{\textbf{Supplementary Figure}}
\renewcommand{\theequation}{S.\arabic{equation}}
\setcounter{figure}{0} 
\setcounter{equation}{0}

\section*{\Large{Supplementary Information}}

\subsection{Descriptions of the Supplementary movies}\label{sec3}

$\bullet$ \textbf{Supplementary Movie 1}: The ``phase-locked'' regime of a magnetic particle subjected to a rotating magnetic field in a quiescent fluid, corresponding to Fig.~1f in the main text.\\

$\bullet$ \textbf{Supplementary Movie 2}: The ``back-and-forth'' regime of a magnetic particle subjected to a rotating magnetic field in a quiescent fluid, corresponding to Fig.~1h in the main text.\\

$\bullet$ \textbf{Supplementary Movie 3}: The experimental results of the particles immersed in turbulence subjected to a rotating magnetic field with a weak strength, corresponding to Fig.~2b in the main text.\\

$\bullet$ \textbf{Supplementary Movie 4}: The experimental results of the particles immersed in turbulence subjected to a rotating magnetic field with an intermediate strength, corresponding to Fig.~2c in the main text.\\

$\bullet$ \textbf{Supplementary Movie 5}: The experimental results of the particles immersed in turbulence subjected to a rotating magnetic field with a strong strength, corresponding to Fig.~2d in the main text.\\

$\bullet$ \textbf{Supplementary Movie 6}: In the ``turbulent-dynamics'' regime, the evolution of the tip of the preferred magnetization direction of the particle, $\bm{n}(t)$, is visualized in space, corresponding to Fig.~3d in the main text.\\

$\bullet$ \textbf{Supplementary Movie 7}: In the ``phase-locked'' regime, the evolution of the tip of the preferred magnetization direction of the particle, $\bm{n}(t)$, is visualized in space, corresponding to Fig.~3h in the main text.\\

$\bullet$ \textbf{Supplementary Movie 8}: In the ``back-and-forth'' regime, the evolution of the tip of the preferred magnetization direction of the particle, $\bm{n}(t)$, is visualized in space, corresponding to Fig.~3l in the main text.\\

$\bullet$ \textbf{Supplementary Movie 9}: The experimental results of the particles immersed in turbulence without an external magnetic field applied, corresponding to Supplementary Fig.~1a.\\

$\bullet$ \textbf{Supplementary Movie 10}: The experimental results of the particles immersed in a stronger turbulence subjected to a rotating magnetic field, corresponding to Supplementary Fig.~1b.\\

\section{Experimental setup}
\subsection{Particle Characterization}
In optical images as shown in Fig.~\ref{fig:fig1}, the particle surfaces may appear ``rough.'' This appearance originates from the contrast between the bright Styrofoam core and the darker magnetic coating, rather than actual geometric protrusions. Variations in paint deposition, the side of the particle in contact with the substrate during spraying, and small inconsistencies in the magnetic paint spraying angle result in non-uniform coating patterns. These patterns enhance optical contrast, creating the visual impression of surface roughness. However, the actual geometric irregularities are negligible compared to the particle diameter, and the surfaces remain effectively smooth on the hydrodynamic scale. Therefore, while such patterns may introduce minor perturbations to local flow, they do not significantly alter translational drag or rotational resistance beyond the effects already captured by particle size and magnetic torque.

The particles are designed to probe rotational dynamics under the combined influence of turbulence and a rotating magnetic field. Their size are comparable to the Kolmogorov scale, ensuring interactions with the smallest turbulent structures while remaining much smaller than the integral scale and Taylor microscale. Consequently, they cannot be regarded as point-like; however, finite-size effects on the rotation are limited, and the dominant particle–flow interactions occur at dissipative scales. The particles’ magnetic properties were characterized using a Superconducting Quantum Interference Device (SQUID) magnetometer (Quantum Design Inc.). Batch measurements yielded a mean volume susceptibility of $|\overline{\chi}_V| = 0.136 \pm 0.046$, while single-particle measurements revealed an average anisotropy of $\Delta \chi = 0.011 \pm 0.002$. 

\subsection{Magnetic field calibration}
The rotating planar magnetic field was calibrated using a SENIS F3A magnetic-field-to-voltage transducer with an integrated 3D Hall probe. Across the frequency range of 1–50 Hz, the maximum magnetic flux density remained stable at $1.60 \pm 0.02$ mT. The driving frequency is set to be 20 Hz for Fig.~\ref{fig:fig2} and 10 Hz for Fig.~\ref{fig:fig4}, which illustrates the stochastic resonance. Further details of the field characterization are reported in our experimental study \citep{wu2025tracking}.

\subsection{Flow field characterization}
Our experiments are conducted in a Von K\'arm\'an-type turbulent flow, characterized by a Taylor-scale Reynolds number of up to $Re_\lambda = 447$, ensuring fully developed homogeneous and isotropic turbulence in the measurement volume. It is noteworthy that the particle rotational dynamics is dominated by the turbulent dynamics at the small scales, i.e. the scales of vortex filaments. It is widely conjectured, motivated by numerical and experimental observations, that the small scales of turbulence are universal \cite{ghira2022characteristics, kang2007dynamics, tanahashi2001appearance, da2011intense,ganapathisubramani2008investigation}. Therefore, it is expected that, provided the turbulence is fully developed, the strength of the large scale forcing, i.e., different values of the Reynolds number, poses negligible effects on the particle rotational dynamics. While the background flow is turbulent, the particle rotational dynamics are primarily governed by interactions with small-scale eddies. The particle rotational Reynolds number, $Re_p = \rho_f r^2|\omega_p - \omega_\eta|$, is estimated to be $\mathcal{O}(10^{-1})$ (below 1) based on typical maximum relative angular velocities, justifying the use of the laminar (Stokes) rotational drag coefficient, $\xi_r$ in out theoretical modelling. This ensures that viscous forces dominate inertial effects at the particle scale, providing an accurate leading-order description of particle-fluid interaction.

While the Von K\'{a}rm\'{a}n-type turbulent flow is not perfectly homogeneous and isotropic, a fully developed turbulent flow can be generated in the central region of the container with small-scale statistics close to local isotropy \citep{mordant2004experimental,volk2008acceleration,zocchi1994measurement,voth2002measurement,mordant2001measurement,la2001fluid,volk2011dynamics}. This region, measuring approximately $20$ mm $\times$ $20$ mm $\times$ $20$ mm, serves as the primary domain for measurements. 
Since our small particles ($r \sim 5\eta$) primarily interact with dissipative-scale velocity gradients in this central region, the observed stochastic resonance and symmetry breaking are governed by particle-scale turbulence and magnetic forcing rather than large-scale flow anisotropy.

The measured dissipation rate $ \varepsilon $ is determined from the driving torque on the two impellers as $\varepsilon = T \Omega/(\rho V_\text{w})$ \cite{Mordant1997, Labbe1996}, where $ T $ is the sum of the driving torques from the impellers at the top and bottom of the water tank and $ V_\text{w} = 2.8 \times 10^{-3} $ m\textsuperscript{3} is the volume of the water tank. The driving torques are measured using two strain gauge torque meters mounted on the motors. 
In our experiments, we investigated two turbulence intensities:
(1) the weak-turbulence condition, for which the results are shown in Fig.~2(b–d) of the main text; and
(2) the strong-turbulence condition, for which the results are presented in Supplementary Fig.~\ref{fig:fig1}(b).
For the weak-turbulence case, the impellers rotate at 0.83 Hz. The corresponding mean energy dissipation rate is $\varepsilon = 0.037 \pm 0.002~\text{m}^2/\text{s}^3$, which yields a Taylor-scale Reynolds number of $Re_\lambda = 398$, a Kolmogorov length scale of $\eta = 0.072~\text{mm}$, and a Kolmogorov time scale of $\tau_\eta = 5.20~\text{ms}$.
For the strong-turbulence case, the impellers rotate at 1.21 Hz. The corresponding mean energy dissipation rate is $\varepsilon = 0.075 \pm 0.001~\text{m}^2/\text{s}^3$, which yields a Taylor-scale Reynolds number of $Re_\lambda = 447$, a Kolmogorov length scale of $\eta = 0.061~\text{mm}$, and a Kolmogorov time scale of $\tau_\eta = 3.66~\text{ms}$.

\section{Comparison between experimental and numerical results}\label{sec1}

Previous studies have indeed explored the rotational dynamics of magnetically driven objects in quiescent environments \citep{croquette1981cascade, poye2018bipolar, kaiser2017flocking,yan2012linking,cimurs2019stability,cebers2006dynamics,erglis2007dynamics,cimurs2013dynamics,morozov2017dynamics}, demonstrating phenomena such as phase locking and oscillatory motion. However, a fundamental distinction of our work lies in investigating these dynamics within the stochastic and chaotic environment of fully developed turbulence. This critical inclusion reveals a new class of phenomena, particularly the emergence of stochastic resonance where turbulent fluctuations play an active role in enhancing the particle response to an external periodic magnetic field. Furthermore, we uncover a novel symmetry-breaking mechanism for inducing net particle rotation in zero-mean turbulent vorticity.

In the main paper, we have presented selected comparisons of the PDFs of particle angular velocity, $\omega_{p,z}$, between experimental and numerical results (Fig.~2 in the main paper), for weak ($1.2~\text{mT}$), \textbf{c} intermediate ($1.4~\text{mT}$), and \textbf{d} strong ($1.6~\text{mT}$) magnetic intensities with constant rotational frequency of the magnetic field ($20~\text{Hz}$), to illustrate the key findings. The PDFs are obtained using 3000 (in experiments) and 512 (in simulations) particle trajectories.
Here, we provide additional comparisons to further support our conclusions.

Supplementary Fig.~\ref{fig:fig1} presents the probability density distributions of the particle angular velocity components $\omega_{p,i}$ ($i=x,y,z$) obtained from both experiments and numerical simulations. In Supplementary Fig.~\ref{fig:fig1}a (Supplementary Movie 9), where no magnetic field is applied, the particle rotation is purely driven by turbulence, exhibiting isotropic behavior. To facilitate a direct comparison with the experimental results (circles), the numerical results (squares) are flattened over all directions (i.e., reshape the angular velocity vector into a one-dimensional (1D) array while preserving the original data order). In Supplementary Fig.~\ref{fig:fig1}b (Supplementary Movie 10), when a magnetic field of $1.6~\text{mT}$ (same intensity as that of Fig.~2d in the main paper) is applied and the turbulence intensity is increased ($Re_{\lambda} = 447$), the numerical and experimental results show good agreement, demonstrating that the numerical simulation is capable of capturing the essential physics of the experiment. 
Supplementary Fig.~\ref{fig:fig1}c further provides a phase diagram illustrating different particle rotational dynamic regimes, which is reproduced from Fig.~3m in the main paper. The experimental conditions corresponding to Supplementary Fig.~\ref{fig:fig1}b are marked as a green triangle. As shown in Supplementary Figures \ref{fig:fig1}a and b, the simulation results agree well with the experimental results. The slight discrepancies are due to the polydispersity of the particles. Specifically, these particles exhibit anisotropy in three key aspects: (1) Shape: The particles are nearly spherical, with an average diameter of $0.762 \pm 0.066$ mm. However, minor deviations from perfect sphericity exist due to natural variations in the manufacturing process. These variations can introduce small differences in rotational dynamics, particularly when particles experience complex hydrodynamic and magnetic interactions in turbulence.
(2) Density: The mean density of the magnetic particles is measured to be $\rho_{\text{particle}} = 0.208 \pm 0.014$ kg/m$^3$. This value is significantly lower than that of water, indicating that the particles remain buoyant in the experimental fluid. The density variation across different particles arises from heterogeneities in the composition of the Styrofoam core and the magnetic coating.
(3) Magnetic Properties: The magnetic characteristics of the particles are inherently non-uniform due to the way the magnetic paint adheres to the surface during fabrication. Unlike ideal dipoles, these particles exhibit variability in three crucial aspects: the magnitude of the magnetic moment $|\bm{m}|$, the orientation of the magnetic moment relative to the particle’s geometric axes, and the spatial distribution of magnetization across the surface.
This non-uniformity means that different particles respond to an externally applied magnetic field in distinct ways, leading to variations in their rotational dynamics. Furthermore, the magnetization patterns on the particle surface serve as markers for tracking their rotational motion in experiments.
All these three aspects of particle polydispersity can introduce minor differences between the simulation results w.r.t the experimental observations. However, these differences remain small, and overall, the simulations show good agreement with the experiments.

In all experiments, the particle volume fraction is maintained at $0.17\%$, ensuring that the system remains in the dilute regime. This low concentration minimizes inter-particle interactions, allowing us to focus on the response of individual particles to turbulent flow and external magnetic fields without significant collision effects.

These additional comparisons reinforce the validity of our numerical model in capturing the stochastic rotational dynamics of the anisotropic magnetic particles under turbulence and external magnetic fields.

\section{Theoretical model}

\subsection{Statistics of direction angle}

When analyzing the mechanism of stochastic resonance, we employ a simplified model of the particle rotational dynamics. In this model, the governing equation is reformulated in terms of the phase lag angle $\beta$, under the assumption that the angle ($\alpha$) between the orientation direction ($\bm{n}$) and $z$-axis ($\bm{\hat{n}}$) satisfies $\alpha \to \pi/2$, which corresponds to $\kappa=\sin^2\alpha \to 1$. Based on this assumption, the equation simplifies to

\begin{align}\label{eq:simplified_angle_eqn} 
    \dot{\beta} & = \omega_H - \frac{\omega_{a}}{2} \sin{2 \beta} - \omega_{f,z}. 
\end{align}

This simplification is justified by the statistical behavior of $\kappa = \sin^2{\alpha}$, whose probability density function is sharply peaked at 1 and rapidly decays for values smaller than 1. To validate this assumption, we plot the probability density distributions of $\sin^2\alpha$, denoted as $P(\sin^2\alpha)$, for different noise intensities, $\omega_\eta/\omega_a$, in both the subcritical ($\omega_H/\omega_a = 0.1 < 1/2$) and supercritical ($\omega_H/\omega_a = 1.1 > 1/2$) regimes. The results confirm that $\kappa$ is indeed strongly concentrated around 1 and the probability decreases rapidly for smaller values of $\alpha$ ($<\pi/2$), supporting the validity of our simplification.

\subsection{Overdamped approximation justification}
To justify the validity of the overdamped model, we carefully assessed inertial effects. For a spherical particle, the rotational response time can be estimated as $\tau_{\text{rot}} = \frac{I}{\xi_r}$, where $I=\frac{2}{5}\mathcal{V}_p \rho_p r^2$ is the particle moment of inertia, and $\xi_r = 8 \pi \mu r^3$ is the rotational drag coefficient. Substituting these into the equation for $\tau_{\text{rot}}$ simplifies to $\tau_{\text{rot}} = \frac{\rho_p r^2}{15 \mu}$. Using our experimental parameters of $r_p = 0.381 \text{ mm}$, $\rho_p = 208 \text{ kg/m}^3$, $\mu \approx 1.0 \times 10^{-3} \text{ Pa}\cdot\text{s}$, and $\tau_{\eta} = 7.56 \text{ ms}$. The rotational response time is $\tau_{\text{rot}} = 2.01 \text{ ms}$. We observe that $\tau_{\text{rot}}$ is approximately $2.01 / 7.56 = 0.266$ or about one-quarter of $\tau_{\eta}$, indicating that the particle rotational dynamics are significantly faster than the characteristic changes in the flow vorticity. Therefore, the particle can respond quickly to the flow vorticity, making the assumption of an overdamped regime a reasonable first approximation.
Note that our experimental frequencies are chosen to ensure the particle can follow the driving field within the overdamped regime, i.e., the frequency of the magnetic field $\omega_H$ is slower than $1/\tau_{rot}\approx 2000~\text{Hz}$, so that the particle rotational motion will remain in the overdamped regime.

\subsection{Neglecting rotation-translation coupling}
In the model of particle angular dynamics, we neglect rotation–translation coupling by assuming that the particle center of mass and hydrodynamic center are approximately aligned. This assumption is justified by the nearly spherical shape, uniform mass distribution, and small size of the particles (compared to the Kolmogorov scale), which collectively minimize translation-induced hydrodynamic torque. Consequently, the hydrodynamic torque induced by translational motion is expected to be negligible.
Even if weak coupling exists, its effect can be interpreted as an additional source of rotational noise that is uncorrelated with the external magnetic field. This may slightly modify the amplitude of angular fluctuations or slightly shift the resonance frequency, but it does not alter the underlying mechanism of stochastic resonance. Since the emergence of stochastic resonance in our system fundamentally arises from the balance between external magnetic forcing and intrinsic rotational noise, the presence of small translation–rotation coupling does not qualitatively affect our main conclusions.

\subsection{Validity of the point-particle approximation}
Despite neglecting finite-size hydrodynamic effects, the point-particle approximation offers a leading-order description of particle–fluid interaction and is widely adopted in studies of particle-laden flows \citep{maxey1983equation,toschi2009lagrangian}. In our system, the particle radius is comparable to the Kolmogorov length scale, ensuring locally linear flow gradients that primarily govern rotational dynamics via local vorticity. This, coupled with a dominant external magnetic torque and a dilute suspension (volume fraction of $\Phi_V=0.17\%$ that minimizes particle-particle interactions \citep{balachandar2010turbulent}), provides a robust physical justification for our model. Crucially, the good agreement between our simulations and experimental measurements further validates the point-particle approximation for this system.

\subsection{Linear stability analysis of the particle rotational dynamics in quiescent fluid}
Here we derive theoretically the ideal critical ratio of $\omega_H/\omega_a$ in a noiseless system and estimate the relaxation time $\tau^*$. 
We consider the simplified nonlinear equation without the noise term, given by
\begin{equation}
    \dot{\beta} = \omega_H - \frac{\omega_a}{2} \sin 2\beta.
\end{equation}

At equilibrium, we set \(\dot{\beta} = 0\), which gives $\omega_H - \frac{\omega_a}{2} \sin 2\beta = 0$.
If \(\left| \frac{2\omega_H}{\omega_a} \right| \leq 1\), the equilibrium points are
\begin{equation}
\beta^* = \frac{1}{2} \arcsin \left( \frac{2\omega_H}{\omega_a} \right) + k \frac{\pi}{2}, \quad k \in \mathbb{Z}.
\end{equation}

Then we introduce a small perturbation around the equilibrium as $\beta = \beta^* + \delta$.
Substituting into the governing equation and linearizing the system as
$\dot{\delta} = -\omega_a \cos 2\beta^* \cdot \delta$. The solution is $\delta(t) = \delta(0) e^{-\omega_a \cos{2 \beta^* t}}$.

The stability of the equilibrium is determined by the eigenvalue $\lambda^* = -\omega_a \cos 2\beta^*$.
The equilibrium is stable if \(\lambda^* < 0\), which requires $\cos 2\beta^* > 0$.
Substituting the expression for \(\beta^*\), we get:
\begin{equation}
\cos \left( \arcsin \frac{2\omega_H}{\omega_a} \right) > 0.
\end{equation}

Using the identity \(\cos(\arcsin x) = \sqrt{1 - x^2}\), we obtain:
\begin{equation}
\sqrt{1 - \frac{4\omega_H^2}{\omega_a^2}} > 0,
\end{equation}
which implies $|\omega_H| < \frac{\omega_a}{2}$.

The corresponding relaxation time is 
\begin{align}
\label{eq:relax_time}
\tau^* &= \frac{1}{|\lambda^*|} 
% &= \frac{1}{\omega_a|\cos{2\beta^*}|}\\
 = \frac{1}{\omega_a \sqrt{1-4\left(\frac{\omega_H}{\omega_a}\right)^2}}.
\end{align}

Physically, the relaxation time $\tau^*$ means the characteristic time it takes for a perturbed rotational motion of the particle to return to its phase-locked state, i.e., $\dot{\beta}=0$, after a small disturbance of $\delta$.

So if $|\omega_H| < \omega_a / 2$, the equilibrium points are stable. If $ |\omega_H| > \omega_a / 2 $, the equilibrium points disappear and the system becomes unstable \cite{cimurs2019stability}. Therefore, the critical ratio is $\omega_H/\omega_a = 1/2$.
If we use $\omega^* = 1/\tau^*$ as the characteristic relaxation frequency, then we can reformulate Eq.~\eqref{eq:relax_time} as 
\begin{align}
\label{eq:relax_freq}
\frac{\omega^*}{\omega_a} = \sqrt{1-4\left(\frac{\omega_H}{\omega_a}\right)^2},
\end{align}
which is useful when analyzing the mode of the resonance curve later.

% \section{Numerical simulation}
% \subsection{Direct numerical simulation}
% Pseudo-spectral method

% \subsection{Numerical integration of theoretical model}
% Third-order Runge-Kutta method

\section{Waiting time statistics and equivalent noise intensity estimation}
To gain deeper insight into the stochastic nature of particle rotation in turbulence, we analyze the waiting time, $\tau$, between transitions of the phase lag angle, $\beta$, in the absence of an external forcing ($\omega_H = 0$). This allows us to isolate the effect of turbulent vorticity (stochastic fluctuations) on the rotational dynamics of the particle. 
The waiting time, $\tau$, is defined as the duration (marked by red segments in Supplementary Fig.~\ref{fig:fig3}a) between successive transitions of $\beta$ between its metastable states. 

For the light particle case, as shown in Supplementary Fig.~\ref{fig:fig3}b, the PDF of $\tau$ follows an exponential distribution, $P(\tau) \sim e^{-\lambda \tau}$, where $\lambda$ represents the characteristic transition rate. This result is consistent with the expected behavior of a noise-driven escape process over an energy barrier, where transitions occur randomly with a Poisson-like distribution. Such Poisson-like distribution behavior is robust for light, neutral, and heavy particles (Supplementary Fig.~\ref{fig:fig3}c), with different transition rates, $\lambda$, for different types of particles indicating exploring different regions (and thus experiencing different vorticity) in turbulence. The light particles preferentially concentrate in the high vorticity regions \cite{mathai2020bubbly,calzavarini2008dimensionality, wang2024localization}, with stronger turbulence vorticity fluctuations driving $\beta$ across the energy barrier, leading to a faster transition (higher $\lambda$). While the heavy particles tend to concentrate in the low vorticity region and thus a slower transition rate. 

The transition rate $\lambda$ varies with the noise intensity $\omega_a$, as shown in Supplementary Fig.~\ref{fig:fig3}d for different types of particles (light, neutral, and heavy). Specifically, $\lambda$ exhibits an exponential dependence on the normalized potential barrier $\omega_a/\omega_\eta$, namely $\lambda \sim \exp(-\omega_a/\epsilon)$. This behavior is reminiscent of Kramers' escape rate in a double-well potential subjected to thermal noise: given by $\lambda \sim e^{-\Delta U/\epsilon}$ \cite{benzi2010stochastic, gammaitoni1998stochastic, coffey2012langevin}, where $\Delta U$ is the potential barrier and $\epsilon$ represents the noise strength. In our case, the effective energy barrier between stable states of $\beta$ is proportional to $\omega_a$. 
% The parameter $\epsilon$ characterizes the effective noise strength induced by turbulence governing the system stochastic transitions.
Physically, $\epsilon$ quantifies the relative strength of turbulent fluctuations compared to the deterministic restoring forces acting on the particle. A larger $\epsilon$ implies a stronger stochastic influence, facilitating more frequent transitions between metastable states, while a lower $\epsilon$ implies that the system remains trapped in its metastable states for longer durations. This aligns with the earlier observation that particles of different densities explore distinct regions of the turbulence and thus different transition rates (slopes) are observed (Supplementary Fig.~\ref{fig:fig3}c). The estimated equivalent noise intensity shown in Supplementary Fig.~\ref{fig:fig3}e provides a direct means to quantify turbulence-induced noise intensity from observable transition dynamics. Within the investigated parameter space, turbulent fluctuations effectively act as a random forcing term that continuously perturbs the system, leading to memoryless escape events over the effective potential barriers. However, this picture may change if the particle enters the underdamped regime or has a finite size, where inertial effects should be taken into account \cite{toschi2009lagrangian,zimmermann2011rotational,homann2010finite}. 
% offering valuable insights into the interplay between stochastic forcing and rotational response in turbulent environments.  

Using the equivalent noise intensity $\epsilon$ shown in Supplementary Fig.~\ref{fig:fig3}e of the light particle (blue symbol), we simulate the behavior of magnetic particles subjected to a rotating magnetic field in the presence of white noise. Specifically, in the governing equation, we replace the turbulent vorticity signal, $\omega_{f,z}$, with a stochastic noise term $\sqrt{\epsilon}\mathcal{N}(0,1)$, with $\mathcal{N}(0,1)$ representing a standard normal distribution with zero mean and unit variance. 
As shown in Supplementary Fig.~\ref{fig:fig4}, the results of the light particles subjected to white noise (Supplementary Fig.~\ref{fig:fig4}f-j) are qualitatively similar to those of the light particles in turbulence (Supplementary Fig.~\ref{fig:fig4}a-e). This comparison demonstrates that the stochastic transitions observed for the light particles can be effectively captured using a simple white noise model with an appropriately chosen noise intensity $\epsilon$.

\section{Results of heavy and neutral particles}\label{secheavy_neutral}

In addition to the simulation results presented in the main paper on light particles in turbulence subjected to a rotating magnetic field, we also investigate the dynamics of particles with different densities. Specifically, we analyze neutral particles, characterized by a density contrast parameter $\beta_p = \frac{3}{1+2\rho_p / \rho_f}=1$ and heavy particles with $\beta_p = 0.01$.

Supplementary Fig.~\ref{fig:comparison} compares the rotational dynamics and stochastic resonance behavior of light, neutral, and heavy particles in a turbulent flow under an applied rotational magnetic field.

The phase diagrams in Supplementary Fig.~\ref{fig:comparison}a, f, and k illustrate different rotational regimes: phase-locked, back-and-forth, and turbulent dynamics. A key dynamical transition occurs when $\omega_a = 2\omega_H+\omega_\eta$ for the light particles (Supplementary Fig.~\ref{fig:comparison}a) based on the fact that light particles explore the high vorticity region with a fluctuation of the same order as $\omega_\eta$. While for the heavy particles, because they explore the low vorticity region (smaller than $\omega_\eta$), the transition boundary is shifted, indicating a wider parameter space for the phase-locked regime in which the particle maintains a stable rotational response to the external field (the purple colored region in Supplementary Fig.~\ref{fig:comparison}k). Neutral particles, which distribute more uniformly within the turbulence, exhibit a transition boundary that lies between those of light and heavy particles.

To further explore the stochastic resonance effect, Figures~\ref{fig:comparison}b, g, and l present phase diagrams where the color represents the normalized time-averaged derivative of the phase lag, $\langle \dot{\beta} \rangle / \omega_\eta$. A more quantitative analysis of this resonance behavior can be obtained by plotting $\langle \dot{\beta} \rangle / \omega_\eta$ as a function of $\omega_\eta / \omega_a$ for different values of $\omega_H / \omega_a$. A pronounced resonance peak is observed at $\omega_\eta / \omega_a \approx 1$ for light particles (Supplementary Fig.~\ref{fig:comparison}c), $\approx 1.5$ for neutral particles (Supplementary Fig.~\ref{fig:comparison}h), and $\approx 2$ for heavy particles (Supplementary Fig.~\ref{fig:comparison}m) when $\omega_H / \omega_a < 1/2$. 
This shift in resonance frequency reflects the underlying preferential concentration effects \cite{mathai2020bubbly,calzavarini2008dimensionality, wang2024localization}: lighter particles predominantly experience high-vorticity fluctuations, whereas heavier particles are more likely to sample low-vorticity regions. The resonance peak occurs when the applied magnetic field intensity $\omega_a$ is comparable to the turbulence vorticity experienced by the particles. If we define $\omega_\mathrm{light}$, $\omega_\mathrm{neutral}$, and $\omega_\mathrm{heavy}$ as the turbulence vorticity experienced by light, neutral, and heavy particles, respectively, and denote $\omega_{a,\mathrm{light}}$, $\omega_{a,\mathrm{neutral}}$, and $\omega_{a,\mathrm{heavy}}$ as the magnetic field intensities at which resonance peaks appear, we find that $\omega_\mathrm{light}>\omega_\mathrm{neutral}>\omega_\mathrm{heavy}$, leading to $\omega_{a,\mathrm{light}}>\omega_{a,\mathrm{neutral}}>\omega_{a,\mathrm{heavy}}$. Given that the resonance curves are normalized by $\omega_\eta$, the peak positions satisfy 
\begin{equation}
    \frac{\omega_\eta}{\omega_{a,\mathrm{light}}}<\frac{\omega_\eta}{\omega_{a,\mathrm{neutral}}}<\frac{\omega_\eta}{\omega_{a,\mathrm{heavy}}}.
\end{equation}

For sufficiently large values of $\omega_\eta / \omega_a$, the data collapses onto a universal curve when normalized by $\omega_H / \omega_a$, as shown in Supplementary Figures~\ref{fig:comparison}d, i, and n. This collapse suggests that the linear response regime, where the turbulence vorticity dominates, is robust across different particle types.

Finally, Supplementary Figures~\ref{fig:comparison}e, j, and o show the dependence of the resonance peak location $\left[\frac{\omega_\eta}{\omega_a} \right]_r$ on the ratio $\omega_H / \omega_a$. The dashed line represents a fit based on the relaxation frequency, see Eq.~\eqref{eq:relax_freq}, giving
\begin{equation}
    \left[\frac{\omega_\eta}{\omega_a} \right]_r = C \sqrt{1 - 4 \left( \frac{\omega_H}{\omega_a} \right)^2},
    \label{eq:relax_freq_fit}
\end{equation}
where $C$ is a dimensionless fitting parameter. The excellent agreement between the resonance peak location and this fit underscores the predictive power of the relaxation frequency model.

The resonance peak location $\left[\frac{\omega_\eta}{\omega_a} \right]_r$ is strongly influenced by the proximity to the critical ratio $\omega_H/\omega_a=1/2$, highlighting the selection of the corresponding relaxation time.

When $\omega_H/\omega_a$ is significantly smaller than this critical value, the resonance peak location $\left[\frac{\omega_\eta}{\omega_a} \right]_r$ remains almost constant, which is determined by the turbulence vorticity fluctuations experienced by the particles. However, as $\omega_H/\omega_a$ approaches the critical value, $\left[\frac{\omega_\eta}{\omega_a} \right]_r$ decreases rapidly as the relaxation time diverges.
A longer relaxation time allows particles to diffuse out of the potential well even at lower noise levels $\omega_\eta/\omega_a$, explaining the observed shift in resonance peak location.

\begin{figure*}[!h]
    \centering
    \includegraphics[width=0.8\textwidth]{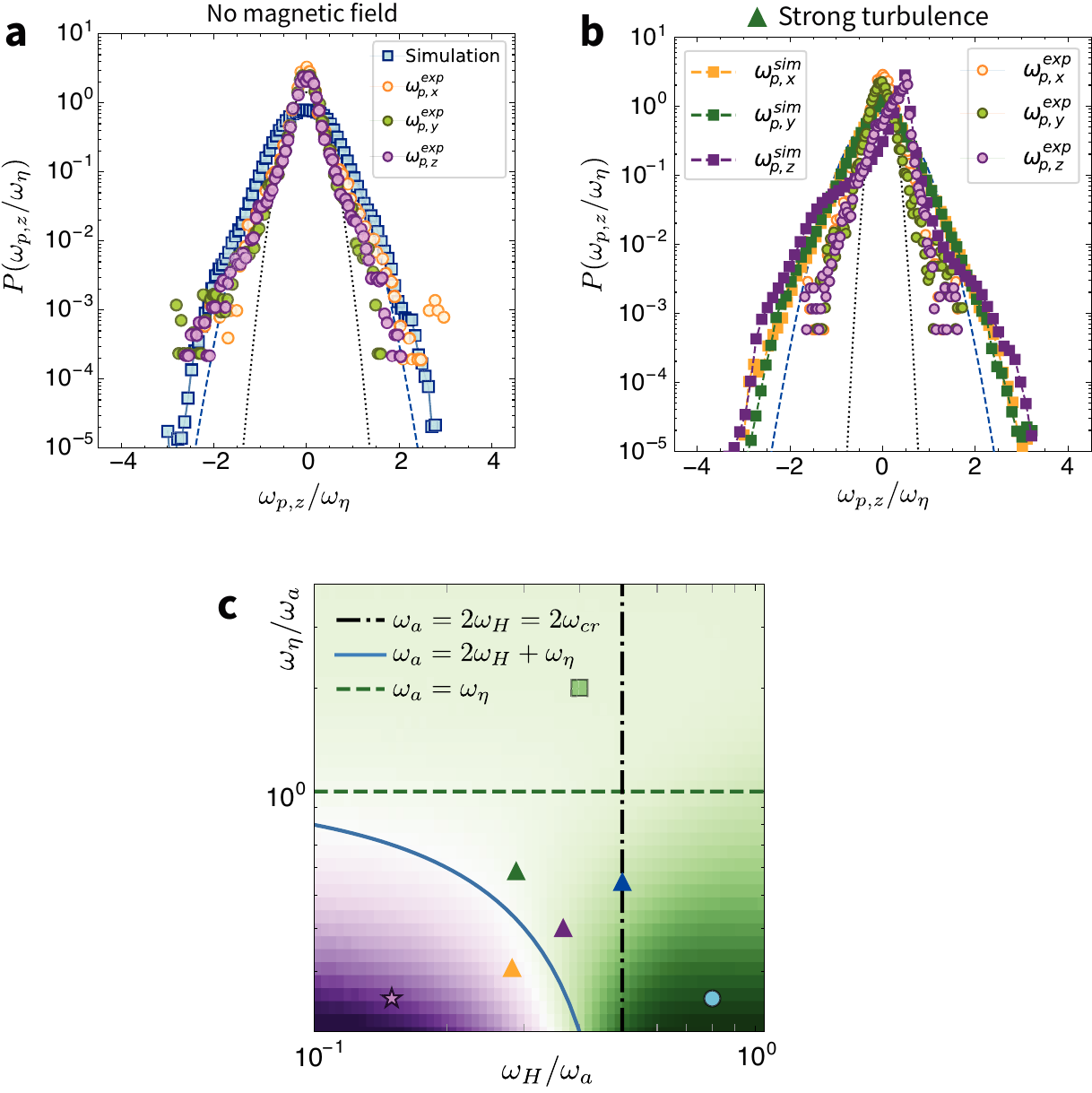}
    \caption{\textbf{Comparison between experimental and numerical results.}
    \textbf{a} and \textbf{b}: The probability density distribution of particle angular velocity $\omega_{p,i}$ with $i=x,y,z$ representing the three components.
    \textbf{a}, In the absence of a magnetic field, the particle's rotation is purely driven by turbulence, exhibiting isotropic behavior. The simulation results (squares) are flattened over all directions (i.e., reshape the angular velocity vector into a one-dimensional (1D) array while preserving the original data order) for direct comparison with the experimental results (circles). 
    \textbf{b}, With an applied magnetic field ($1.6~\text{mT}$) and increased turbulence (here $Re_{\lambda}\approx 360$, while the turbulence used in Fig.~2 of the main text is Re$_{\lambda}\approx 320$), there is a good agreement indicating the numerical simulation can indeed capture the main physics of the experiments when the physical parameters are normalized by the Kolmogrov scales, i.e, $\omega_\eta$.
    % and keep the magnetic field intensity ($1.6~\text{mT}$) the same as that of Fig.2d in the main text.
    \textbf{c}, Phase diagram of particle rotation dynamic regimes (reproduced from Fig.~3m of the main text). The experimental setting of \textbf{b} is marked as a green triangle. 
    }
    \label{fig:fig1}
\end{figure*}

\begin{figure*}[!h]
    \centering
    \includegraphics[width=0.9\textwidth]{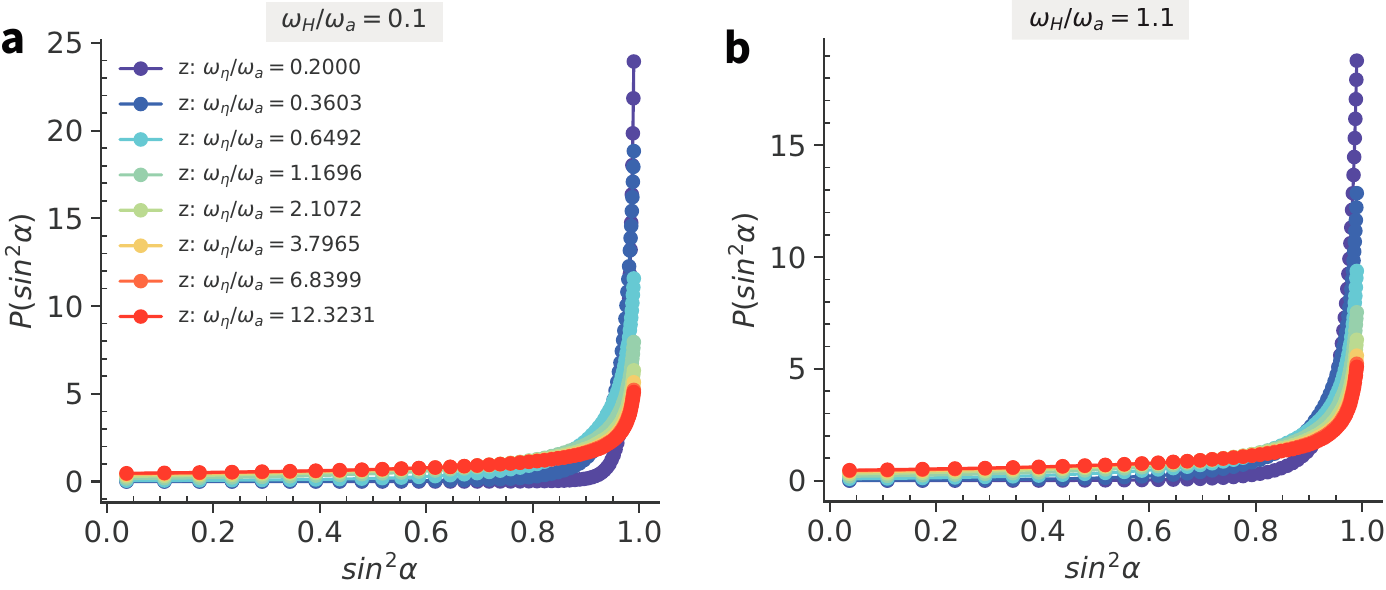}
    \caption{\textbf{Probability density distribution of $\kappa$.}
    \textbf{a}, In the subcritical regime ($\omega_H/\omega_a=0.1<1/2$), the PDFs of $\kappa = \sin^2{\alpha}$ for different levels of noise intensities ($\omega_\eta/\omega_a$).
    \textbf{b}, In the supercritical regime ($\omega_H/\omega_a=1.1>1/2$), the PDFs of $\sin^2\alpha$ for different levels of noise intensities ($\omega_\eta/\omega_a$). Symbols with the same color correspond to the same value of $\omega_\eta/\omega_a$ in \textbf{a}.
    }
    \label{fig:fig2}
\end{figure*}

\begin{figure*}[!t]
    \centering
    \includegraphics[width=0.8\textwidth]{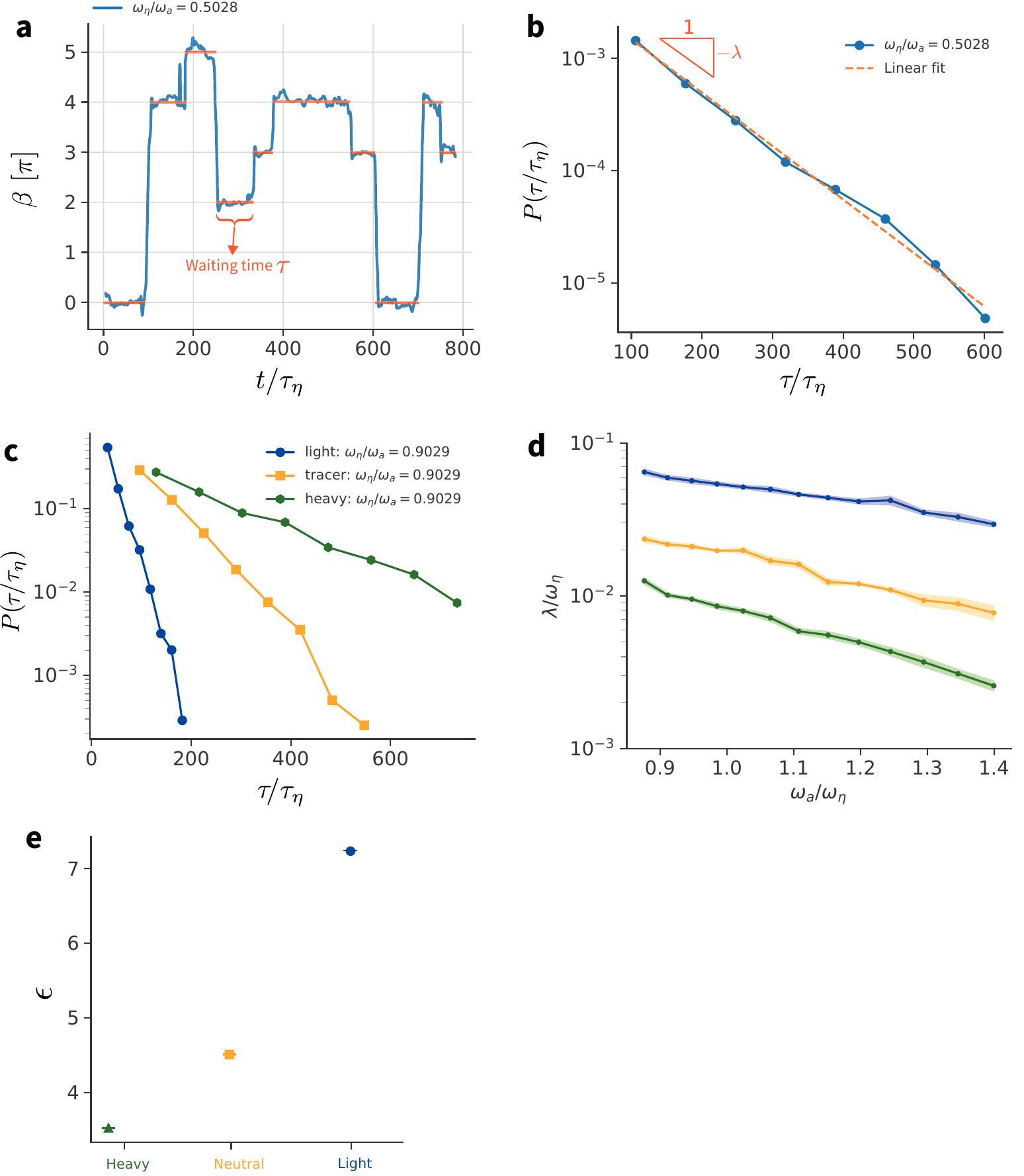}
    \caption{
    \textbf{Statistical analysis of waiting time.}
    The statistical properties of transition time, $\tau$, are analyzed in the absence of external forcing ($\omega_H = 0$), where particle rotational dynamics are solely influenced by turbulent vorticity (stochastic fluctuations).
    \textbf{a}, A typical time series of the phase lag angle, $\beta$ (blue solid line), with waiting times $\tau$ (red segments) marked between transitions. Parameter setting: $\omega_a/\omega_\eta = 2.0$ for a light particle.  
    \textbf{b}, The PDF of the normalized waiting time, $\tau/\tau_\eta$. The red dashed line represents an exponential fit, $P(\tau) \sim e^{-\lambda \tau}$ where $\lambda$ is the characteristic transition rate.
    \textbf{c}, Comparison of the PDF of normalized waiting time, $\tau/\tau_\eta$, for light (blue), neutral (yellow), and heavy (green) particles. The data are collective results of 512 particles for a duration of $12 T_L$ (with $T_L$ representing the turbulence integral time scale). The results confirm that $P(\tau) \sim e^{-\lambda \tau}$ holds across different particle types. Parameter setting: $\omega_a/\omega_\eta = 1.1$.
    \textbf{d}, The characteristic transition rate, $\lambda$, as a function of the applied noise intensity, $\omega_a/\omega_\eta$, for light (blue), neutral (yellow), and heavy (green) particles. The error bar is indicated by the shaded area, which is the covariance from the fitting process of $\lambda$ for the relation $P(\tau) \sim e^{-\lambda \tau}$.
    \textbf{e}, The fitted noise intensity, $\epsilon$. Here, $\epsilon$ means the equivalent noise intensity in the white noise scenario, i.e., $\omega_{p,z} \sim \sqrt{\epsilon} \mathcal{N}(0,1)$ with $\mathcal{N}(0,1)$ representing the standard normal distribution which has a mean of zero and a variance of one, satisfying the relationship $\lambda \sim e^{-\omega_a/\epsilon}$. The error bar is the covariance from the fitting process of $\epsilon$ for the relation $\lambda \sim e^{-\omega_a/\epsilon}$.
    }
    \label{fig:fig3}
\end{figure*}

\begin{figure*}[!t]
    \centering
    \includegraphics[width=0.72\textwidth]{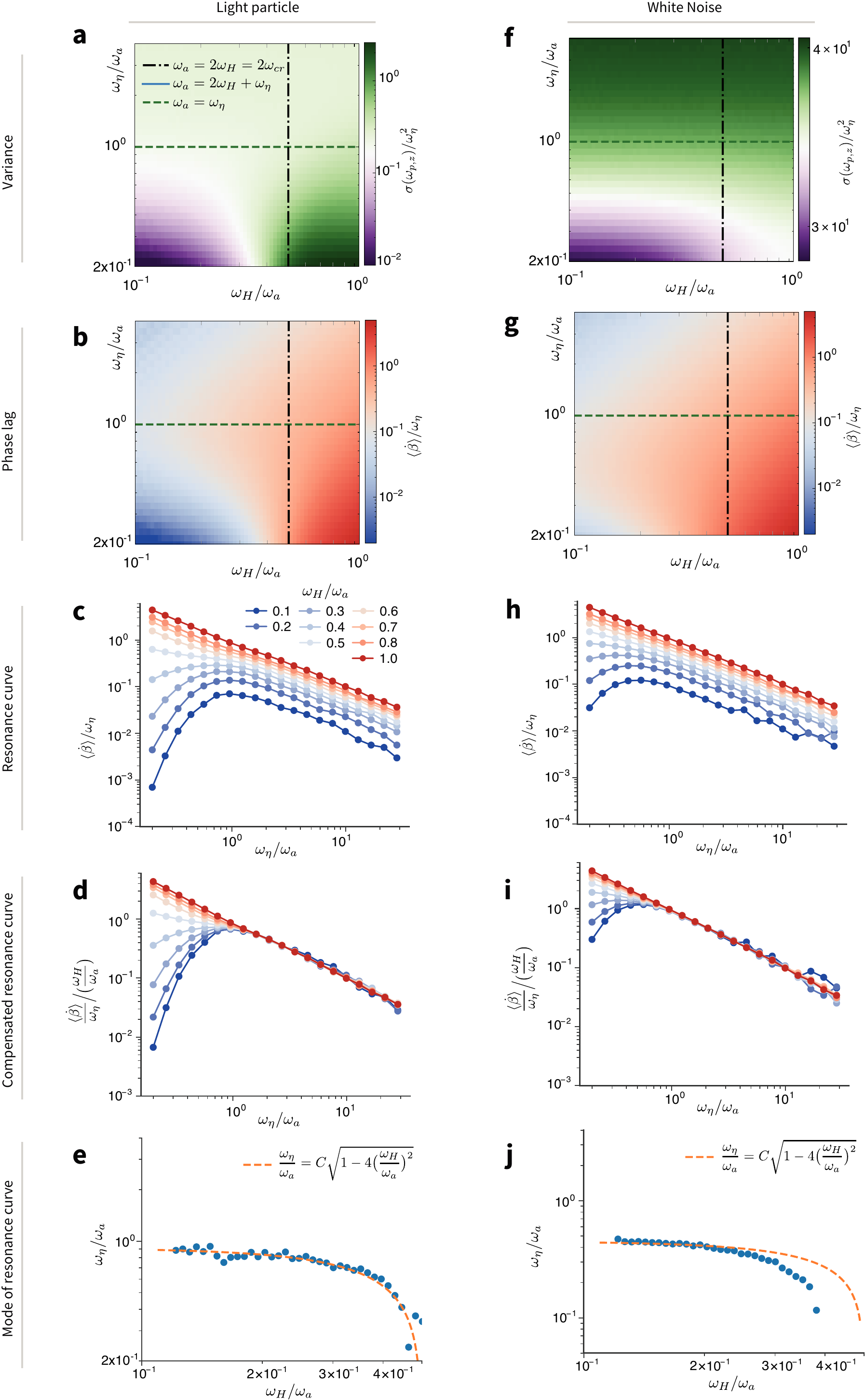}
    \caption{\textbf{Comparison between the turbulence vorticity signal and white noise.}
    The white noise term, $\sqrt{\epsilon}\mathcal{N}(0,1)$, in the governing equation of the numerical integration has one key parameter, $\epsilon$, which is the noise intensity. We extract $\epsilon$ from the light particle results (namely the value of blue symbol in Supplementary Fig.~\ref{fig:fig3}e) and use it as the input parameter for the white noise simulation.
    The white noise simulation results are shown in \textbf{f}-\textbf{j}.
    For the convenience of comparison, we also show in \textbf{a}-\textbf{e} the corresponding results of light particles (namely the results shown in the main text). 
    \textbf{e} and \textbf{j}, the mode of the resonance curve shown in \textbf{c} and \textbf{h} as a function of the ratio $\omega_H/\omega_a$. The dashed line is a fitting by Eq.~\eqref{eq:relax_freq}, i.e., $\omega_\eta/\omega_a = C \sqrt{1-4(\omega_H/\omega_a)^2}$ with C a fitting parameter. C=0.91 (\textbf{e}) and 0.45 (\textbf{j}).
    }
    \label{fig:fig4}
\end{figure*}

\begin{figure*}[!t]
    \centering
    \includegraphics[width=0.9\textwidth]{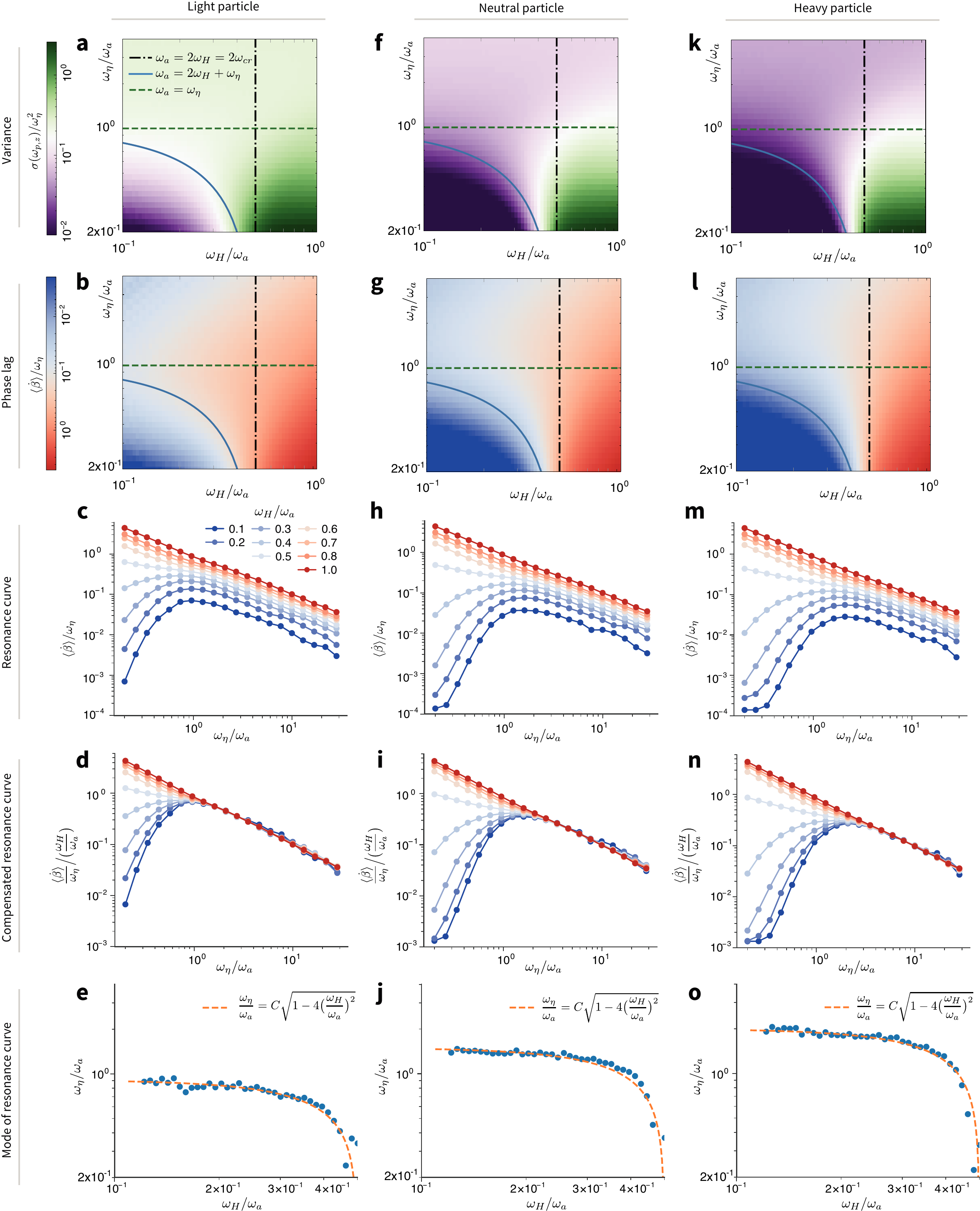}
    \caption{\textbf{Comparison between the light, neutral, and heavy particles.}
    \textbf{a}, \textbf{f}, and \textbf{k}: Phase diagrams of particle rotational dynamics, colored by the normalized variance of the particle's angular velocity in the $z$-direction, $\langle \omega_{p,z}^2 - \langle \omega_{p,z}\rangle^2\rangle/\omega_\eta^2$.
    The black dot-dashed line indicates the critical frequency, $\omega_{cr} = \omega_a/2$. 
    The green dashed line marks the balance between turbulent fluctuations and magnetic strength, i.e., $\omega_a = \omega_\eta$. 
    The phase-locked regime, predicted by scaling analysis, is enclosed by the thick blue line, satisfying $\omega_a = 2\omega_H + \omega_\eta$. 
    \textbf{b}, \textbf{g}, and \textbf{l}: Phase diagrams of stochastic resonance. The color represents the normalized time-averaged derivative of the phase lag, $\langle \dot{\beta}\rangle/\omega_\eta$. 
    \textbf{c}, \textbf{h}, and \textbf{m}: Plots of
    $\langle \dot{\beta}\rangle/\omega_\eta$ against $\omega_\eta/\omega_a$ for various values of $\omega_H/\omega_a$.
    A pronounced resonant peak emerges at $\omega_\eta/\omega_a \approx 1$ for light particles (\textbf{c}), $\approx 1.5$ for neutral particles (\textbf{h}), and $\approx 2$ for heavy particles (\textbf{m}) when $\omega_H/\omega_a<1/2$. 
    A perfect collapse is observed for the linear regime (high $\omega_\eta/\omega_a$) when normalized by $\omega_H/\omega_a$ (\textbf{d}, \textbf{i}, and \textbf{n}).
    \textbf{e}, \textbf{j}, and \textbf{o}: The mode of the resonance curve from \textbf{d}, \textbf{i}, and \textbf{n} as a function of the $\omega_H/\omega_a$. The dashed line represents a fit base on Eq.~\eqref{eq:relax_freq}, given by $\omega_\eta/\omega_a = C \sqrt{1-4(\omega_H/\omega_a)^2}$ with fitting parameters: C=0.91 (\textbf{e}), 1.5 (\textbf{j}), and 2 (\textbf{o}).
    }
    \label{fig:comparison}
\end{figure*}

\clearpage
\bibliography{main}

\begin{thebibliography}{95}%
\makeatletter
\providecommand \@ifxundefined [1]{%
 \@ifx{#1\undefined}
}%
\providecommand \@ifnum [1]{%
 \ifnum #1\expandafter \@firstoftwo
 \else \expandafter \@secondoftwo
 \fi
}%
\providecommand \@ifx [1]{%
 \ifx #1\expandafter \@firstoftwo
 \else \expandafter \@secondoftwo
 \fi
}%
\providecommand \natexlab [1]{#1}%
\providecommand \enquote  [1]{``#1''}%
\providecommand \bibnamefont  [1]{#1}%
\providecommand \bibfnamefont [1]{#1}%
\providecommand \citenamefont [1]{#1}%
\providecommand \href@noop [0]{\@secondoftwo}%
\providecommand \href [0]{\begingroup \@sanitize@url \@href}%
\providecommand \@href[1]{\@@startlink{#1}\@@href}%
\providecommand \@@href[1]{\endgroup#1\@@endlink}%
\providecommand \@sanitize@url [0]{\catcode `\\12\catcode `\$12\catcode
  `\&12\catcode `\#12\catcode `\^12\catcode `\_12\catcode `\%12\relax}%
\providecommand \@@startlink[1]{}%
\providecommand \@@endlink[0]{}%
\providecommand \url  [0]{\begingroup\@sanitize@url \@url }%
\providecommand \@url [1]{\endgroup\@href {#1}{\urlprefix }}%
\providecommand \urlprefix  [0]{URL }%
\providecommand \Eprint [0]{\href }%
\providecommand \doibase [0]{https://doi.org/}%
\providecommand \selectlanguage [0]{\@gobble}%
\providecommand \bibinfo  [0]{\@secondoftwo}%
\providecommand \bibfield  [0]{\@secondoftwo}%
\providecommand \translation [1]{[#1]}%
\providecommand \BibitemOpen [0]{}%
\providecommand \bibitemStop [0]{}%
\providecommand \bibitemNoStop [0]{.\EOS\space}%
\providecommand \EOS [0]{\spacefactor3000\relax}%
\providecommand \BibitemShut  [1]{\csname bibitem#1\endcsname}%
\let\auto@bib@innerbib\@empty
%</preamble>
\bibitem [{\citenamefont {Frisch}(1995)}]{frisch1995turbulence}%
  \BibitemOpen
  \bibfield  {author} {\bibinfo {author} {\bibfnamefont {U.}~\bibnamefont
  {Frisch}},\ }\href@noop {} {\emph {\bibinfo {title} {Turbulence: The Legacy
  of AN Kolmogorov}}}\ (\bibinfo  {publisher} {Cambridge University Press},\
  \bibinfo {year} {1995})\BibitemShut {NoStop}%
\bibitem [{\citenamefont {Pope}(2000)}]{Pope_2000}%
  \BibitemOpen
  \bibfield  {author} {\bibinfo {author} {\bibfnamefont {S.~B.}\ \bibnamefont
  {Pope}},\ }\href@noop {} {\emph {\bibinfo {title} {Turbulent Flows}}}\
  (\bibinfo  {publisher} {Cambridge University Press},\ \bibinfo {year}
  {2000})\BibitemShut {NoStop}%
\bibitem [{\citenamefont {Jimenez}\ and\ \citenamefont
  {Wray}(1998)}]{jimenez1998characteristics}%
  \BibitemOpen
  \bibfield  {author} {\bibinfo {author} {\bibfnamefont {J.}~\bibnamefont
  {Jimenez}}\ and\ \bibinfo {author} {\bibfnamefont {A.~A.}\ \bibnamefont
  {Wray}},\ }\bibfield  {title} {\bibinfo {title} {On the characteristics of
  vortex filaments in isotropic turbulence},\ }\href@noop {} {\bibfield
  {journal} {\bibinfo  {journal} {Journal of Fluid Mechanics}\ }\textbf
  {\bibinfo {volume} {373}},\ \bibinfo {pages} {255} (\bibinfo {year}
  {1998})}\BibitemShut {NoStop}%
\bibitem [{\citenamefont {Sreenivasan}\ and\ \citenamefont
  {Antonia}(1997)}]{sreenivasan1997phenomenology}%
  \BibitemOpen
  \bibfield  {author} {\bibinfo {author} {\bibfnamefont {K.~R.}\ \bibnamefont
  {Sreenivasan}}\ and\ \bibinfo {author} {\bibfnamefont {R.}~\bibnamefont
  {Antonia}},\ }\bibfield  {title} {\bibinfo {title} {The phenomenology of
  small-scale turbulence},\ }\href@noop {} {\bibfield  {journal} {\bibinfo
  {journal} {Annual Review of Fluid Mechanics}\ }\textbf {\bibinfo {volume}
  {29}},\ \bibinfo {pages} {435} (\bibinfo {year} {1997})}\BibitemShut
  {NoStop}%
\bibitem [{\citenamefont {Wallace}\ and\ \citenamefont
  {Foss}(1995)}]{wallace1995measurement}%
  \BibitemOpen
  \bibfield  {author} {\bibinfo {author} {\bibfnamefont {J.~M.}\ \bibnamefont
  {Wallace}}\ and\ \bibinfo {author} {\bibfnamefont {J.~F.}\ \bibnamefont
  {Foss}},\ }\bibfield  {title} {\bibinfo {title} {The measurement of vorticity
  in turbulent flows},\ }\href@noop {} {\bibfield  {journal} {\bibinfo
  {journal} {Annual Review of Fluid Mechanics}\ }\textbf {\bibinfo {volume}
  {27}},\ \bibinfo {pages} {469} (\bibinfo {year} {1995})}\BibitemShut
  {NoStop}%
\bibitem [{\citenamefont {Marcus}\ \emph {et~al.}(2014)\citenamefont {Marcus},
  \citenamefont {Parsa}, \citenamefont {Kramel}, \citenamefont {Ni},\ and\
  \citenamefont {Voth}}]{marcus2014measurements}%
  \BibitemOpen
  \bibfield  {author} {\bibinfo {author} {\bibfnamefont {G.~G.}\ \bibnamefont
  {Marcus}}, \bibinfo {author} {\bibfnamefont {S.}~\bibnamefont {Parsa}},
  \bibinfo {author} {\bibfnamefont {S.}~\bibnamefont {Kramel}}, \bibinfo
  {author} {\bibfnamefont {R.}~\bibnamefont {Ni}},\ and\ \bibinfo {author}
  {\bibfnamefont {G.~A.}\ \bibnamefont {Voth}},\ }\bibfield  {title} {\bibinfo
  {title} {Measurements of the solid-body rotation of anisotropic particles in
  3d turbulence},\ }\href@noop {} {\bibfield  {journal} {\bibinfo  {journal}
  {New Journal of Physics}\ }\textbf {\bibinfo {volume} {16}},\ \bibinfo
  {pages} {102001} (\bibinfo {year} {2014})}\BibitemShut {NoStop}%
\bibitem [{\citenamefont {Mathai}\ \emph {et~al.}(2020)\citenamefont {Mathai},
  \citenamefont {Lohse},\ and\ \citenamefont {Sun}}]{mathai2020bubbly}%
  \BibitemOpen
  \bibfield  {author} {\bibinfo {author} {\bibfnamefont {V.}~\bibnamefont
  {Mathai}}, \bibinfo {author} {\bibfnamefont {D.}~\bibnamefont {Lohse}},\ and\
  \bibinfo {author} {\bibfnamefont {C.}~\bibnamefont {Sun}},\ }\bibfield
  {title} {\bibinfo {title} {Bubbly and buoyant particle--laden turbulent
  flows},\ }\href@noop {} {\bibfield  {journal} {\bibinfo  {journal} {Annual
  Review of Condensed Matter Physics}\ }\textbf {\bibinfo {volume} {11}},\
  \bibinfo {pages} {529} (\bibinfo {year} {2020})}\BibitemShut {NoStop}%
\bibitem [{\citenamefont {Tierno}\ \emph {et~al.}(2009)\citenamefont {Tierno},
  \citenamefont {Claret}, \citenamefont {Sagu{\'e}s},\ and\ \citenamefont
  {C{\=e}bers}}]{tierno2009overdamped}%
  \BibitemOpen
  \bibfield  {author} {\bibinfo {author} {\bibfnamefont {P.}~\bibnamefont
  {Tierno}}, \bibinfo {author} {\bibfnamefont {J.}~\bibnamefont {Claret}},
  \bibinfo {author} {\bibfnamefont {F.}~\bibnamefont {Sagu{\'e}s}},\ and\
  \bibinfo {author} {\bibfnamefont {A.}~\bibnamefont {C{\=e}bers}},\ }\bibfield
   {title} {\bibinfo {title} {Overdamped dynamics of paramagnetic ellipsoids in
  a precessing magnetic field},\ }\href@noop {} {\bibfield  {journal} {\bibinfo
   {journal} {Physical Review E}\ }\textbf {\bibinfo {volume} {79}},\ \bibinfo
  {pages} {021501} (\bibinfo {year} {2009})}\BibitemShut {NoStop}%
\bibitem [{\citenamefont {Guazzelli}\ and\ \citenamefont
  {Morris}(2011)}]{guazzelli2011physical}%
  \BibitemOpen
  \bibfield  {author} {\bibinfo {author} {\bibfnamefont {E.}~\bibnamefont
  {Guazzelli}}\ and\ \bibinfo {author} {\bibfnamefont {J.~F.}\ \bibnamefont
  {Morris}},\ }\href@noop {} {\emph {\bibinfo {title} {A Physical Introduction
  to Suspension Dynamics}}},\ Vol.~\bibinfo {volume} {45}\ (\bibinfo
  {publisher} {Cambridge University Press},\ \bibinfo {year}
  {2011})\BibitemShut {NoStop}%
\bibitem [{\citenamefont {Calzavarini}\ \emph {et~al.}(2008)\citenamefont
  {Calzavarini}, \citenamefont {Kerscher}, \citenamefont {Lohse},\ and\
  \citenamefont {Toschi}}]{calzavarini2008dimensionality}%
  \BibitemOpen
  \bibfield  {author} {\bibinfo {author} {\bibfnamefont {E.}~\bibnamefont
  {Calzavarini}}, \bibinfo {author} {\bibfnamefont {M.}~\bibnamefont
  {Kerscher}}, \bibinfo {author} {\bibfnamefont {D.}~\bibnamefont {Lohse}},\
  and\ \bibinfo {author} {\bibfnamefont {F.}~\bibnamefont {Toschi}},\
  }\bibfield  {title} {\bibinfo {title} {Dimensionality and morphology of
  particle and bubble clusters in turbulent flow},\ }\href@noop {} {\bibfield
  {journal} {\bibinfo  {journal} {Journal of Fluid Mechanics}\ }\textbf
  {\bibinfo {volume} {607}},\ \bibinfo {pages} {13} (\bibinfo {year}
  {2008})}\BibitemShut {NoStop}%
\bibitem [{\citenamefont {Wang}\ \emph {et~al.}(2024)\citenamefont {Wang},
  \citenamefont {de~Wit},\ and\ \citenamefont {Toschi}}]{wang2024localization}%
  \BibitemOpen
  \bibfield  {author} {\bibinfo {author} {\bibfnamefont {Z.}~\bibnamefont
  {Wang}}, \bibinfo {author} {\bibfnamefont {X.~M.}\ \bibnamefont {de~Wit}},\
  and\ \bibinfo {author} {\bibfnamefont {F.}~\bibnamefont {Toschi}},\
  }\bibfield  {title} {\bibinfo {title} {Localization--delocalization
  transition for light particles in turbulence},\ }\href@noop {} {\bibfield
  {journal} {\bibinfo  {journal} {Proceedings of the National Academy of
  Sciences}\ }\textbf {\bibinfo {volume} {121}},\ \bibinfo {pages}
  {e2405459121} (\bibinfo {year} {2024})}\BibitemShut {NoStop}%
\bibitem [{\citenamefont {Martin}(2009)}]{martin2009theory}%
  \BibitemOpen
  \bibfield  {author} {\bibinfo {author} {\bibfnamefont {J.~E.}\ \bibnamefont
  {Martin}},\ }\bibfield  {title} {\bibinfo {title} {Theory of strong intrinsic
  mixing of particle suspensions in vortex magnetic fields},\ }\href@noop {}
  {\bibfield  {journal} {\bibinfo  {journal} {Physical Review E}\ }\textbf
  {\bibinfo {volume} {79}},\ \bibinfo {pages} {011503} (\bibinfo {year}
  {2009})}\BibitemShut {NoStop}%
\bibitem [{\citenamefont {Shanko}\ \emph {et~al.}(2019)\citenamefont {Shanko},
  \citenamefont {van~de Burgt}, \citenamefont {Anderson},\ and\ \citenamefont
  {den Toonder}}]{shanko2019microfluidic}%
  \BibitemOpen
  \bibfield  {author} {\bibinfo {author} {\bibfnamefont {E.-S.}\ \bibnamefont
  {Shanko}}, \bibinfo {author} {\bibfnamefont {Y.}~\bibnamefont {van~de
  Burgt}}, \bibinfo {author} {\bibfnamefont {P.~D.}\ \bibnamefont {Anderson}},\
  and\ \bibinfo {author} {\bibfnamefont {J.~M.}\ \bibnamefont {den Toonder}},\
  }\bibfield  {title} {\bibinfo {title} {Microfluidic magnetic mixing at low
  reynolds numbers and in stagnant fluids},\ }\href@noop {} {\bibfield
  {journal} {\bibinfo  {journal} {Micromachines}\ }\textbf {\bibinfo {volume}
  {10}},\ \bibinfo {pages} {731} (\bibinfo {year} {2019})}\BibitemShut
  {NoStop}%
\bibitem [{\citenamefont {Yan}\ \emph {et~al.}(2012)\citenamefont {Yan},
  \citenamefont {Bloom}, \citenamefont {Bae}, \citenamefont {Luijten},\ and\
  \citenamefont {Granick}}]{yan2012linking}%
  \BibitemOpen
  \bibfield  {author} {\bibinfo {author} {\bibfnamefont {J.}~\bibnamefont
  {Yan}}, \bibinfo {author} {\bibfnamefont {M.}~\bibnamefont {Bloom}}, \bibinfo
  {author} {\bibfnamefont {S.~C.}\ \bibnamefont {Bae}}, \bibinfo {author}
  {\bibfnamefont {E.}~\bibnamefont {Luijten}},\ and\ \bibinfo {author}
  {\bibfnamefont {S.}~\bibnamefont {Granick}},\ }\bibfield  {title} {\bibinfo
  {title} {Linking synchronization to self-assembly using magnetic janus
  colloids},\ }\href@noop {} {\bibfield  {journal} {\bibinfo  {journal}
  {Nature}\ }\textbf {\bibinfo {volume} {491}},\ \bibinfo {pages} {578}
  (\bibinfo {year} {2012})}\BibitemShut {NoStop}%
\bibitem [{\citenamefont {Piet}\ \emph {et~al.}(2013)\citenamefont {Piet},
  \citenamefont {Straube}, \citenamefont {Snezhko},\ and\ \citenamefont
  {Aranson}}]{piet2013viscosity}%
  \BibitemOpen
  \bibfield  {author} {\bibinfo {author} {\bibfnamefont {D.}~\bibnamefont
  {Piet}}, \bibinfo {author} {\bibfnamefont {A.}~\bibnamefont {Straube}},
  \bibinfo {author} {\bibfnamefont {A.}~\bibnamefont {Snezhko}},\ and\ \bibinfo
  {author} {\bibfnamefont {I.}~\bibnamefont {Aranson}},\ }\bibfield  {title}
  {\bibinfo {title} {Viscosity control of the dynamic self-assembly in
  ferromagnetic suspensions},\ }\href@noop {} {\bibfield  {journal} {\bibinfo
  {journal} {Physical Review Letters}\ }\textbf {\bibinfo {volume} {110}},\
  \bibinfo {pages} {198001} (\bibinfo {year} {2013})}\BibitemShut {NoStop}%
\bibitem [{\citenamefont {Wittbracht}\ \emph {et~al.}(2012)\citenamefont
  {Wittbracht}, \citenamefont {Weddemann}, \citenamefont {Eickenberg},
  \citenamefont {Zahn},\ and\ \citenamefont
  {H{\"u}tten}}]{wittbracht2012enhanced}%
  \BibitemOpen
  \bibfield  {author} {\bibinfo {author} {\bibfnamefont {F.}~\bibnamefont
  {Wittbracht}}, \bibinfo {author} {\bibfnamefont {A.}~\bibnamefont
  {Weddemann}}, \bibinfo {author} {\bibfnamefont {B.}~\bibnamefont
  {Eickenberg}}, \bibinfo {author} {\bibfnamefont {M.}~\bibnamefont {Zahn}},\
  and\ \bibinfo {author} {\bibfnamefont {A.}~\bibnamefont {H{\"u}tten}},\
  }\bibfield  {title} {\bibinfo {title} {Enhanced fluid mixing and separation
  of magnetic bead agglomerates based on dipolar interaction in rotating
  magnetic fields},\ }\href@noop {} {\bibfield  {journal} {\bibinfo  {journal}
  {Applied Physics Letters}\ }\textbf {\bibinfo {volume} {100}} (\bibinfo
  {year} {2012})}\BibitemShut {NoStop}%
\bibitem [{\citenamefont {Falcon}\ \emph {et~al.}(2017)\citenamefont {Falcon},
  \citenamefont {Bacri},\ and\ \citenamefont {Laroche}}]{falcon2017dissipated}%
  \BibitemOpen
  \bibfield  {author} {\bibinfo {author} {\bibfnamefont {E.}~\bibnamefont
  {Falcon}}, \bibinfo {author} {\bibfnamefont {J.-C.}\ \bibnamefont {Bacri}},\
  and\ \bibinfo {author} {\bibfnamefont {C.}~\bibnamefont {Laroche}},\
  }\bibfield  {title} {\bibinfo {title} {Dissipated power within a turbulent
  flow forced homogeneously by magnetic particles},\ }\href@noop {} {\bibfield
  {journal} {\bibinfo  {journal} {Physical Review Fluids}\ }\textbf {\bibinfo
  {volume} {2}},\ \bibinfo {pages} {102601} (\bibinfo {year}
  {2017})}\BibitemShut {NoStop}%
\bibitem [{\citenamefont {Cazaubiel}\ \emph {et~al.}(2021)\citenamefont
  {Cazaubiel}, \citenamefont {Gorce}, \citenamefont {Bacri}, \citenamefont
  {Berhanu}, \citenamefont {Laroche},\ and\ \citenamefont
  {Falcon}}]{cazaubiel2021three}%
  \BibitemOpen
  \bibfield  {author} {\bibinfo {author} {\bibfnamefont {A.}~\bibnamefont
  {Cazaubiel}}, \bibinfo {author} {\bibfnamefont {J.-B.}\ \bibnamefont
  {Gorce}}, \bibinfo {author} {\bibfnamefont {J.-C.}\ \bibnamefont {Bacri}},
  \bibinfo {author} {\bibfnamefont {M.}~\bibnamefont {Berhanu}}, \bibinfo
  {author} {\bibfnamefont {C.}~\bibnamefont {Laroche}},\ and\ \bibinfo {author}
  {\bibfnamefont {E.}~\bibnamefont {Falcon}},\ }\bibfield  {title} {\bibinfo
  {title} {Three-dimensional turbulence generated homogeneously by magnetic
  particles},\ }\href@noop {} {\bibfield  {journal} {\bibinfo  {journal}
  {Physical Review Fluids}\ }\textbf {\bibinfo {volume} {6}},\ \bibinfo {pages}
  {L112601} (\bibinfo {year} {2021})}\BibitemShut {NoStop}%
\bibitem [{\citenamefont {Gorce}\ and\ \citenamefont
  {Falcon}(2024)}]{gorce2024freely}%
  \BibitemOpen
  \bibfield  {author} {\bibinfo {author} {\bibfnamefont {J.-B.}\ \bibnamefont
  {Gorce}}\ and\ \bibinfo {author} {\bibfnamefont {E.}~\bibnamefont {Falcon}},\
  }\bibfield  {title} {\bibinfo {title} {Freely decaying saffman turbulence
  experimentally generated by magnetic stirrers},\ }\href@noop {} {\bibfield
  {journal} {\bibinfo  {journal} {Physical Review Letters}\ }\textbf {\bibinfo
  {volume} {132}},\ \bibinfo {pages} {264001} (\bibinfo {year}
  {2024})}\BibitemShut {NoStop}%
\bibitem [{\citenamefont {C{\=\i}murs}\ \emph {et~al.}(2019)\citenamefont
  {C{\=\i}murs}, \citenamefont {Brasovs},\ and\ \citenamefont
  {{\=E}rglis}}]{cimurs2019stability}%
  \BibitemOpen
  \bibfield  {author} {\bibinfo {author} {\bibfnamefont {J.}~\bibnamefont
  {C{\=\i}murs}}, \bibinfo {author} {\bibfnamefont {A.}~\bibnamefont
  {Brasovs}},\ and\ \bibinfo {author} {\bibfnamefont {K.}~\bibnamefont
  {{\=E}rglis}},\ }\bibfield  {title} {\bibinfo {title} {Stability analysis of
  a paramagnetic spheroid in a precessing field},\ }\href@noop {} {\bibfield
  {journal} {\bibinfo  {journal} {Journal of Magnetism and Magnetic Materials}\
  }\textbf {\bibinfo {volume} {491}},\ \bibinfo {pages} {165630} (\bibinfo
  {year} {2019})}\BibitemShut {NoStop}%
\bibitem [{\citenamefont {C{\=e}bers}\ and\ \citenamefont
  {Ozols}(2006)}]{cebers2006dynamics}%
  \BibitemOpen
  \bibfield  {author} {\bibinfo {author} {\bibfnamefont {A.}~\bibnamefont
  {C{\=e}bers}}\ and\ \bibinfo {author} {\bibfnamefont {M.}~\bibnamefont
  {Ozols}},\ }\bibfield  {title} {\bibinfo {title} {Dynamics of an active
  magnetic particle in a rotating magnetic field},\ }\href@noop {} {\bibfield
  {journal} {\bibinfo  {journal} {Physical Review E}\ }\textbf {\bibinfo
  {volume} {73}},\ \bibinfo {pages} {021505} (\bibinfo {year}
  {2006})}\BibitemShut {NoStop}%
\bibitem [{\citenamefont {{\=E}rglis}\ \emph {et~al.}(2007)\citenamefont
  {{\=E}rglis}, \citenamefont {Wen}, \citenamefont {Ose}, \citenamefont
  {Zeltins}, \citenamefont {Sharipo}, \citenamefont {Janmey},\ and\
  \citenamefont {C{\=e}bers}}]{erglis2007dynamics}%
  \BibitemOpen
  \bibfield  {author} {\bibinfo {author} {\bibfnamefont {K.}~\bibnamefont
  {{\=E}rglis}}, \bibinfo {author} {\bibfnamefont {Q.}~\bibnamefont {Wen}},
  \bibinfo {author} {\bibfnamefont {V.}~\bibnamefont {Ose}}, \bibinfo {author}
  {\bibfnamefont {A.}~\bibnamefont {Zeltins}}, \bibinfo {author} {\bibfnamefont
  {A.}~\bibnamefont {Sharipo}}, \bibinfo {author} {\bibfnamefont {P.~A.}\
  \bibnamefont {Janmey}},\ and\ \bibinfo {author} {\bibfnamefont
  {A.}~\bibnamefont {C{\=e}bers}},\ }\bibfield  {title} {\bibinfo {title}
  {Dynamics of magnetotactic bacteria in a rotating magnetic field},\
  }\href@noop {} {\bibfield  {journal} {\bibinfo  {journal} {Biophysical
  Journal}\ }\textbf {\bibinfo {volume} {93}},\ \bibinfo {pages} {1402}
  (\bibinfo {year} {2007})}\BibitemShut {NoStop}%
\bibitem [{\citenamefont {Sinn}\ \emph {et~al.}(2011)\citenamefont {Sinn},
  \citenamefont {Kinnunen}, \citenamefont {Pei}, \citenamefont {Clarke},
  \citenamefont {McNaughton},\ and\ \citenamefont
  {Kopelman}}]{Sinn2011Magnetically}%
  \BibitemOpen
  \bibfield  {author} {\bibinfo {author} {\bibfnamefont {I.}~\bibnamefont
  {Sinn}}, \bibinfo {author} {\bibfnamefont {P.}~\bibnamefont {Kinnunen}},
  \bibinfo {author} {\bibfnamefont {S.~N.}\ \bibnamefont {Pei}}, \bibinfo
  {author} {\bibfnamefont {R.}~\bibnamefont {Clarke}}, \bibinfo {author}
  {\bibfnamefont {B.~H.}\ \bibnamefont {McNaughton}},\ and\ \bibinfo {author}
  {\bibfnamefont {R.}~\bibnamefont {Kopelman}},\ }\bibfield  {title} {\bibinfo
  {title} {Magnetically uniform and tunable janus particles},\ }\href@noop {}
  {\bibfield  {journal} {\bibinfo  {journal} {Applied Physics Letters}\
  }\textbf {\bibinfo {volume} {98}},\ \bibinfo {pages} {024101} (\bibinfo
  {year} {2011})}\BibitemShut {NoStop}%
\bibitem [{\citenamefont {C{\=\i}murs}\ and\ \citenamefont
  {C{\=e}bers}(2013)}]{cimurs2013dynamics}%
  \BibitemOpen
  \bibfield  {author} {\bibinfo {author} {\bibfnamefont {J.}~\bibnamefont
  {C{\=\i}murs}}\ and\ \bibinfo {author} {\bibfnamefont {A.}~\bibnamefont
  {C{\=e}bers}},\ }\bibfield  {title} {\bibinfo {title} {Dynamics of
  anisotropic superparamagnetic particles in a precessing magnetic field},\
  }\href@noop {} {\bibfield  {journal} {\bibinfo  {journal} {Physical Review
  E}\ }\textbf {\bibinfo {volume} {87}},\ \bibinfo {pages} {062318} (\bibinfo
  {year} {2013})}\BibitemShut {NoStop}%
\bibitem [{\citenamefont {Morozov}\ \emph {et~al.}(2017)\citenamefont
  {Morozov}, \citenamefont {Mirzae}, \citenamefont {Kenneth},\ and\
  \citenamefont {Leshansky}}]{morozov2017dynamics}%
  \BibitemOpen
  \bibfield  {author} {\bibinfo {author} {\bibfnamefont {K.~I.}\ \bibnamefont
  {Morozov}}, \bibinfo {author} {\bibfnamefont {Y.}~\bibnamefont {Mirzae}},
  \bibinfo {author} {\bibfnamefont {O.}~\bibnamefont {Kenneth}},\ and\ \bibinfo
  {author} {\bibfnamefont {A.~M.}\ \bibnamefont {Leshansky}},\ }\bibfield
  {title} {\bibinfo {title} {Dynamics of arbitrary shaped propellers driven by
  a rotating magnetic field},\ }\href@noop {} {\bibfield  {journal} {\bibinfo
  {journal} {Physical Review Fluids}\ }\textbf {\bibinfo {volume} {2}},\
  \bibinfo {pages} {044202} (\bibinfo {year} {2017})}\BibitemShut {NoStop}%
\bibitem [{\citenamefont {Francois}\ \emph {et~al.}(2014)\citenamefont
  {Francois}, \citenamefont {Xia}, \citenamefont {Punzmann}, \citenamefont
  {Ramsden},\ and\ \citenamefont {Shats}}]{francois2014three}%
  \BibitemOpen
  \bibfield  {author} {\bibinfo {author} {\bibfnamefont {N.}~\bibnamefont
  {Francois}}, \bibinfo {author} {\bibfnamefont {H.}~\bibnamefont {Xia}},
  \bibinfo {author} {\bibfnamefont {H.}~\bibnamefont {Punzmann}}, \bibinfo
  {author} {\bibfnamefont {S.}~\bibnamefont {Ramsden}},\ and\ \bibinfo {author}
  {\bibfnamefont {M.}~\bibnamefont {Shats}},\ }\bibfield  {title} {\bibinfo
  {title} {Three-dimensional fluid motion in faraday waves: creation of
  vorticity and generation of two-dimensional turbulence},\ }\href@noop {}
  {\bibfield  {journal} {\bibinfo  {journal} {Physical Review X}\ }\textbf
  {\bibinfo {volume} {4}},\ \bibinfo {pages} {021021} (\bibinfo {year}
  {2014})}\BibitemShut {NoStop}%
\bibitem [{\citenamefont {Kokot}\ \emph {et~al.}(2017)\citenamefont {Kokot},
  \citenamefont {Das}, \citenamefont {Winkler}, \citenamefont {Gompper},
  \citenamefont {Aranson},\ and\ \citenamefont {Snezhko}}]{kokot2017active}%
  \BibitemOpen
  \bibfield  {author} {\bibinfo {author} {\bibfnamefont {G.}~\bibnamefont
  {Kokot}}, \bibinfo {author} {\bibfnamefont {S.}~\bibnamefont {Das}}, \bibinfo
  {author} {\bibfnamefont {R.~G.}\ \bibnamefont {Winkler}}, \bibinfo {author}
  {\bibfnamefont {G.}~\bibnamefont {Gompper}}, \bibinfo {author} {\bibfnamefont
  {I.~S.}\ \bibnamefont {Aranson}},\ and\ \bibinfo {author} {\bibfnamefont
  {A.}~\bibnamefont {Snezhko}},\ }\bibfield  {title} {\bibinfo {title} {Active
  turbulence in a gas of self-assembled spinners},\ }\href@noop {} {\bibfield
  {journal} {\bibinfo  {journal} {Proceedings of the National Academy of
  Sciences}\ }\textbf {\bibinfo {volume} {114}},\ \bibinfo {pages} {12870}
  (\bibinfo {year} {2017})}\BibitemShut {NoStop}%
\bibitem [{\citenamefont {Bourgoin}\ \emph {et~al.}(2020)\citenamefont
  {Bourgoin}, \citenamefont {Kervil}, \citenamefont {Cottin-Bizonne},
  \citenamefont {Raynal}, \citenamefont {Volk},\ and\ \citenamefont
  {Ybert}}]{bourgoin2020kolmogorovian}%
  \BibitemOpen
  \bibfield  {author} {\bibinfo {author} {\bibfnamefont {M.}~\bibnamefont
  {Bourgoin}}, \bibinfo {author} {\bibfnamefont {R.}~\bibnamefont {Kervil}},
  \bibinfo {author} {\bibfnamefont {C.}~\bibnamefont {Cottin-Bizonne}},
  \bibinfo {author} {\bibfnamefont {F.}~\bibnamefont {Raynal}}, \bibinfo
  {author} {\bibfnamefont {R.}~\bibnamefont {Volk}},\ and\ \bibinfo {author}
  {\bibfnamefont {C.}~\bibnamefont {Ybert}},\ }\bibfield  {title} {\bibinfo
  {title} {Kolmogorovian active turbulence of a sparse assembly of interacting
  marangoni surfers},\ }\href@noop {} {\bibfield  {journal} {\bibinfo
  {journal} {Physical Review X}\ }\textbf {\bibinfo {volume} {10}},\ \bibinfo
  {pages} {021065} (\bibinfo {year} {2020})}\BibitemShut {NoStop}%
\bibitem [{\citenamefont {Croquette}\ and\ \citenamefont
  {Poitou}(1981)}]{croquette1981cascade}%
  \BibitemOpen
  \bibfield  {author} {\bibinfo {author} {\bibfnamefont {V.}~\bibnamefont
  {Croquette}}\ and\ \bibinfo {author} {\bibfnamefont {C.}~\bibnamefont
  {Poitou}},\ }\bibfield  {title} {\bibinfo {title} {Cascade of period doubling
  bifurcations and large stochasticity in the motions of a compass},\
  }\href@noop {} {\bibfield  {journal} {\bibinfo  {journal} {Journal de
  Physique Lettres}\ }\textbf {\bibinfo {volume} {42}},\ \bibinfo {pages} {537}
  (\bibinfo {year} {1981})}\BibitemShut {NoStop}%
\bibitem [{\citenamefont {Poy{\'e}}\ \emph {et~al.}(2018)\citenamefont
  {Poy{\'e}}, \citenamefont {D{\'e}sangles}, \citenamefont {Jim{\'e}nez},
  \citenamefont {Martin},\ and\ \citenamefont {Proto}}]{poye2018bipolar}%
  \BibitemOpen
  \bibfield  {author} {\bibinfo {author} {\bibfnamefont {A.}~\bibnamefont
  {Poy{\'e}}}, \bibinfo {author} {\bibfnamefont {V.}~\bibnamefont
  {D{\'e}sangles}}, \bibinfo {author} {\bibfnamefont {X.}~\bibnamefont
  {Jim{\'e}nez}}, \bibinfo {author} {\bibfnamefont {M.}~\bibnamefont
  {Martin}},\ and\ \bibinfo {author} {\bibfnamefont {Y.}~\bibnamefont
  {Proto}},\ }\bibfield  {title} {\bibinfo {title} {Bipolar motor: rotation,
  parametric instabilities and chaos},\ }\href@noop {} {\bibfield  {journal}
  {\bibinfo  {journal} {Physica Scripta}\ }\textbf {\bibinfo {volume} {94}},\
  \bibinfo {pages} {015002} (\bibinfo {year} {2018})}\BibitemShut {NoStop}%
\bibitem [{\citenamefont {Kaiser}\ \emph {et~al.}(2017)\citenamefont {Kaiser},
  \citenamefont {Snezhko},\ and\ \citenamefont {Aranson}}]{kaiser2017flocking}%
  \BibitemOpen
  \bibfield  {author} {\bibinfo {author} {\bibfnamefont {A.}~\bibnamefont
  {Kaiser}}, \bibinfo {author} {\bibfnamefont {A.}~\bibnamefont {Snezhko}},\
  and\ \bibinfo {author} {\bibfnamefont {I.~S.}\ \bibnamefont {Aranson}},\
  }\bibfield  {title} {\bibinfo {title} {Flocking ferromagnetic colloids},\
  }\href@noop {} {\bibfield  {journal} {\bibinfo  {journal} {Science advances}\
  }\textbf {\bibinfo {volume} {3}},\ \bibinfo {pages} {e1601469} (\bibinfo
  {year} {2017})}\BibitemShut {NoStop}%
\bibitem [{\citenamefont {Benzi}\ \emph {et~al.}(1981)\citenamefont {Benzi},
  \citenamefont {Sutera},\ and\ \citenamefont {Vulpiani}}]{benzi1981mechanism}%
  \BibitemOpen
  \bibfield  {author} {\bibinfo {author} {\bibfnamefont {R.}~\bibnamefont
  {Benzi}}, \bibinfo {author} {\bibfnamefont {A.}~\bibnamefont {Sutera}},\ and\
  \bibinfo {author} {\bibfnamefont {A.}~\bibnamefont {Vulpiani}},\ }\bibfield
  {title} {\bibinfo {title} {The mechanism of stochastic resonance},\
  }\href@noop {} {\bibfield  {journal} {\bibinfo  {journal} {Journal of Physics
  A: Mathematical and General}\ }\textbf {\bibinfo {volume} {14}},\ \bibinfo
  {pages} {L453} (\bibinfo {year} {1981})}\BibitemShut {NoStop}%
\bibitem [{\citenamefont {Nicolis}(1982)}]{nicolis1982stochastic}%
  \BibitemOpen
  \bibfield  {author} {\bibinfo {author} {\bibfnamefont {C.}~\bibnamefont
  {Nicolis}},\ }\bibfield  {title} {\bibinfo {title} {Stochastic aspects of
  climatic transitions--response to a periodic forcing},\ }\href@noop {}
  {\bibfield  {journal} {\bibinfo  {journal} {Tellus}\ }\textbf {\bibinfo
  {volume} {34}},\ \bibinfo {pages} {1} (\bibinfo {year} {1982})}\BibitemShut
  {NoStop}%
\bibitem [{\citenamefont {Gammaitoni}\ \emph {et~al.}(1998)\citenamefont
  {Gammaitoni}, \citenamefont {H{\"a}nggi}, \citenamefont {Jung},\ and\
  \citenamefont {Marchesoni}}]{gammaitoni1998stochastic}%
  \BibitemOpen
  \bibfield  {author} {\bibinfo {author} {\bibfnamefont {L.}~\bibnamefont
  {Gammaitoni}}, \bibinfo {author} {\bibfnamefont {P.}~\bibnamefont
  {H{\"a}nggi}}, \bibinfo {author} {\bibfnamefont {P.}~\bibnamefont {Jung}},\
  and\ \bibinfo {author} {\bibfnamefont {F.}~\bibnamefont {Marchesoni}},\
  }\bibfield  {title} {\bibinfo {title} {Stochastic resonance},\ }\href@noop {}
  {\bibfield  {journal} {\bibinfo  {journal} {Reviews of Modern Physics}\
  }\textbf {\bibinfo {volume} {70}},\ \bibinfo {pages} {223} (\bibinfo {year}
  {1998})}\BibitemShut {NoStop}%
\bibitem [{\citenamefont {Simonotto}\ \emph {et~al.}(1997)\citenamefont
  {Simonotto}, \citenamefont {Riani}, \citenamefont {Seife}, \citenamefont
  {Roberts}, \citenamefont {Twitty},\ and\ \citenamefont
  {Moss}}]{simonotto1997visual}%
  \BibitemOpen
  \bibfield  {author} {\bibinfo {author} {\bibfnamefont {E.}~\bibnamefont
  {Simonotto}}, \bibinfo {author} {\bibfnamefont {M.}~\bibnamefont {Riani}},
  \bibinfo {author} {\bibfnamefont {C.}~\bibnamefont {Seife}}, \bibinfo
  {author} {\bibfnamefont {M.}~\bibnamefont {Roberts}}, \bibinfo {author}
  {\bibfnamefont {J.}~\bibnamefont {Twitty}},\ and\ \bibinfo {author}
  {\bibfnamefont {F.}~\bibnamefont {Moss}},\ }\bibfield  {title} {\bibinfo
  {title} {Visual perception of stochastic resonance},\ }\href@noop {}
  {\bibfield  {journal} {\bibinfo  {journal} {Physical Review Letters}\
  }\textbf {\bibinfo {volume} {78}},\ \bibinfo {pages} {1186} (\bibinfo {year}
  {1997})}\BibitemShut {NoStop}%
\bibitem [{\citenamefont {Benzi}(2010)}]{benzi2010stochastic}%
  \BibitemOpen
  \bibfield  {author} {\bibinfo {author} {\bibfnamefont {R.}~\bibnamefont
  {Benzi}},\ }\bibfield  {title} {\bibinfo {title} {Stochastic resonance: from
  climate to biology},\ }\href@noop {} {\bibfield  {journal} {\bibinfo
  {journal} {Nonlinear Processes in Geophysics}\ }\textbf {\bibinfo {volume}
  {17}},\ \bibinfo {pages} {431} (\bibinfo {year} {2010})}\BibitemShut
  {NoStop}%
\bibitem [{\citenamefont {Budrikis}(2021)}]{budrikis2021forty}%
  \BibitemOpen
  \bibfield  {author} {\bibinfo {author} {\bibfnamefont {Z.}~\bibnamefont
  {Budrikis}},\ }\bibfield  {title} {\bibinfo {title} {Forty years of
  stochastic resonance},\ }\href@noop {} {\bibfield  {journal} {\bibinfo
  {journal} {Nature Reviews Physics}\ }\textbf {\bibinfo {volume} {3}},\
  \bibinfo {pages} {771} (\bibinfo {year} {2021})}\BibitemShut {NoStop}%
\bibitem [{\citenamefont {Lu}\ \emph {et~al.}(2015)\citenamefont {Lu},
  \citenamefont {He},\ and\ \citenamefont {Kong}}]{lu2015effects}%
  \BibitemOpen
  \bibfield  {author} {\bibinfo {author} {\bibfnamefont {S.}~\bibnamefont
  {Lu}}, \bibinfo {author} {\bibfnamefont {Q.}~\bibnamefont {He}},\ and\
  \bibinfo {author} {\bibfnamefont {F.}~\bibnamefont {Kong}},\ }\bibfield
  {title} {\bibinfo {title} {Effects of underdamped step-varying second-order
  stochastic resonance for weak signal detection},\ }\href@noop {} {\bibfield
  {journal} {\bibinfo  {journal} {Digital Signal Processing}\ }\textbf
  {\bibinfo {volume} {36}},\ \bibinfo {pages} {93} (\bibinfo {year}
  {2015})}\BibitemShut {NoStop}%
\bibitem [{\citenamefont {Wu}\ \emph {et~al.}(2024)\citenamefont {Wu},
  \citenamefont {Xie}, \citenamefont {Li}, \citenamefont {Guo}, \citenamefont
  {Zou},\ and\ \citenamefont {Xiang}}]{wu2024nonlinearity}%
  \BibitemOpen
  \bibfield  {author} {\bibinfo {author} {\bibfnamefont {K.-D.}\ \bibnamefont
  {Wu}}, \bibinfo {author} {\bibfnamefont {C.}~\bibnamefont {Xie}}, \bibinfo
  {author} {\bibfnamefont {C.-F.}\ \bibnamefont {Li}}, \bibinfo {author}
  {\bibfnamefont {G.-C.}\ \bibnamefont {Guo}}, \bibinfo {author} {\bibfnamefont
  {C.-L.}\ \bibnamefont {Zou}},\ and\ \bibinfo {author} {\bibfnamefont {G.-Y.}\
  \bibnamefont {Xiang}},\ }\bibfield  {title} {\bibinfo {title}
  {Nonlinearity-enhanced continuous microwave detection based on stochastic
  resonance},\ }\href@noop {} {\bibfield  {journal} {\bibinfo  {journal}
  {Science Advances}\ }\textbf {\bibinfo {volume} {10}},\ \bibinfo {pages}
  {eado8130} (\bibinfo {year} {2024})}\BibitemShut {NoStop}%
\bibitem [{\citenamefont {Peters}\ \emph {et~al.}(2021)\citenamefont {Peters},
  \citenamefont {Geng}, \citenamefont {Malmir}, \citenamefont {Smith},\ and\
  \citenamefont {Rodriguez}}]{peters2021extremely}%
  \BibitemOpen
  \bibfield  {author} {\bibinfo {author} {\bibfnamefont {K.}~\bibnamefont
  {Peters}}, \bibinfo {author} {\bibfnamefont {Z.}~\bibnamefont {Geng}},
  \bibinfo {author} {\bibfnamefont {K.}~\bibnamefont {Malmir}}, \bibinfo
  {author} {\bibfnamefont {J.}~\bibnamefont {Smith}},\ and\ \bibinfo {author}
  {\bibfnamefont {S.}~\bibnamefont {Rodriguez}},\ }\bibfield  {title} {\bibinfo
  {title} {Extremely broadband stochastic resonance of light and enhanced
  energy harvesting enabled by memory effects in the nonlinear response},\
  }\href@noop {} {\bibfield  {journal} {\bibinfo  {journal} {Physical Review
  Letters}\ }\textbf {\bibinfo {volume} {126}},\ \bibinfo {pages} {213901}
  (\bibinfo {year} {2021})}\BibitemShut {NoStop}%
\bibitem [{\citenamefont {Douglass}\ \emph {et~al.}(1993)\citenamefont
  {Douglass}, \citenamefont {Wilkens}, \citenamefont {Pantazelou},\ and\
  \citenamefont {Moss}}]{douglass1993noise}%
  \BibitemOpen
  \bibfield  {author} {\bibinfo {author} {\bibfnamefont {J.~K.}\ \bibnamefont
  {Douglass}}, \bibinfo {author} {\bibfnamefont {L.}~\bibnamefont {Wilkens}},
  \bibinfo {author} {\bibfnamefont {E.}~\bibnamefont {Pantazelou}},\ and\
  \bibinfo {author} {\bibfnamefont {F.}~\bibnamefont {Moss}},\ }\bibfield
  {title} {\bibinfo {title} {Noise enhancement of information transfer in
  crayfish mechanoreceptors by stochastic resonance},\ }\href@noop {}
  {\bibfield  {journal} {\bibinfo  {journal} {Nature}\ }\textbf {\bibinfo
  {volume} {365}},\ \bibinfo {pages} {337} (\bibinfo {year}
  {1993})}\BibitemShut {NoStop}%
\bibitem [{\citenamefont {Braun}\ \emph {et~al.}(1994)\citenamefont {Braun},
  \citenamefont {Wissing}, \citenamefont {Sch{\"a}fer},\ and\ \citenamefont
  {Hirsch}}]{braun1994oscillation}%
  \BibitemOpen
  \bibfield  {author} {\bibinfo {author} {\bibfnamefont {H.~A.}\ \bibnamefont
  {Braun}}, \bibinfo {author} {\bibfnamefont {H.}~\bibnamefont {Wissing}},
  \bibinfo {author} {\bibfnamefont {K.}~\bibnamefont {Sch{\"a}fer}},\ and\
  \bibinfo {author} {\bibfnamefont {M.~C.}\ \bibnamefont {Hirsch}},\ }\bibfield
   {title} {\bibinfo {title} {Oscillation and noise determine signal
  transduction in shark multimodal sensory cells},\ }\href@noop {} {\bibfield
  {journal} {\bibinfo  {journal} {Nature}\ }\textbf {\bibinfo {volume} {367}},\
  \bibinfo {pages} {270} (\bibinfo {year} {1994})}\BibitemShut {NoStop}%
\bibitem [{\citenamefont {Wiesenfeld}\ and\ \citenamefont
  {Moss}(1995)}]{wiesenfeld1995stochastic}%
  \BibitemOpen
  \bibfield  {author} {\bibinfo {author} {\bibfnamefont {K.}~\bibnamefont
  {Wiesenfeld}}\ and\ \bibinfo {author} {\bibfnamefont {F.}~\bibnamefont
  {Moss}},\ }\bibfield  {title} {\bibinfo {title} {Stochastic resonance and the
  benefits of noise: from ice ages to crayfish and squids},\ }\href@noop {}
  {\bibfield  {journal} {\bibinfo  {journal} {Nature}\ }\textbf {\bibinfo
  {volume} {373}},\ \bibinfo {pages} {33} (\bibinfo {year} {1995})}\BibitemShut
  {NoStop}%
\bibitem [{\citenamefont {McDonnell}\ and\ \citenamefont
  {Abbott}(2009)}]{mcdonnell2009stochastic}%
  \BibitemOpen
  \bibfield  {author} {\bibinfo {author} {\bibfnamefont {M.~D.}\ \bibnamefont
  {McDonnell}}\ and\ \bibinfo {author} {\bibfnamefont {D.}~\bibnamefont
  {Abbott}},\ }\bibfield  {title} {\bibinfo {title} {What is stochastic
  resonance? definitions, misconceptions, debates, and its relevance to
  biology},\ }\href@noop {} {\bibfield  {journal} {\bibinfo  {journal} {PLOS
  Computational Biology}\ }\textbf {\bibinfo {volume} {5}},\ \bibinfo {pages}
  {e1000348} (\bibinfo {year} {2009})}\BibitemShut {NoStop}%
\bibitem [{\citenamefont {McDonnell}\ and\ \citenamefont
  {Ward}(2011)}]{mcdonnell2011benefits}%
  \BibitemOpen
  \bibfield  {author} {\bibinfo {author} {\bibfnamefont {M.~D.}\ \bibnamefont
  {McDonnell}}\ and\ \bibinfo {author} {\bibfnamefont {L.~M.}\ \bibnamefont
  {Ward}},\ }\bibfield  {title} {\bibinfo {title} {The benefits of noise in
  neural systems: bridging theory and experiment},\ }\href@noop {} {\bibfield
  {journal} {\bibinfo  {journal} {Nature Reviews Neuroscience}\ }\textbf
  {\bibinfo {volume} {12}},\ \bibinfo {pages} {415} (\bibinfo {year}
  {2011})}\BibitemShut {NoStop}%
\bibitem [{\citenamefont {Pisarchik}\ and\ \citenamefont
  {Hramov}(2023)}]{pisarchik2023coherence}%
  \BibitemOpen
  \bibfield  {author} {\bibinfo {author} {\bibfnamefont {A.~N.}\ \bibnamefont
  {Pisarchik}}\ and\ \bibinfo {author} {\bibfnamefont {A.~E.}\ \bibnamefont
  {Hramov}},\ }\bibfield  {title} {\bibinfo {title} {Coherence resonance in
  neural networks: Theory and experiments},\ }\href@noop {} {\bibfield
  {journal} {\bibinfo  {journal} {Physics Reports}\ }\textbf {\bibinfo {volume}
  {1000}},\ \bibinfo {pages} {1} (\bibinfo {year} {2023})}\BibitemShut
  {NoStop}%
\bibitem [{\citenamefont {Mordant}\ \emph {et~al.}(2004)\citenamefont
  {Mordant}, \citenamefont {L{\'e}v{\^e}que},\ and\ \citenamefont
  {Pinton}}]{mordant2004experimental}%
  \BibitemOpen
  \bibfield  {author} {\bibinfo {author} {\bibfnamefont {N.}~\bibnamefont
  {Mordant}}, \bibinfo {author} {\bibfnamefont {E.}~\bibnamefont
  {L{\'e}v{\^e}que}},\ and\ \bibinfo {author} {\bibfnamefont {J.-F.}\
  \bibnamefont {Pinton}},\ }\bibfield  {title} {\bibinfo {title} {Experimental
  and numerical study of the lagrangian dynamics of high reynolds turbulence},\
  }\href@noop {} {\bibfield  {journal} {\bibinfo  {journal} {New Journal of
  Physics}\ }\textbf {\bibinfo {volume} {6}},\ \bibinfo {pages} {116} (\bibinfo
  {year} {2004})}\BibitemShut {NoStop}%
\bibitem [{\citenamefont {Volk}\ \emph {et~al.}(2008)\citenamefont {Volk},
  \citenamefont {Calzavarini}, \citenamefont {Verhille}, \citenamefont {Lohse},
  \citenamefont {Mordant}, \citenamefont {Pinton},\ and\ \citenamefont
  {Toschi}}]{volk2008acceleration}%
  \BibitemOpen
  \bibfield  {author} {\bibinfo {author} {\bibfnamefont {R.}~\bibnamefont
  {Volk}}, \bibinfo {author} {\bibfnamefont {E.}~\bibnamefont {Calzavarini}},
  \bibinfo {author} {\bibfnamefont {G.}~\bibnamefont {Verhille}}, \bibinfo
  {author} {\bibfnamefont {D.}~\bibnamefont {Lohse}}, \bibinfo {author}
  {\bibfnamefont {N.}~\bibnamefont {Mordant}}, \bibinfo {author} {\bibfnamefont
  {J.-F.}\ \bibnamefont {Pinton}},\ and\ \bibinfo {author} {\bibfnamefont
  {F.}~\bibnamefont {Toschi}},\ }\bibfield  {title} {\bibinfo {title}
  {Acceleration of heavy and light particles in turbulence: comparison between
  experiments and direct numerical simulations},\ }\href@noop {} {\bibfield
  {journal} {\bibinfo  {journal} {Physica D: Nonlinear Phenomena}\ }\textbf
  {\bibinfo {volume} {237}},\ \bibinfo {pages} {2084} (\bibinfo {year}
  {2008})}\BibitemShut {NoStop}%
\bibitem [{\citenamefont {Voth}\ \emph {et~al.}(2002)\citenamefont {Voth},
  \citenamefont {La~Porta}, \citenamefont {Crawford}, \citenamefont
  {Alexander},\ and\ \citenamefont {Bodenschatz}}]{voth2002measurement}%
  \BibitemOpen
  \bibfield  {author} {\bibinfo {author} {\bibfnamefont {G.~A.}\ \bibnamefont
  {Voth}}, \bibinfo {author} {\bibfnamefont {A.}~\bibnamefont {La~Porta}},
  \bibinfo {author} {\bibfnamefont {A.~M.}\ \bibnamefont {Crawford}}, \bibinfo
  {author} {\bibfnamefont {J.}~\bibnamefont {Alexander}},\ and\ \bibinfo
  {author} {\bibfnamefont {E.}~\bibnamefont {Bodenschatz}},\ }\bibfield
  {title} {\bibinfo {title} {Measurement of particle accelerations in fully
  developed turbulence},\ }\href@noop {} {\bibfield  {journal} {\bibinfo
  {journal} {Journal of Fluid Mechanics}\ }\textbf {\bibinfo {volume} {469}},\
  \bibinfo {pages} {121} (\bibinfo {year} {2002})}\BibitemShut {NoStop}%
\bibitem [{\citenamefont {Volk}\ \emph {et~al.}(2011)\citenamefont {Volk},
  \citenamefont {Calzavarini}, \citenamefont {Leveque},\ and\ \citenamefont
  {Pinton}}]{volk2011dynamics}%
  \BibitemOpen
  \bibfield  {author} {\bibinfo {author} {\bibfnamefont {R.}~\bibnamefont
  {Volk}}, \bibinfo {author} {\bibfnamefont {E.}~\bibnamefont {Calzavarini}},
  \bibinfo {author} {\bibfnamefont {E.}~\bibnamefont {Leveque}},\ and\ \bibinfo
  {author} {\bibfnamefont {J.-F.}\ \bibnamefont {Pinton}},\ }\bibfield  {title}
  {\bibinfo {title} {Dynamics of inertial particles in a turbulent von
  k{\'a}rm{\'a}n flow},\ }\href@noop {} {\bibfield  {journal} {\bibinfo
  {journal} {Journal of Fluid Mechanics}\ }\textbf {\bibinfo {volume} {668}},\
  \bibinfo {pages} {223} (\bibinfo {year} {2011})}\BibitemShut {NoStop}%
\bibitem [{\citenamefont {Osborn}(1945)}]{osborn1945demagnetizing}%
  \BibitemOpen
  \bibfield  {author} {\bibinfo {author} {\bibfnamefont {J.~A.}\ \bibnamefont
  {Osborn}},\ }\bibfield  {title} {\bibinfo {title} {Demagnetizing factors of
  the general ellipsoid},\ }\href@noop {} {\bibfield  {journal} {\bibinfo
  {journal} {Physical Review}\ }\textbf {\bibinfo {volume} {67}},\ \bibinfo
  {pages} {351} (\bibinfo {year} {1945})}\BibitemShut {NoStop}%
\bibitem [{\citenamefont {Peyret}(2002)}]{peyret2002spectral}%
  \BibitemOpen
  \bibfield  {author} {\bibinfo {author} {\bibfnamefont {R.}~\bibnamefont
  {Peyret}},\ }\href@noop {} {\emph {\bibinfo {title} {Spectral Methods for
  Incompressible Viscous Flow}}},\ Vol.\ \bibinfo {volume} {148}\ (\bibinfo
  {publisher} {Springer},\ \bibinfo {year} {2002})\BibitemShut {NoStop}%
\bibitem [{\citenamefont {Bec}\ \emph {et~al.}(2006)\citenamefont {Bec},
  \citenamefont {Biferale}, \citenamefont {Boffetta}, \citenamefont {Celani},
  \citenamefont {Cencini}, \citenamefont {Lanotte}, \citenamefont {Musacchio},\
  and\ \citenamefont {Toschi}}]{bec2006acceleration}%
  \BibitemOpen
  \bibfield  {author} {\bibinfo {author} {\bibfnamefont {J.}~\bibnamefont
  {Bec}}, \bibinfo {author} {\bibfnamefont {L.}~\bibnamefont {Biferale}},
  \bibinfo {author} {\bibfnamefont {G.}~\bibnamefont {Boffetta}}, \bibinfo
  {author} {\bibfnamefont {A.}~\bibnamefont {Celani}}, \bibinfo {author}
  {\bibfnamefont {M.}~\bibnamefont {Cencini}}, \bibinfo {author} {\bibfnamefont
  {A.}~\bibnamefont {Lanotte}}, \bibinfo {author} {\bibfnamefont
  {S.}~\bibnamefont {Musacchio}},\ and\ \bibinfo {author} {\bibfnamefont
  {F.}~\bibnamefont {Toschi}},\ }\bibfield  {title} {\bibinfo {title}
  {Acceleration statistics of heavy particles in turbulence},\ }\href@noop {}
  {\bibfield  {journal} {\bibinfo  {journal} {Journal of Fluid Mechanics}\
  }\textbf {\bibinfo {volume} {550}},\ \bibinfo {pages} {349} (\bibinfo {year}
  {2006})}\BibitemShut {NoStop}%
\bibitem [{\citenamefont {Freitas}\ \emph {et~al.}(2025)\citenamefont
  {Freitas}, \citenamefont {de~Wit}, \citenamefont {Wang}, \citenamefont
  {Biferale},\ and\ \citenamefont {Toschi}}]{freitas2025statistical}%
  \BibitemOpen
  \bibfield  {author} {\bibinfo {author} {\bibfnamefont {A.}~\bibnamefont
  {Freitas}}, \bibinfo {author} {\bibfnamefont {X.~M.}\ \bibnamefont {de~Wit}},
  \bibinfo {author} {\bibfnamefont {Z.}~\bibnamefont {Wang}}, \bibinfo {author}
  {\bibfnamefont {L.}~\bibnamefont {Biferale}},\ and\ \bibinfo {author}
  {\bibfnamefont {F.}~\bibnamefont {Toschi}},\ }\bibfield  {title} {\bibinfo
  {title} {Statistical properties of turbulence under a smart lagrangian
  forcing},\ }\href@noop {} {\bibfield  {journal} {\bibinfo  {journal} {arXiv
  preprint arXiv:2508.06660}\ } (\bibinfo {year} {2025})}\BibitemShut {NoStop}%
\bibitem [{\citenamefont {Scholz}\ \emph {et~al.}(2018)\citenamefont {Scholz},
  \citenamefont {Engel},\ and\ \citenamefont
  {P{\"o}schel}}]{scholz2018rotating}%
  \BibitemOpen
  \bibfield  {author} {\bibinfo {author} {\bibfnamefont {C.}~\bibnamefont
  {Scholz}}, \bibinfo {author} {\bibfnamefont {M.}~\bibnamefont {Engel}},\ and\
  \bibinfo {author} {\bibfnamefont {T.}~\bibnamefont {P{\"o}schel}},\
  }\bibfield  {title} {\bibinfo {title} {Rotating robots move collectively and
  self-organize},\ }\href@noop {} {\bibfield  {journal} {\bibinfo  {journal}
  {Nature Communications}\ }\textbf {\bibinfo {volume} {9}},\ \bibinfo {pages}
  {931} (\bibinfo {year} {2018})}\BibitemShut {NoStop}%
\bibitem [{\citenamefont {Yang}\ and\ \citenamefont
  {Zhang}(2021)}]{yang2021motion}%
  \BibitemOpen
  \bibfield  {author} {\bibinfo {author} {\bibfnamefont {L.}~\bibnamefont
  {Yang}}\ and\ \bibinfo {author} {\bibfnamefont {L.}~\bibnamefont {Zhang}},\
  }\bibfield  {title} {\bibinfo {title} {Motion control in magnetic
  microrobotics: From individual and multiple robots to swarms},\ }\href@noop
  {} {\bibfield  {journal} {\bibinfo  {journal} {Annual Review of Control,
  Robotics, and Autonomous Systems}\ }\textbf {\bibinfo {volume} {4}},\
  \bibinfo {pages} {509} (\bibinfo {year} {2021})}\BibitemShut {NoStop}%
\bibitem [{\citenamefont {Jiang}\ \emph {et~al.}(2023)\citenamefont {Jiang},
  \citenamefont {Li}, \citenamefont {Zhao}, \citenamefont {Wu}, \citenamefont
  {Zhang}, \citenamefont {Zhao}, \citenamefont {Zhang}, \citenamefont {Yu},
  \citenamefont {Shao}, \citenamefont {Zhang} \emph
  {et~al.}}]{jiang2023magnetic}%
  \BibitemOpen
  \bibfield  {author} {\bibinfo {author} {\bibfnamefont {S.}~\bibnamefont
  {Jiang}}, \bibinfo {author} {\bibfnamefont {B.}~\bibnamefont {Li}}, \bibinfo
  {author} {\bibfnamefont {J.}~\bibnamefont {Zhao}}, \bibinfo {author}
  {\bibfnamefont {D.}~\bibnamefont {Wu}}, \bibinfo {author} {\bibfnamefont
  {Y.}~\bibnamefont {Zhang}}, \bibinfo {author} {\bibfnamefont
  {Z.}~\bibnamefont {Zhao}}, \bibinfo {author} {\bibfnamefont {Y.}~\bibnamefont
  {Zhang}}, \bibinfo {author} {\bibfnamefont {H.}~\bibnamefont {Yu}}, \bibinfo
  {author} {\bibfnamefont {K.}~\bibnamefont {Shao}}, \bibinfo {author}
  {\bibfnamefont {C.}~\bibnamefont {Zhang}}, \emph {et~al.},\ }\bibfield
  {title} {\bibinfo {title} {Magnetic janus origami robot for cross-scale
  droplet omni-manipulation},\ }\href@noop {} {\bibfield  {journal} {\bibinfo
  {journal} {Nature Communications}\ }\textbf {\bibinfo {volume} {14}},\
  \bibinfo {pages} {5455} (\bibinfo {year} {2023})}\BibitemShut {NoStop}%
\bibitem [{\citenamefont {Chen}\ \emph {et~al.}(2025)\citenamefont {Chen},
  \citenamefont {Weady}, \citenamefont {Atis}, \citenamefont {Matsuzawa},
  \citenamefont {Shelley},\ and\ \citenamefont {Irvine}}]{chen2025self}%
  \BibitemOpen
  \bibfield  {author} {\bibinfo {author} {\bibfnamefont {P.}~\bibnamefont
  {Chen}}, \bibinfo {author} {\bibfnamefont {S.}~\bibnamefont {Weady}},
  \bibinfo {author} {\bibfnamefont {S.}~\bibnamefont {Atis}}, \bibinfo {author}
  {\bibfnamefont {T.}~\bibnamefont {Matsuzawa}}, \bibinfo {author}
  {\bibfnamefont {M.~J.}\ \bibnamefont {Shelley}},\ and\ \bibinfo {author}
  {\bibfnamefont {W.~T.}\ \bibnamefont {Irvine}},\ }\bibfield  {title}
  {\bibinfo {title} {Self-propulsion, flocking and chiral active phases from
  particles spinning at intermediate reynolds numbers},\ }\href@noop {}
  {\bibfield  {journal} {\bibinfo  {journal} {Nature Physics}\ }\textbf
  {\bibinfo {volume} {21}},\ \bibinfo {pages} {146} (\bibinfo {year}
  {2025})}\BibitemShut {NoStop}%
\bibitem [{\citenamefont {Fruchart}\ \emph {et~al.}(2023)\citenamefont
  {Fruchart}, \citenamefont {Scheibner},\ and\ \citenamefont
  {Vitelli}}]{fruchart2023odd}%
  \BibitemOpen
  \bibfield  {author} {\bibinfo {author} {\bibfnamefont {M.}~\bibnamefont
  {Fruchart}}, \bibinfo {author} {\bibfnamefont {C.}~\bibnamefont
  {Scheibner}},\ and\ \bibinfo {author} {\bibfnamefont {V.}~\bibnamefont
  {Vitelli}},\ }\bibfield  {title} {\bibinfo {title} {Odd viscosity and odd
  elasticity},\ }\href@noop {} {\bibfield  {journal} {\bibinfo  {journal}
  {Annual Review of Condensed Matter Physics}\ }\textbf {\bibinfo {volume}
  {14}},\ \bibinfo {pages} {471} (\bibinfo {year} {2023})}\BibitemShut
  {NoStop}%
\bibitem [{\citenamefont {Wang}\ \emph {et~al.}(2025)\citenamefont {Wang},
  \citenamefont {de~Wit}, \citenamefont {Benzi}, \citenamefont {Wu},
  \citenamefont {Kunnen}, \citenamefont {Clercx},\ and\ \citenamefont
  {Toschi}}]{wang_2025_17076195}%
  \BibitemOpen
  \bibfield  {author} {\bibinfo {author} {\bibfnamefont {Z.}~\bibnamefont
  {Wang}}, \bibinfo {author} {\bibfnamefont {X.}~\bibnamefont {de~Wit}},
  \bibinfo {author} {\bibfnamefont {R.}~\bibnamefont {Benzi}}, \bibinfo
  {author} {\bibfnamefont {C.}~\bibnamefont {Wu}}, \bibinfo {author}
  {\bibfnamefont {R.~P.~J.}\ \bibnamefont {Kunnen}}, \bibinfo {author}
  {\bibfnamefont {H.~J.~H.}\ \bibnamefont {Clercx}},\ and\ \bibinfo {author}
  {\bibfnamefont {F.}~\bibnamefont {Toschi}},\ }\bibfield  {title} {\bibinfo
  {title} {Stochastic resonance of rotating particles in turbulence},\ }\href
  {https://doi.org/10.5281/zenodo.17076195} {10.5281/zenodo.17076195} (\bibinfo
  {year} {2025})\BibitemShut {NoStop}%
\bibitem [{\citenamefont {Wu}\ \emph {et~al.}(2025)\citenamefont {Wu},
  \citenamefont {Kunnen}, \citenamefont {Wang}, \citenamefont {de~Wit},
  \citenamefont {Toschi},\ and\ \citenamefont {Clercx}}]{wu2025tracking}%
  \BibitemOpen
  \bibfield  {author} {\bibinfo {author} {\bibfnamefont {C.}~\bibnamefont
  {Wu}}, \bibinfo {author} {\bibfnamefont {R.~P.~J.}\ \bibnamefont {Kunnen}},
  \bibinfo {author} {\bibfnamefont {Z.}~\bibnamefont {Wang}}, \bibinfo {author}
  {\bibfnamefont {X.~M.}\ \bibnamefont {de~Wit}}, \bibinfo {author}
  {\bibfnamefont {F.}~\bibnamefont {Toschi}},\ and\ \bibinfo {author}
  {\bibfnamefont {H.~J.~H.}\ \bibnamefont {Clercx}},\ }\bibfield  {title}
  {\bibinfo {title} {Tracking the rotation of light magnetic particles in
  turbulence},\ }\href@noop {} {\bibfield  {journal} {\bibinfo  {journal}
  {arXiv preprint arXiv:2506.21769}\ } (\bibinfo {year} {2025})}\BibitemShut
  {NoStop}%
\bibitem [{\citenamefont {Zocchi}\ \emph {et~al.}(1994)\citenamefont {Zocchi},
  \citenamefont {Tabeling}, \citenamefont {Maurer},\ and\ \citenamefont
  {Willaime}}]{zocchi1994measurement}%
  \BibitemOpen
  \bibfield  {author} {\bibinfo {author} {\bibfnamefont {G.}~\bibnamefont
  {Zocchi}}, \bibinfo {author} {\bibfnamefont {P.}~\bibnamefont {Tabeling}},
  \bibinfo {author} {\bibfnamefont {J.}~\bibnamefont {Maurer}},\ and\ \bibinfo
  {author} {\bibfnamefont {H.}~\bibnamefont {Willaime}},\ }\bibfield  {title}
  {\bibinfo {title} {Measurement of the scaling of the dissipation at high
  reynolds numbers},\ }\href@noop {} {\bibfield  {journal} {\bibinfo  {journal}
  {Physical Review E}\ }\textbf {\bibinfo {volume} {50}},\ \bibinfo {pages}
  {3693} (\bibinfo {year} {1994})}\BibitemShut {NoStop}%
\bibitem [{\citenamefont {Mordant}\ \emph {et~al.}(2001)\citenamefont
  {Mordant}, \citenamefont {Metz}, \citenamefont {Michel},\ and\ \citenamefont
  {Pinton}}]{mordant2001measurement}%
  \BibitemOpen
  \bibfield  {author} {\bibinfo {author} {\bibfnamefont {N.}~\bibnamefont
  {Mordant}}, \bibinfo {author} {\bibfnamefont {P.}~\bibnamefont {Metz}},
  \bibinfo {author} {\bibfnamefont {O.}~\bibnamefont {Michel}},\ and\ \bibinfo
  {author} {\bibfnamefont {J.-F.}\ \bibnamefont {Pinton}},\ }\bibfield  {title}
  {\bibinfo {title} {Measurement of lagrangian velocity in fully developed
  turbulence},\ }\href@noop {} {\bibfield  {journal} {\bibinfo  {journal}
  {Physical Review Letters}\ }\textbf {\bibinfo {volume} {87}},\ \bibinfo
  {pages} {214501} (\bibinfo {year} {2001})}\BibitemShut {NoStop}%
\bibitem [{\citenamefont {La~Porta}\ \emph {et~al.}(2001)\citenamefont
  {La~Porta}, \citenamefont {Voth}, \citenamefont {Crawford}, \citenamefont
  {Alexander},\ and\ \citenamefont {Bodenschatz}}]{la2001fluid}%
  \BibitemOpen
  \bibfield  {author} {\bibinfo {author} {\bibfnamefont {A.}~\bibnamefont
  {La~Porta}}, \bibinfo {author} {\bibfnamefont {G.~A.}\ \bibnamefont {Voth}},
  \bibinfo {author} {\bibfnamefont {A.~M.}\ \bibnamefont {Crawford}}, \bibinfo
  {author} {\bibfnamefont {J.}~\bibnamefont {Alexander}},\ and\ \bibinfo
  {author} {\bibfnamefont {E.}~\bibnamefont {Bodenschatz}},\ }\bibfield
  {title} {\bibinfo {title} {Fluid particle accelerations in fully developed
  turbulence},\ }\href@noop {} {\bibfield  {journal} {\bibinfo  {journal}
  {Nature}\ }\textbf {\bibinfo {volume} {409}},\ \bibinfo {pages} {1017}
  (\bibinfo {year} {2001})}\BibitemShut {NoStop}%
\bibitem [{\citenamefont {Naso}\ and\ \citenamefont
  {Prosperetti}(2010)}]{naso2010interaction}%
  \BibitemOpen
  \bibfield  {author} {\bibinfo {author} {\bibfnamefont {A.}~\bibnamefont
  {Naso}}\ and\ \bibinfo {author} {\bibfnamefont {A.}~\bibnamefont
  {Prosperetti}},\ }\bibfield  {title} {\bibinfo {title} {The interaction
  between a solid particle and a turbulent flow},\ }\href@noop {} {\bibfield
  {journal} {\bibinfo  {journal} {New Journal of Physics}\ }\textbf {\bibinfo
  {volume} {12}},\ \bibinfo {pages} {033040} (\bibinfo {year}
  {2010})}\BibitemShut {NoStop}%
\bibitem [{\citenamefont {Coffey}\ and\ \citenamefont
  {Kalmykov}(2012)}]{coffey2012langevin}%
  \BibitemOpen
  \bibfield  {author} {\bibinfo {author} {\bibfnamefont {W.}~\bibnamefont
  {Coffey}}\ and\ \bibinfo {author} {\bibfnamefont {Y.~P.}\ \bibnamefont
  {Kalmykov}},\ }\href@noop {} {\emph {\bibinfo {title} {The Langevin equation:
  with applications to stochastic problems in physics, chemistry and electrical
  engineering}}},\ Vol.~\bibinfo {volume} {27}\ (\bibinfo  {publisher} {World
  Scientific},\ \bibinfo {year} {2012})\BibitemShut {NoStop}%
\bibitem [{\citenamefont {Bouchaud}\ and\ \citenamefont
  {Cont}(1998)}]{bouchaud1998langevin}%
  \BibitemOpen
  \bibfield  {author} {\bibinfo {author} {\bibfnamefont {J.-P.}\ \bibnamefont
  {Bouchaud}}\ and\ \bibinfo {author} {\bibfnamefont {R.}~\bibnamefont
  {Cont}},\ }\bibfield  {title} {\bibinfo {title} {A langevin approach to stock
  market fluctuations and crashes},\ }\href@noop {} {\bibfield  {journal}
  {\bibinfo  {journal} {The European Physical Journal B-Condensed Matter and
  Complex Systems}\ }\textbf {\bibinfo {volume} {6}},\ \bibinfo {pages} {543}
  (\bibinfo {year} {1998})}\BibitemShut {NoStop}%
\bibitem [{\citenamefont {Jia}\ \emph {et~al.}(2001)\citenamefont {Jia},
  \citenamefont {Yu},\ and\ \citenamefont {Li}}]{jia2001effects}%
  \BibitemOpen
  \bibfield  {author} {\bibinfo {author} {\bibfnamefont {Y.}~\bibnamefont
  {Jia}}, \bibinfo {author} {\bibfnamefont {S.}~\bibnamefont {Yu}},\ and\
  \bibinfo {author} {\bibfnamefont {J.}~\bibnamefont {Li}},\ }\bibfield
  {title} {\bibinfo {title} {Effects of random potential on transport},\
  }\href@noop {} {\bibfield  {journal} {\bibinfo  {journal} {Physical Review
  E}\ }\textbf {\bibinfo {volume} {63}},\ \bibinfo {pages} {052101} (\bibinfo
  {year} {2001})}\BibitemShut {NoStop}%
\bibitem [{\citenamefont {Berdichevsky}\ and\ \citenamefont
  {Gitterman}(1997)}]{berdichevsky1997josephson}%
  \BibitemOpen
  \bibfield  {author} {\bibinfo {author} {\bibfnamefont {V.}~\bibnamefont
  {Berdichevsky}}\ and\ \bibinfo {author} {\bibfnamefont {M.}~\bibnamefont
  {Gitterman}},\ }\bibfield  {title} {\bibinfo {title} {Josephson junction with
  noise},\ }\href@noop {} {\bibfield  {journal} {\bibinfo  {journal} {Physical
  Review E}\ }\textbf {\bibinfo {volume} {56}},\ \bibinfo {pages} {6340}
  (\bibinfo {year} {1997})}\BibitemShut {NoStop}%
\bibitem [{\citenamefont {De~Santis}\ \emph {et~al.}(2022)\citenamefont
  {De~Santis}, \citenamefont {Guarcello}, \citenamefont {Spagnolo},
  \citenamefont {Carollo},\ and\ \citenamefont {Valenti}}]{de2022generation}%
  \BibitemOpen
  \bibfield  {author} {\bibinfo {author} {\bibfnamefont {D.}~\bibnamefont
  {De~Santis}}, \bibinfo {author} {\bibfnamefont {C.}~\bibnamefont
  {Guarcello}}, \bibinfo {author} {\bibfnamefont {B.}~\bibnamefont {Spagnolo}},
  \bibinfo {author} {\bibfnamefont {A.}~\bibnamefont {Carollo}},\ and\ \bibinfo
  {author} {\bibfnamefont {D.}~\bibnamefont {Valenti}},\ }\bibfield  {title}
  {\bibinfo {title} {Generation of travelling sine-gordon breathers in noisy
  long josephson junctions},\ }\href@noop {} {\bibfield  {journal} {\bibinfo
  {journal} {Chaos, Solitons \& Fractals}\ }\textbf {\bibinfo {volume} {158}},\
  \bibinfo {pages} {112039} (\bibinfo {year} {2022})}\BibitemShut {NoStop}%
\bibitem [{\citenamefont {Pountougnigni}\ \emph {et~al.}(2023)\citenamefont
  {Pountougnigni}, \citenamefont {Yamapi}, \citenamefont {Filatrella},\ and\
  \citenamefont {Tchawoua}}]{pountougnigni2023detection}%
  \BibitemOpen
  \bibfield  {author} {\bibinfo {author} {\bibfnamefont {O.~V.}\ \bibnamefont
  {Pountougnigni}}, \bibinfo {author} {\bibfnamefont {R.}~\bibnamefont
  {Yamapi}}, \bibinfo {author} {\bibfnamefont {G.}~\bibnamefont {Filatrella}},\
  and\ \bibinfo {author} {\bibfnamefont {C.}~\bibnamefont {Tchawoua}},\
  }\bibfield  {title} {\bibinfo {title} {Detection of colored noise correlation
  time with josephson junctions},\ }\href@noop {} {\bibfield  {journal}
  {\bibinfo  {journal} {Physica C: Superconductivity and its Applications}\
  }\textbf {\bibinfo {volume} {614}},\ \bibinfo {pages} {1354379} (\bibinfo
  {year} {2023})}\BibitemShut {NoStop}%
\bibitem [{\citenamefont {Wand}\ \emph {et~al.}(2024)\citenamefont {Wand},
  \citenamefont {Wiedemann}, \citenamefont {Harren},\ and\ \citenamefont
  {Kamps}}]{wand2024estimating}%
  \BibitemOpen
  \bibfield  {author} {\bibinfo {author} {\bibfnamefont {T.}~\bibnamefont
  {Wand}}, \bibinfo {author} {\bibfnamefont {T.}~\bibnamefont {Wiedemann}},
  \bibinfo {author} {\bibfnamefont {J.}~\bibnamefont {Harren}},\ and\ \bibinfo
  {author} {\bibfnamefont {O.}~\bibnamefont {Kamps}},\ }\bibfield  {title}
  {\bibinfo {title} {Estimating stable fixed points and langevin potentials for
  financial dynamics},\ }\href@noop {} {\bibfield  {journal} {\bibinfo
  {journal} {Physical Review E}\ }\textbf {\bibinfo {volume} {109}},\ \bibinfo
  {pages} {024226} (\bibinfo {year} {2024})}\BibitemShut {NoStop}%
\bibitem [{\citenamefont {Inui}\ \emph {et~al.}(1989)\citenamefont {Inui},
  \citenamefont {Littlewood},\ and\ \citenamefont
  {Coppersmith}}]{inui1989pinning}%
  \BibitemOpen
  \bibfield  {author} {\bibinfo {author} {\bibfnamefont {M.}~\bibnamefont
  {Inui}}, \bibinfo {author} {\bibfnamefont {P.}~\bibnamefont {Littlewood}},\
  and\ \bibinfo {author} {\bibfnamefont {S.}~\bibnamefont {Coppersmith}},\
  }\bibfield  {title} {\bibinfo {title} {Pinning and thermal fluctuations of a
  flux line in high-temperature superconductors},\ }\href@noop {} {\bibfield
  {journal} {\bibinfo  {journal} {Physical Review Letters}\ }\textbf {\bibinfo
  {volume} {63}},\ \bibinfo {pages} {2421} (\bibinfo {year}
  {1989})}\BibitemShut {NoStop}%
\bibitem [{\citenamefont {Soroka}\ \emph {et~al.}(2007)\citenamefont {Soroka},
  \citenamefont {Shklovskij},\ and\ \citenamefont {Huth}}]{soroka2007guiding}%
  \BibitemOpen
  \bibfield  {author} {\bibinfo {author} {\bibfnamefont {O.~K.}\ \bibnamefont
  {Soroka}}, \bibinfo {author} {\bibfnamefont {V.~A.}\ \bibnamefont
  {Shklovskij}},\ and\ \bibinfo {author} {\bibfnamefont {M.}~\bibnamefont
  {Huth}},\ }\bibfield  {title} {\bibinfo {title} {Guiding of vortices under
  competing isotropic and anisotropic pinning conditions: Theory and
  experiment},\ }\href@noop {} {\bibfield  {journal} {\bibinfo  {journal}
  {Physical Review B}\ }\textbf {\bibinfo {volume} {76}},\ \bibinfo {pages}
  {014504} (\bibinfo {year} {2007})}\BibitemShut {NoStop}%
\bibitem [{\citenamefont {H{\"a}nggi}\ \emph {et~al.}(1990)\citenamefont
  {H{\"a}nggi}, \citenamefont {Talkner},\ and\ \citenamefont
  {Borkovec}}]{hanggi1990reaction}%
  \BibitemOpen
  \bibfield  {author} {\bibinfo {author} {\bibfnamefont {P.}~\bibnamefont
  {H{\"a}nggi}}, \bibinfo {author} {\bibfnamefont {P.}~\bibnamefont
  {Talkner}},\ and\ \bibinfo {author} {\bibfnamefont {M.}~\bibnamefont
  {Borkovec}},\ }\bibfield  {title} {\bibinfo {title} {Reaction-rate theory:
  fifty years after kramers},\ }\href@noop {} {\bibfield  {journal} {\bibinfo
  {journal} {Reviews of Modern Physics}\ }\textbf {\bibinfo {volume} {62}},\
  \bibinfo {pages} {251} (\bibinfo {year} {1990})}\BibitemShut {NoStop}%
\bibitem [{\citenamefont {Reimann}(2002)}]{reimann2002brownian}%
  \BibitemOpen
  \bibfield  {author} {\bibinfo {author} {\bibfnamefont {P.}~\bibnamefont
  {Reimann}},\ }\bibfield  {title} {\bibinfo {title} {Brownian motors: noisy
  transport far from equilibrium},\ }\href@noop {} {\bibfield  {journal}
  {\bibinfo  {journal} {Physics Reports}\ }\textbf {\bibinfo {volume} {361}},\
  \bibinfo {pages} {57} (\bibinfo {year} {2002})}\BibitemShut {NoStop}%
\bibitem [{\citenamefont {Calzavarini}\ \emph {et~al.}(2009)\citenamefont
  {Calzavarini}, \citenamefont {Volk}, \citenamefont {Bourgoin}, \citenamefont
  {L{\'e}v{\^e}que}, \citenamefont {Pinton},\ and\ \citenamefont
  {Toschi}}]{Calzavarinibook}%
  \BibitemOpen
  \bibfield  {author} {\bibinfo {author} {\bibfnamefont {E.}~\bibnamefont
  {Calzavarini}}, \bibinfo {author} {\bibfnamefont {R.}~\bibnamefont {Volk}},
  \bibinfo {author} {\bibfnamefont {M.}~\bibnamefont {Bourgoin}}, \bibinfo
  {author} {\bibfnamefont {E.}~\bibnamefont {L{\'e}v{\^e}que}}, \bibinfo
  {author} {\bibfnamefont {J.-F.}\ \bibnamefont {Pinton}},\ and\ \bibinfo
  {author} {\bibfnamefont {F.}~\bibnamefont {Toschi}},\ }\bibfield  {title}
  {\bibinfo {title} {Effect of fax{\'e}n forces on acceleration statistics of
  material particles in turbulent flow},\ }in\ \href@noop {} {\emph {\bibinfo
  {booktitle} {Advances in Turbulence XII}}},\ \bibinfo {editor} {edited by\
  \bibinfo {editor} {\bibfnamefont {B.}~\bibnamefont {Eckhardt}}}\ (\bibinfo
  {publisher} {Springer Berlin Heidelberg},\ \bibinfo {year} {2009})\ pp.\
  \bibinfo {pages} {11--14}\BibitemShut {NoStop}%
\bibitem [{\citenamefont {Maxey}\ and\ \citenamefont
  {Riley}(1983)}]{maxey1983equation}%
  \BibitemOpen
  \bibfield  {author} {\bibinfo {author} {\bibfnamefont {M.~R.}\ \bibnamefont
  {Maxey}}\ and\ \bibinfo {author} {\bibfnamefont {J.~J.}\ \bibnamefont
  {Riley}},\ }\bibfield  {title} {\bibinfo {title} {Equation of motion for a
  small rigid sphere in a nonuniform flow},\ }\href@noop {} {\bibfield
  {journal} {\bibinfo  {journal} {Physics of Fluids}\ }\textbf {\bibinfo
  {volume} {26}},\ \bibinfo {pages} {883} (\bibinfo {year} {1983})}\BibitemShut
  {NoStop}%
\bibitem [{\citenamefont {Auton}\ \emph {et~al.}(1988)\citenamefont {Auton},
  \citenamefont {Hunt},\ and\ \citenamefont {Prud'Homme}}]{auton1988force}%
  \BibitemOpen
  \bibfield  {author} {\bibinfo {author} {\bibfnamefont {T.}~\bibnamefont
  {Auton}}, \bibinfo {author} {\bibfnamefont {J.}~\bibnamefont {Hunt}},\ and\
  \bibinfo {author} {\bibfnamefont {M.}~\bibnamefont {Prud'Homme}},\ }\bibfield
   {title} {\bibinfo {title} {The force exerted on a body in inviscid unsteady
  non-uniform rotational flow},\ }\href@noop {} {\bibfield  {journal} {\bibinfo
   {journal} {Journal of Fluid Mechanics}\ }\textbf {\bibinfo {volume} {197}},\
  \bibinfo {pages} {241} (\bibinfo {year} {1988})}\BibitemShut {NoStop}%
\bibitem [{\citenamefont {Biferale}\ \emph {et~al.}(2010)\citenamefont
  {Biferale}, \citenamefont {Scagliarini}, \citenamefont {Toschi} \emph
  {et~al.}}]{biferale2010measurement}%
  \BibitemOpen
  \bibfield  {author} {\bibinfo {author} {\bibfnamefont {L.}~\bibnamefont
  {Biferale}}, \bibinfo {author} {\bibfnamefont {A.}~\bibnamefont
  {Scagliarini}}, \bibinfo {author} {\bibfnamefont {F.}~\bibnamefont {Toschi}},
  \emph {et~al.},\ }\bibfield  {title} {\bibinfo {title} {On the measurement of
  vortex filament lifetime statistics in turbulence},\ }\href@noop {}
  {\bibfield  {journal} {\bibinfo  {journal} {Physics of Fluids}\ }\textbf
  {\bibinfo {volume} {22}} (\bibinfo {year} {2010})}\BibitemShut {NoStop}%
\bibitem [{\citenamefont {Toschi}\ and\ \citenamefont
  {Bodenschatz}(2009)}]{toschi2009lagrangian}%
  \BibitemOpen
  \bibfield  {author} {\bibinfo {author} {\bibfnamefont {F.}~\bibnamefont
  {Toschi}}\ and\ \bibinfo {author} {\bibfnamefont {E.}~\bibnamefont
  {Bodenschatz}},\ }\bibfield  {title} {\bibinfo {title} {Lagrangian properties
  of particles in turbulence},\ }\href@noop {} {\bibfield  {journal} {\bibinfo
  {journal} {Annual Review of Fluid Mechanics}\ }\textbf {\bibinfo {volume}
  {41}},\ \bibinfo {pages} {375} (\bibinfo {year} {2009})}\BibitemShut
  {NoStop}%
\bibitem [{\citenamefont {de~Wit}\ \emph {et~al.}(2024)\citenamefont {de~Wit},
  \citenamefont {Kunnen}, \citenamefont {Clercx},\ and\ \citenamefont
  {Toschi}}]{dewit2024}%
  \BibitemOpen
  \bibfield  {author} {\bibinfo {author} {\bibfnamefont {X.~M.}\ \bibnamefont
  {de~Wit}}, \bibinfo {author} {\bibfnamefont {R.~P.~J.}\ \bibnamefont
  {Kunnen}}, \bibinfo {author} {\bibfnamefont {H.~J.~H.}\ \bibnamefont
  {Clercx}},\ and\ \bibinfo {author} {\bibfnamefont {F.}~\bibnamefont
  {Toschi}},\ }\bibfield  {title} {\bibinfo {title} {Efficient point-based
  simulation of four-way coupled particles in turbulence at high number
  density},\ }\href@noop {} {\bibfield  {journal} {\bibinfo  {journal}
  {Physical Review E}\ }\textbf {\bibinfo {volume} {110}},\ \bibinfo {pages}
  {015301} (\bibinfo {year} {2024})}\BibitemShut {NoStop}%
\bibitem [{\citenamefont {Maxey}\ \emph {et~al.}(1994)\citenamefont {Maxey},
  \citenamefont {Chang},\ and\ \citenamefont {Wang}}]{maxey1994simulation}%
  \BibitemOpen
  \bibfield  {author} {\bibinfo {author} {\bibfnamefont {M.}~\bibnamefont
  {Maxey}}, \bibinfo {author} {\bibfnamefont {E.}~\bibnamefont {Chang}},\ and\
  \bibinfo {author} {\bibfnamefont {L.-P.}\ \bibnamefont {Wang}},\ }\bibfield
  {title} {\bibinfo {title} {Simulation of interactions between microbubbles
  and turbulent flows},\ }\href@noop {} {\bibfield  {journal} {\bibinfo
  {journal} {Applied Mechanics Reviews}\ }\textbf {\bibinfo {volume} {47}},\
  \bibinfo {pages} {S70} (\bibinfo {year} {1994})}\BibitemShut {NoStop}%
\bibitem [{\citenamefont {Wang}\ \emph {et~al.}(2019)\citenamefont {Wang},
  \citenamefont {Sierakowski},\ and\ \citenamefont
  {Prosperetti}}]{wang2019rotational}%
  \BibitemOpen
  \bibfield  {author} {\bibinfo {author} {\bibfnamefont {Y.}~\bibnamefont
  {Wang}}, \bibinfo {author} {\bibfnamefont {A.}~\bibnamefont {Sierakowski}},\
  and\ \bibinfo {author} {\bibfnamefont {A.}~\bibnamefont {Prosperetti}},\
  }\bibfield  {title} {\bibinfo {title} {Rotational dynamics of a particle in a
  turbulent stream},\ }\href@noop {} {\bibfield  {journal} {\bibinfo  {journal}
  {Physical Review Fluids}\ }\textbf {\bibinfo {volume} {4}},\ \bibinfo {pages}
  {064304} (\bibinfo {year} {2019})}\BibitemShut {NoStop}%
\bibitem [{\citenamefont {Zimmermann}\ \emph
  {et~al.}(2011{\natexlab{a}})\citenamefont {Zimmermann}, \citenamefont
  {Gasteuil}, \citenamefont {Bourgoin}, \citenamefont {Volk}, \citenamefont
  {Pumir}, \citenamefont {Pinton} \emph {et~al.}}]{zimmermann2011tracking}%
  \BibitemOpen
  \bibfield  {author} {\bibinfo {author} {\bibfnamefont {R.}~\bibnamefont
  {Zimmermann}}, \bibinfo {author} {\bibfnamefont {Y.}~\bibnamefont
  {Gasteuil}}, \bibinfo {author} {\bibfnamefont {M.}~\bibnamefont {Bourgoin}},
  \bibinfo {author} {\bibfnamefont {R.}~\bibnamefont {Volk}}, \bibinfo {author}
  {\bibfnamefont {A.}~\bibnamefont {Pumir}}, \bibinfo {author} {\bibfnamefont
  {J.-F.}\ \bibnamefont {Pinton}}, \emph {et~al.},\ }\bibfield  {title}
  {\bibinfo {title} {Tracking the dynamics of translation and absolute
  orientation of a sphere in a turbulent flow},\ }\href@noop {} {\bibfield
  {journal} {\bibinfo  {journal} {Review of Scientific Instruments}\ }\textbf
  {\bibinfo {volume} {82}} (\bibinfo {year} {2011}{\natexlab{a}})}\BibitemShut
  {NoStop}%
\bibitem [{\citenamefont {Ghira}\ \emph {et~al.}(2022)\citenamefont {Ghira},
  \citenamefont {Elsinga},\ and\ \citenamefont
  {Da~Silva}}]{ghira2022characteristics}%
  \BibitemOpen
  \bibfield  {author} {\bibinfo {author} {\bibfnamefont {A.}~\bibnamefont
  {Ghira}}, \bibinfo {author} {\bibfnamefont {G.}~\bibnamefont {Elsinga}},\
  and\ \bibinfo {author} {\bibfnamefont {C.}~\bibnamefont {Da~Silva}},\
  }\bibfield  {title} {\bibinfo {title} {Characteristics of the intense
  vorticity structures in isotropic turbulence at high reynolds numbers},\
  }\href@noop {} {\bibfield  {journal} {\bibinfo  {journal} {Physical Review
  Fluids}\ }\textbf {\bibinfo {volume} {7}},\ \bibinfo {pages} {104605}
  (\bibinfo {year} {2022})}\BibitemShut {NoStop}%
\bibitem [{\citenamefont {Kang}\ \emph {et~al.}(2007)\citenamefont {Kang},
  \citenamefont {Tanahashi},\ and\ \citenamefont
  {Miyauchi}}]{kang2007dynamics}%
  \BibitemOpen
  \bibfield  {author} {\bibinfo {author} {\bibfnamefont {S.-J.}\ \bibnamefont
  {Kang}}, \bibinfo {author} {\bibfnamefont {M.}~\bibnamefont {Tanahashi}},\
  and\ \bibinfo {author} {\bibfnamefont {T.}~\bibnamefont {Miyauchi}},\
  }\bibfield  {title} {\bibinfo {title} {Dynamics of fine scale eddy clusters
  in turbulent channel flows},\ }\href@noop {} {\bibfield  {journal} {\bibinfo
  {journal} {Journal of Turbulence}\ }\textbf {\bibinfo {volume} {8}},\
  \bibinfo {pages} {N52} (\bibinfo {year} {2007})}\BibitemShut {NoStop}%
\bibitem [{\citenamefont {Tanahashi}\ \emph {et~al.}(2001)\citenamefont
  {Tanahashi}, \citenamefont {Iwase},\ and\ \citenamefont
  {Miyauchi}}]{tanahashi2001appearance}%
  \BibitemOpen
  \bibfield  {author} {\bibinfo {author} {\bibfnamefont {M.}~\bibnamefont
  {Tanahashi}}, \bibinfo {author} {\bibfnamefont {S.}~\bibnamefont {Iwase}},\
  and\ \bibinfo {author} {\bibfnamefont {T.}~\bibnamefont {Miyauchi}},\
  }\bibfield  {title} {\bibinfo {title} {Appearance and alignment with strain
  rate of coherent fine scale eddies in turbulent mixing layer},\ }\href@noop
  {} {\bibfield  {journal} {\bibinfo  {journal} {Journal of Turbulence}\
  }\textbf {\bibinfo {volume} {2}},\ \bibinfo {pages} {006} (\bibinfo {year}
  {2001})}\BibitemShut {NoStop}%
\bibitem [{\citenamefont {da~Silva}\ \emph {et~al.}(2011)\citenamefont
  {da~Silva}, \citenamefont {Dos~Reis},\ and\ \citenamefont
  {Pereira}}]{da2011intense}%
  \BibitemOpen
  \bibfield  {author} {\bibinfo {author} {\bibfnamefont {C.~B.}\ \bibnamefont
  {da~Silva}}, \bibinfo {author} {\bibfnamefont {R.~J.}\ \bibnamefont
  {Dos~Reis}},\ and\ \bibinfo {author} {\bibfnamefont {J.~C.}\ \bibnamefont
  {Pereira}},\ }\bibfield  {title} {\bibinfo {title} {The intense vorticity
  structures near the turbulent/non-turbulent interface in a jet},\ }\href@noop
  {} {\bibfield  {journal} {\bibinfo  {journal} {Journal of Fluid Mechanics}\
  }\textbf {\bibinfo {volume} {685}},\ \bibinfo {pages} {165} (\bibinfo {year}
  {2011})}\BibitemShut {NoStop}%
\bibitem [{\citenamefont {Ganapathisubramani}\ \emph
  {et~al.}(2008)\citenamefont {Ganapathisubramani}, \citenamefont
  {Lakshminarasimhan},\ and\ \citenamefont
  {Clemens}}]{ganapathisubramani2008investigation}%
  \BibitemOpen
  \bibfield  {author} {\bibinfo {author} {\bibfnamefont {B.}~\bibnamefont
  {Ganapathisubramani}}, \bibinfo {author} {\bibfnamefont {K.}~\bibnamefont
  {Lakshminarasimhan}},\ and\ \bibinfo {author} {\bibfnamefont
  {N.}~\bibnamefont {Clemens}},\ }\bibfield  {title} {\bibinfo {title}
  {Investigation of three-dimensional structure of fine scales in a turbulent
  jet by using cinematographic stereoscopic particle image velocimetry},\
  }\href@noop {} {\bibfield  {journal} {\bibinfo  {journal} {Journal of Fluid
  Mechanics}\ }\textbf {\bibinfo {volume} {598}},\ \bibinfo {pages} {141}
  (\bibinfo {year} {2008})}\BibitemShut {NoStop}%
\bibitem [{\citenamefont {Mordant}(1997)}]{Mordant1997}%
  \BibitemOpen
  \bibfield  {author} {\bibinfo {author} {\bibfnamefont {N.}~\bibnamefont
  {Mordant}},\ }\bibfield  {title} {\bibinfo {title} {Characterization of
  turbulence in a closed flow},\ }\href {https://doi.org/10.1051/jp2:1997212}
  {\bibfield  {journal} {\bibinfo  {journal} {J. Phys. II France}\ }\textbf
  {\bibinfo {volume} {7}},\ \bibinfo {pages} {1729} (\bibinfo {year}
  {1997})}\BibitemShut {NoStop}%
\bibitem [{\citenamefont {Labbé}\ \emph {et~al.}(1996)\citenamefont {Labbé},
  \citenamefont {Pinton},\ and\ \citenamefont {Fauve}}]{Labbe1996}%
  \BibitemOpen
  \bibfield  {author} {\bibinfo {author} {\bibfnamefont {R.}~\bibnamefont
  {Labbé}}, \bibinfo {author} {\bibfnamefont {J.-F.}\ \bibnamefont {Pinton}},\
  and\ \bibinfo {author} {\bibfnamefont {S.}~\bibnamefont {Fauve}},\ }\bibfield
   {title} {\bibinfo {title} {Study of the von kármán flow between coaxial
  corotating disks},\ }\href {https://doi.org/10.1063/1.868871} {\bibfield
  {journal} {\bibinfo  {journal} {Phys. Fluids}\ }\textbf {\bibinfo {volume}
  {8}},\ \bibinfo {pages} {914} (\bibinfo {year} {1996})}\BibitemShut {NoStop}%
\bibitem [{\citenamefont {Balachandar}\ and\ \citenamefont
  {Eaton}(2010)}]{balachandar2010turbulent}%
  \BibitemOpen
  \bibfield  {author} {\bibinfo {author} {\bibfnamefont {S.}~\bibnamefont
  {Balachandar}}\ and\ \bibinfo {author} {\bibfnamefont {J.~K.}\ \bibnamefont
  {Eaton}},\ }\bibfield  {title} {\bibinfo {title} {Turbulent dispersed
  multiphase flow},\ }\href@noop {} {\bibfield  {journal} {\bibinfo  {journal}
  {Annual review of fluid mechanics}\ }\textbf {\bibinfo {volume} {42}},\
  \bibinfo {pages} {111} (\bibinfo {year} {2010})}\BibitemShut {NoStop}%
\bibitem [{\citenamefont {Zimmermann}\ \emph
  {et~al.}(2011{\natexlab{b}})\citenamefont {Zimmermann}, \citenamefont
  {Gasteuil}, \citenamefont {Bourgoin}, \citenamefont {Volk}, \citenamefont
  {Pumir},\ and\ \citenamefont {Pinton}}]{zimmermann2011rotational}%
  \BibitemOpen
  \bibfield  {author} {\bibinfo {author} {\bibfnamefont {R.}~\bibnamefont
  {Zimmermann}}, \bibinfo {author} {\bibfnamefont {Y.}~\bibnamefont
  {Gasteuil}}, \bibinfo {author} {\bibfnamefont {M.}~\bibnamefont {Bourgoin}},
  \bibinfo {author} {\bibfnamefont {R.}~\bibnamefont {Volk}}, \bibinfo {author}
  {\bibfnamefont {A.}~\bibnamefont {Pumir}},\ and\ \bibinfo {author}
  {\bibfnamefont {J.-F.}\ \bibnamefont {Pinton}},\ }\bibfield  {title}
  {\bibinfo {title} {Rotational intermittency and turbulence induced lift
  experienced by large particles in a turbulent flow},\ }\href@noop {}
  {\bibfield  {journal} {\bibinfo  {journal} {Physical Review Letters}\
  }\textbf {\bibinfo {volume} {106}},\ \bibinfo {pages} {154501} (\bibinfo
  {year} {2011}{\natexlab{b}})}\BibitemShut {NoStop}%
\bibitem [{\citenamefont {Homann}\ and\ \citenamefont
  {Bec}(2010)}]{homann2010finite}%
  \BibitemOpen
  \bibfield  {author} {\bibinfo {author} {\bibfnamefont {H.}~\bibnamefont
  {Homann}}\ and\ \bibinfo {author} {\bibfnamefont {J.}~\bibnamefont {Bec}},\
  }\bibfield  {title} {\bibinfo {title} {Finite-size effects in the dynamics of
  neutrally buoyant particles in turbulent flow},\ }\href@noop {} {\bibfield
  {journal} {\bibinfo  {journal} {Journal of Fluid Mechanics}\ }\textbf
  {\bibinfo {volume} {651}},\ \bibinfo {pages} {81} (\bibinfo {year}
  {2010})}\BibitemShut {NoStop}%
\end{thebibliography}%

\end{document}